\documentclass[aps,prx,twocolumn,nofootinbib,10pt]{revtex4-2}

\usepackage{amsmath,amssymb,mathtools,bm}
\usepackage{graphicx}
\usepackage{tikz}
\usepackage{physics}
\usepackage[colorlinks=true,linkcolor=blue,citecolor=blue,urlcolor=blue]{hyperref}

\renewcommand{\Tr}{\operatorname{Tr}}
\newcommand{\sgn}{\operatorname{sgn}}
\renewcommand{\dd}{\mathrm{d}}
\newcommand{\ii}{\mathrm{i}}
\newcommand{\kk}{\mathbf{k}}
\renewcommand{\acomm}[2]{\left\{#1,#2\right\}}
\renewcommand{\comm}[2]{\left[#1,#2\right]}
\newcommand{\nablamixture}{\nabla^{(m)}}
\newcommand{\nablaexp}{\nabla^{(e)}}

\newcommand{\Id}{\mathbb{I}}

\begin{document}

\title{Multi-State Geometry of Density Matrices and Rectification Sum Rules}
\author{Barry Bradlyn}
\email{bbradlyn@illinois.edu}
\affiliation{Department of Physics, University of Illinois Urbana-Champaign, Urbana IL 61801, USA}
\affiliation{Anthony J.~Leggett Institute for Condensed Matter Theory, University of Illinois Urbana-Champaign, Urbana IL 61801, USA}
\date{\today}

\begin{abstract}
The geometry of quantum states has emerged as a key ingredient in understanding the linear and nonlinear responses of quantum materials. To date, however, the connection between geometry and nonlinear response is best understood for clean, noninteracting systems at zero temperature. In this work, we develop a theory of multi-state geometry for density matrices and use it to derive sum rules for second-order rectification, making no assumptions about the strength of disorder or interactions. We first show that perturbation theory for thermal density matrices gives rise to two dual information-theoretic connections and an almost complex structure. We introduce a complex, quantum generalization of the Amari-Chentsov tensor of classical information theory, the cQAC tensor, which captures the multi-state geometry of the perturbed density matrix. We derive a zero-temperature sum rule for the frequency-integrated DC rectification response of an insulator as a difference between a ground state third cumulant and the complex distortion tensor, a multi-state geometric quantity built from the cQAC tensor. This generalizes known single-particle sum rules for the shift and nonlinear Hall currents to many-body systems and general perturbations. Specializing to the shift current, we resolve the geometric contribution for multiband insulators into particle-like and hole-like terms. We verify the sum rule numerically in a generalized Kane-Mele model, finding that the geometric contribution can dominate the integrated response. Finally, we show that although the splitting of the sum rule into cumulant and geometric contributions does not survive at nonzero temperature, the measured sum rule for insulators differs from its zero-temperature form by corrections exponentially small in the gap, allowing low-temperature rectification measurements to probe the multi-state geometry of insulators.
\end{abstract}

\maketitle

\section{Introduction}
\label{sec:introduction}

Recent developments in condensed matter physics have highlighted the important role that the topology and geometry of electronic states can play in determining material properties~\cite{hasan2010colloquium}. The clearest example of this is in topological insulators, where the Berry phases and curvature of the ground state wavefunctions determine exotic observable properties such as quantized Hall conductance and the existence of gapless edge states. Even beyond topology, the modern theory of polarization demonstrated that the geometry of quantum states can determine observable properties like electronic polarization and localizability~\cite{kingsmith1993theory,resta1994macroscopic,marzari1997maximally}. Building on these works, geometry (primarily as encoded in the quantum metric and quantum geometric tensor) has been shown to govern sum rules for optical conductivity and other linear response functions~\cite{souza2000polarization,onishi2025quantum,verma2024instantaneous,balut2025quantum,balut2026fundamental,ji2025density,chowdhury2026information,mao2025lowenergy,mendez2024low}, superfluidity in flat-band systems~\cite{peotta2015superfluidity,liang2017band,torma2022superconductivity,herzog2022superfluid,huhtinen2022revisiting}, and the tendency towards the formation of fractionalized phases~\cite{roy2014band,wang2021exact,neupert2015fractional,ledwith2023vortexability,yu2024quantum}.

For noninteracting electrons, there has been further evidence that quantum geometry may play a role in determining nonlinear optical and photogalvanic responses as well. Studies of circular photocurrents in topological semimetals~\cite{dejuan2017quantized,flicker2018chiral,morimoto2016topological,rees2020helicitydependent,kaushik2021magnetic,ni2020linear,chang2017unconventional,ahn2020lowfrequency,parker2019diagrammatic} first hinted that the shift and injection currents in noninteracting electron systems~\cite{aversa1995nonlinear,sipe2000second} could have a geometric interpretation. Parallel work also provided a new geometric lens on nonlinear Hall effects in terms of the single-particle Berry curvature distribution~\cite{sodemann2015quantum,matsyshyn2019nonlinear,morimoto2016semiclassical,du2021nonlinear,ortix2021nonlinear,nandy2019symmetry}. A unifying framework for all of these effects was provided by Ref.~\cite{ahn2022riemannian}, which showed that the nonlinear optical responses of clean free fermion systems could be organized within the language of complex Riemannian geometry of the Bloch bands. Several subsequent works have furthered the study of nonlinear optical responses from the geometric perspective~\cite{bouhon2023quantum,jankowski2024quantized,alexandradinata2024quantization,zhu2024anomalous,mitscherling2024gauge,avdoshkin2023extrinsic,avdoshkin2024multi}. In particular, Ref.~\cite{avdoshkin2024multi} derived a sum rule relating the shift current response of clean, noninteracting systems to a multi-state geometric quantity involving both the occupied and unoccupied bands.

These geometric developments raise several questions. First, it seems natural to extend this framework to other nonlinear response functions beyond optics. While recent work has taken some steps in this direction~\cite{jain2026topological,jain2026nonlinear,guan2026exploring}, a general framework is still lacking. Second, the connection between quantum geometry and nonlinear optical response is most clear for clean, noninteracting systems at zero temperature. It is known, however, that scattering from disorder and interactions can have a large effect on measured photocurrents~\cite{yu2026photogalvanic,sukhachov2026vortex}. This motivates the need for a geometric theory of photogalvanic effects that goes beyond the single-particle picture. 

In this work, we take steps to address these questions by examining the connection between the quantum geometry of density matrices and nonlinear response theory. Building on the approaches of Refs.~\cite{guan2026exploring,ji2025density,hetenyi2023fluctuations}, we consider the complex geometry of the space of density matrices subject to time-dependent perturbations. We formulate a notion of multi-state geometry for these spaces, generalizing the single-particle approach of Refs.~\cite{mitscherling2024gauge,avdoshkin2023extrinsic}. We derive a general Kubo formula for second-order DC rectification response and show how the multi-state geometry determines a sum rule for the rectification response. Crucially, our approach makes no assumptions about the strength of disorder or interactions in the system. Turning to the shift current specifically, we will show how our formalism generalizes known sum rules for the DC photocurrent. Finally, we examine the fate of the quantum geometric sum rule at nonzero temperature.

The structure of the paper is as follows: First, in Sec.~\ref{sub:guide_to_the_main_results}, we present a roadmap to our main results. Then, in Sec.~\ref{sec:multi_state_geometry} we will develop the theory of multi-state geometry for time-dependent, many-body density matrices. 
We generalize the approach of Refs.~\cite{guan2026exploring,ji2025density} to show that two natural information-theoretic connections and an (almost) complex structure emerge from time-dependent perturbation theory for thermal density matrices. 
We introduce a complex, quantum generalization of the Amari-Chentsov tensor of classical information theory~\cite{amari2000methods} that captures the multi-state geometry of these connections.
Next, in Sec.~\ref{sec:rectification} we will show using second-order response theory that the multi-state geometry encoded in the complex quantum Amari-Chentsov tensor (cQAC) can be measured through DC rectification response in insulators. 
In particular, we derive a zero-temperature sum rule for general rectification response that expresses the positive-frequency integral of the response rate for general currents as the difference between a ground state third cumulant and a complex distortion tensor built from the cQAC tensor. 
This generalizes the single-particle shift and nonlinear Hall sum rules of Refs.~\cite{avdoshkin2024multi,ahn2020lowfrequency,ahn2022riemannian,matsyshyn2019nonlinear} to general rectification response in many-body systems. 
To demonstrate this, we apply our results to the shift current response of electronic systems in the length gauge (with open boundary conditions) in Sec.~\ref{sec:application_shift_current_in_bounded_systems}.
We then specialize to free fermion systems in Sec.~\ref{sec:application_to_free_fermions}, taking the thermodynamic limit in the length gauge.
We recover the sum rule of Ref.~\cite{avdoshkin2024multi} for the case of insulators with one occupied band. 
For more general multiband insulators we resolve the additional correction terms in the sum rule from the complex distortion tensor into particle-like and hole-like contributions, with the hole-like contribution organizing the multiple-occupied-band corrections contained in the general formulas of Ref.~\cite{avdoshkin2024multi}.  
We illustrate this by computing the shift current response in a generalized Kane-Mele model with perturbations that break mirror and time-reversal symmetry. We verify the sum rule numerically and show that the complex distortion tensor can be the dominant contribution to the integrated response. 
Finally, in Sec.~\ref{sec:nonzero_T} we examine the fate of the geometric sum rule for insulators at nonzero temperature. 
We show that the zero-temperature splitting of the sum rule into cumulant and geometric contributions does not survive at nonzero temperature, as the complex distortion tensor acquires a temperature dependence that cannot be extracted from the DC response alone. Nevertheless, the measured sum rule differs from its zero-temperature value by corrections that are exponentially small in the gap, allowing for low-temperature experiments to probe the zero-temperature quantum geometry. 
To conclude, in Sec.~\ref{sec:conclusion} we discuss the implications of our formalism for using quantum geometry as a probe and tuning knob for responses in materials.

Note that throughout this work, we will use natural units $\hbar=c=e=1$. Additionally, we will use the Einstein summation convention for repeated indices unless otherwise specified. 

\subsection{Guide to the main results}\label{sub:guide_to_the_main_results}
Here we present an overview and guide to our main physical results. We start by introducing the mathematical framework for computing the geometry associated to perturbed density matrices, which culminates in our introduction of the complex quantum Amari-Chentsov tensor (cQAC) $Z_{\sigma;\mu\nu}$ in Eq.~\eqref{eq:cubic-z-main}. We show in Eq.~\eqref{eq:cubic-t-phi-main} that the real part of $Z_{\sigma;\mu\nu}$ is the quantum Amari-Chentsov tensor of Refs.~\cite{hasegawa1993alpha,hasegawa1997noncommutative,nagaoka1995differential,amari2000methods} that represents the difference between natural connections on the space of density matrices. Additionally, the imaginary part of $Z_{\sigma;\mu\nu}$ can be related to the dipole of Uhlmann curvature. Combining these results, we show in Eq.~\eqref{eq:z-covariant-derivative-main} that the cQAC tensor is a covariant derivative of a generalization of the quantum geometric tensor to the space of density matrices. By introducing an almost complex structure on the space of perturbed density matrices, we introduce the complex distortion tensor in Eq.~\eqref{eq:same-order-to-response-main} defined in terms of the cQAC tensor. In Eq.~\eqref{eq:ordered-s-main} we show that the zero-temperature limit of the complex distortion tensor captures the many-body multi-state geometry of the density matrix. In particular, by carefully treating the zero-temperature limit we show how the complex distortion tensor and the many-body multi-state geometry depend not just on the properties of the ground state, but also on virtual excited states, in analogy with how single-particle multi-state geometry depends on both occupied and unoccupied Bloch bands.

Using this framework combined with nonlinear response theory, we derive our main results Eqs.~\eqref{eq:primitive-c3-s-main} and \eqref{eq:primitive-absorptive-sumrule-main}: zero-temperature sum rules relating the frequency integral of the DC rectification response in insulators to the difference of a ground state cumulant and the complex distortion tensor. Applying our result to photocurrents in many-body systems in the length gauge, we derive in Eq.~\eqref{eq:obc-sum-rule-main} a sum rule for the zero-temperature shift current in gapped systems in terms of the third cumulant of the position operator and the complex distortion tensor associated to position perturbations. We show in Eq.~\eqref{eq:population-polarization-inst} how the total accumulated polarization after an instantaneous pulse measures the symmetric part of the sum rule. In Eqs.~\eqref{eq:ff-sum-rule-main} and \eqref{eq:ff-s-main} we specialize our sum rule to free fermion systems, showing how our approach generalizes the results of Ref.~\cite{avdoshkin2024multi}. We apply our formalism to compute the shift current and multi-state geometry in a generalized Kane-Mele model in Figs.~\ref{fig:km-bands-conductivity} and \ref{fig:km-sumrule}. Finally, in Eqs.~\eqref{eq:thermal-lehmann-main} and \eqref{eq:s-thermal-main} we show how the correspondence between the sum rule and multi-state geometry breaks down for insulators at nonzero temperature.

\section{Multi-state geometry}\label{sec:multi_state_geometry}

In this section we introduce the multi-state geometric framework for general density matrices. First, in Sec.~\ref{subsec:bures} we examine the quantum geometric tensor defined from unitary perturbations to an equilibrium density matrix. Then, in Sec.~\ref{sub:dual_connections_and_the_quantum_amari_chentsov_tensor} we examine two natural affine connections on this space and introduce the complex quantum Amari-Chentsov tensor that describes the covariant derivative of the quantum geometric tensor. Finally, in Sec.~\ref{sub:the_almost_complex_structure_and_conjugated_tensors} we introduce an almost complex structure on the space of perturbed density matrices which allows us to define the complex distortion tensor that will play a key role in second-order rectification. In Table~\ref{tab:geometry-summary} we give a summary of the key geometric quantities.

\subsection{Bures geometry of density matrix orbits}\label{subsec:bures}

Following Ref.~\cite{guan2026exploring}, let $\rho_0$ be the equilibrium density matrix for our system of interest. We consider perturbing our Hamiltonian with some (either time-dependent or quasistatic) external fields $\lambda^\mu$. From time-dependent perturbation theory, this causes the density matrix to evolve according to
\begin{equation}
    \rho(\lambda)=U(\lambda)\rho_0U^\dagger(\lambda),
    \label{eq:orbit-param-main}
\end{equation}
where $U(\lambda)$ is a unitary evolution operator.
To define the quantum geometric structures associated with this evolution, we can work perturbatively in the limit of small $\lambda^\mu$. We first introduce the shorthand
\begin{equation}
\begin{aligned}
    \partial_\mu&=\frac{\partial}{\partial\lambda^\mu},
    \\
    \rho_\mu&=\partial_\mu\rho,
    \\
    \rho_{\mu\nu}&=\partial_\mu\partial_\nu\rho ,
\end{aligned}
    \label{eq:coordinate-conventions-main}
\end{equation}
for variations of $\rho$ with respect to the parameters $\lambda^\mu$. Note that for time-dependent parameters, $\partial_\mu$ should be interpreted as a variational derivative with respect to $\lambda(t)^\mu$ at a particular time. By differentiating Eq.~\eqref{eq:orbit-param-main} we can define the Hermitian evolution generator
\begin{equation}
\begin{aligned}
    \mathcal A_\mu&=\ii(\partial_\mu U)U^\dagger,
    \\
    \mathcal A_\mu^\dagger&=\mathcal A_\mu,
    \\
    \rho_\mu&=-\ii\comm{\mathcal A_\mu}{\rho}.
\end{aligned}
    \label{eq:orbit-generator-main}
\end{equation}
The generators $\mathcal{A}_\mu$ are basis tangent vectors to the orbit $\rho(\lambda)$. A general tangent vector $u$ can be defined through its action
\begin{equation}
\begin{aligned}
    u(\rho)&=u^\mu\rho_\mu=-\ii\comm{\mathcal{A}_u}{\rho},
    \\
    \mathcal A_u&=u^\mu \mathcal A_\mu,
\end{aligned}
    \label{eq:vector-shorthand-main}
\end{equation}
on $\rho(\lambda)$. 

We can view the action Eq.~\eqref{eq:vector-shorthand-main} of tangent vectors on the density matrix orbit from the point of view of purifications, which leads directly to the notion of Bures geometry. Let $W(\lambda)$ be a purification of the density matrix $\rho(\lambda)$, such that
\begin{equation}
\rho(\lambda) = W(\lambda) W^\dag(\lambda). \label{eq:purification}
\end{equation}
Tangent vectors $u$ to the orbit $\rho(\lambda)$ lift to the purification as
\begin{align}
G_u&= u^\mu G_\mu \\
G_\mu & = G_{\partial_\mu}, \label{eq:lift-component-shorthand-main}
\end{align}
such that 
\begin{equation}
u(W) = G_u W, \label{eq:horizontal}
\end{equation}
with $G_\mu$ Hermitian. Hermiticity ensures that $G_\mu$ are \emph{horizontal} tangent vectors, i.e. that they generate minimal variation of $W$ along directions that do not change $\rho(\lambda)$. By differentiating Eq.~\eqref{eq:purification} and using Eq.~\eqref{eq:horizontal}, we find that the action of a tangent vector $u$ on $\rho$ is given by the Lyapunov equation 
\begin{equation}
    u(\rho)=\acomm{G_u}{\rho},
    \label{eq:sld-lyapunov}
\end{equation}
which defines $G_\mu$ as (half) the symmetric logarithmic derivative (SLD) of $\rho$.
By taking matrix elements of Eq.~\eqref{eq:sld-lyapunov} in the eigenbasis $\{\ket{n}\}$ of $\rho$, we can define the matrix elements of $(G_u)_{mn}\equiv \bra{m}G_u\ket{n}$ as~\cite{braunstein1994statistical} 
\begin{equation}
    (G_u)_{mn}
    =
    \begin{cases}
    \dfrac{u(\rho)_{mn}}{p_m+p_n}, & p_m+p_n\neq 0,\\[1.1em]
    0, & p_m+p_n=0.
    \end{cases}
    \label{eq:sld-spectral}
\end{equation}
Importantly, the exclusion of matrix elements when $p_n+p_m=0$ ensures that $G_u$ is well-defined even for density matrices without full rank. Combining Eq.~\eqref{eq:sld-spectral} with the definition Eq.~\eqref{eq:vector-shorthand-main} of the orbit generators, we can also write $G_u$ as
\begin{equation}
    (G_u)_{mn}
    =
    -\ii\,\frac{p_n-p_m}{p_m+p_n}(\mathcal A_u)_{mn},
    \qquad p_m+p_n\neq 0 .
    \label{eq:sld-orbit-lift}
\end{equation}
Note that, for thermal density matrices, taking the limit of Eq.~\eqref{eq:sld-orbit-lift} as temperature $T\rightarrow 0$ need not in general coincide with setting $T=0$ directly in the definition. This is because for a $T=0$ pure state, Eq.~\eqref{eq:sld-lyapunov} does not uniquely define the matrix elements of $G_u$ between excited states. In this work, we will always define these off-diagonal matrix elements through the $T\rightarrow 0$ limit of a thermal state when needed.

Using the SLD generators $G_u$, we can define geometric structures on the space of density matrices. Generalizing previous works for pure states~\cite{provost1980riemannian,yu2024quantum,verma2025quantum}, we can use the projection Eq.~\eqref{eq:purification} from purifications to density matrices to define a quantum geometric tensor (QGT) operator. Consider the projection
\begin{align}
[u(W)][u(W)]^\dag &= u^\mu u^\nu G_\mu W W^\dag G_\nu
\equiv u^\mu u^\nu Q_{\mu\nu},
\end{align}
where we have defined
\begin{equation}
Q_{\mu\nu}=G_\mu\rho G_\nu.\label{eq:operator-qgt-main}
\end{equation}
$Q_{\mu\nu}$ is the operator generalization of the Bures metric and Uhlmann curvature~\cite{Bengtsson_Zyczkowski_2006,uhlmann1992metric,uhlmann1995geometric,carollo2020geometry}. Indeed, taking the trace of both sides of Eq.~\eqref{eq:operator-qgt-main} yields 
\begin{equation}
    \Tr(Q_{\mu\nu})=g_{\mu\nu}-\ii\mathcal F_{\mu\nu},
    \label{eq:operator-qgt-trace-main}
\end{equation}
where $g_{\mu\nu}\equiv g(\partial_\mu, \partial_\nu)$ are the components of the Bures metric
\begin{equation}
    g(u,v)
    =
    \frac{1}{2}\Tr\!\left(\rho\acomm{G_u}{G_v}\right),
    \label{eq:bures-main}
\end{equation}
and $\mathcal F_{\mu\nu}\equiv \mathcal F(\partial_\mu,\partial_\nu)$ are the components of the Uhlmann curvature
\begin{equation}
    \mathcal F(u,v)
    =
    -\frac{\ii}{2}\Tr\!\left(\rho\comm{G_u}{G_v}\right).
    \label{eq:curvature-form-main}
\end{equation}
This shows that $Q_{\mu\nu}$ reproduces the ordinary quantum geometry of density matrices through its trace, while retaining multi-state geometric data through its off-diagonal matrix elements. By analogy with the pure state case~\cite{yu2024quantum}, in what follows we will refer to $Q_{\mu\nu}$ as the quantum geometric tensor (QGT) operator and $\Tr Q_{\mu\nu}$ as the quantum geometric tensor.
In the remainder of this work, we will use the Lyapunov equation \eqref{eq:sld-lyapunov} to study the geometric objects that can be built from (covariant) derivatives of the QGT operator $Q_{\mu\nu}$.

\subsection{Dual connections and the quantum Amari-Chentsov tensor}\label{sub:dual_connections_and_the_quantum_amari_chentsov_tensor}

In the previous section, we focused on the geometric objects that could be defined from first derivatives of the density matrix $\rho(\lambda)$ in Eq.~\eqref{eq:orbit-param-main}. To construct the geometric objects that will govern second-order response, we will need to look at second derivatives of the density matrix, and hence derivatives of tangent vectors to the orbit. To do so, we must specify an affine connection on the space of tangent vectors. Following Refs.~\cite{amari2000methods,nagaoka1995differential}, we will define the connections by specifying how tangent vectors are parallel transported. 

There are two natural choices of connection given the geometric structures we defined in Sec.~\ref{subsec:bures}. First, we can consider the case where we treat the tangent vectors $u(\rho)$ as flat. This defines the mixture connection
\begin{equation}
    (\nablamixture_u v)(\rho)=u\!\left(v(\rho)\right).
    \label{eq:mixture-main}
\end{equation}
From Eq.~\eqref{eq:mixture-main}, we see that geodesics of the mixture connection are given by mixtures $\rho(t) = (1-t)\rho_0+t \rho_1$ that depend only linearly on parameters; since the second derivatives of $\rho(t)$ vanish, $\nablamixture_{\dot\rho}\dot\rho=0$. 

Alternatively, we can consider an affine connection under which exponential families $\rho(t) = e^{-t H}/\mathcal{Z}(t)$ with constant generators $H$ are flat. This defines the exponential connection
\begin{equation}
    (\nablaexp_u v)(\rho)=\acomm{u(G_v)+2g(u,v)\Id}{\rho}.
    \label{eq:exponential-main}
\end{equation}
Note that the term proportional to the identity in Eq.~\eqref{eq:exponential-main} subtracts out the average of $u(G_v)$, ensuring that 
\begin{equation}
G_{\nablaexp_u v}=u(G_v)+2g(u,v)\Id \label{eq:g-from-exp}
\end{equation}
is a well-defined SLD generator per Eq.~\eqref{eq:sld-lyapunov}. For $\rho(t) = e^{-t H}/\mathcal{Z}(t)$, we can verify that $\dot{G}_{\dot{\rho}} = -1/2(d^2\log \mathcal{Z} / dt^2)\,\Id = -\Tr\!\left(G_{\dot\rho}\dot\rho\right)\Id,$ and so $\nablaexp_{\dot\rho}\dot\rho=0$.

Note that the connections $\nabla^{(e)}$ and $\nabla^{(m)}$ are defined on the ambient space of all density matrices. However, for the purposes of this work we will focus on the orbits Eq.~\eqref{eq:orbit-param-main} parameterized by external fields. We will thus focus on the induced connections on this restricted space. This corresponds to discarding the components of covariant derivatives of tangent vectors that are normal to the orbits, and thus restricting our attention to the intrinsic geometry of the space of orbits. For the remainder of this work, we will only refer to these induced geometric structures.

The (induced) mixture and exponential connection form a dual pair with respect to the Bures metric. To see this first, note from Eq.~\eqref{eq:bures-main} that
\begin{align}
g(u,v) &= \frac{1}{2}\Tr\!\left(\rho\acomm{G_u}{G_v}\right),\nonumber \\
&=\frac{1}{2}\Tr\!\left(\acomm{G_u}{\rho}G_v\right),\nonumber \\
&=\frac{1}{2}\Tr\!\left(u(\rho)G_v\right),\label{eq:metric-rewrite}
\end{align}
where in the last line we used Eq.~\eqref{eq:sld-lyapunov}. Using Eq.~\eqref{eq:metric-rewrite}, we can define the connection coefficients (of the first kind) associated to any affine connection as
\begin{equation}
    g(\nabla_u v,w)=\frac{1}{2}\Tr(G_w(\nabla_u v)(\rho) ).
    \label{eq:first-kind-main}
\end{equation}
For the exponential connection, we directly find the connection coefficients
\begin{align}
    g(\nablaexp_u v,w)
    &=
    \frac{1}{2}\Tr\!\left(
        \rho\acomm{u(G_v)}{G_w}
    \right) \nonumber \\
    &=\frac{1}{2}\Tr\!\left(w(\rho) u(G_v)\right).
    \label{eq:exp-first-kind-main}
\end{align}
On the other hand, for the mixture connection we find
\begin{equation}
g(\nablamixture_u v, w) = \frac{1}{2}\Tr\!\left(G_w u(v(\rho))\right). \label{eq:mixture-main-simple}
\end{equation}
Combining Eqs.~\eqref{eq:exp-first-kind-main} and \eqref{eq:mixture-main-simple}, we find that derivatives of the Bures metric can be expressed in terms of the duality relation
\begin{align}
u(g(v,w)) &=\frac{1}{2}\Tr\!\left(u(v(\rho))G_w\right) + \frac{1}{2}\Tr\!\left(v(\rho)u(G_w)\right) \nonumber \\
&= g(\nablamixture_u v, w) + g(v,\nablaexp_u  w). \label{eq:dual-product-main}
\end{align}

Going further, we can express the mixture connection coefficients in terms of the exponential connection coefficients. We first expand
\begin{align}
    u\!\left(v(\rho)\right)
    &=
    u(G_v)\rho+\rho u(G_v)
    +G_vu(\rho)+u(\rho)G_v \nonumber\\
    &=
    \acomm{u(G_v)}{\rho}
    +G_vG_u\rho
    \nonumber\\
    &\quad
    +G_v\rho G_u
    +G_u\rho G_v+\rho G_uG_v .
    \label{eq:mixture-expanded-main}
\end{align}
Inserting this into Eq.~\eqref{eq:mixture-main-simple} and using Eq.~\eqref{eq:exp-first-kind-main} we find
\begin{equation}
    g(\nablamixture_u v,w)
    =
    g(\nablaexp_u v,w)+T(u,v,w),
    \label{eq:dual-difference-main}
\end{equation}
where $T(u,v,w)$ is given by
\begin{equation}
    T(u,v,w)
    =
    \frac{1}{2}\Tr\!\left(
        \rho\acomm{G_u}{\acomm{G_v}{G_w}}
    \right).
    \label{eq:qac-main}
\end{equation}
Since $T$ is the difference between two connections, it transforms like a tensor. It is a quantum generalization of the Amari-Chentsov tensor of classical information theory. We will thus call $T$ the quantum Amari-Chentsov (QAC) tensor. In classical information theory, the Amari-Chentsov tensor is an invariant that describes the skewness of the log-likelihood of a probability distribution~\cite{amari2000methods}. As we will show, its quantum analog tells us by how much the Bures metric (and hence the quantum Fisher information) fails to be compatible with evolution of the density matrix. Crucially, we will show how this manifests in nonlinear response experiments.

To make contact with Refs.~\cite{guan2026exploring,hetenyi2023fluctuations}, it is helpful to express Eqs.~\eqref{eq:mixture-main} and \eqref{eq:exponential-main} in terms of coordinates. Using our basis $\lambda^\mu$ from Eq.~\eqref{eq:coordinate-conventions-main}, we can define the Christoffel symbols for the mixture and exponential connections as
\begin{equation}
\begin{aligned}
    \Gamma^{(m)}_{\mu\nu\sigma}
    &= g(\nablamixture_{\partial_\mu}\partial_\nu,\partial_\sigma) \\
    &=\frac{1}{2}\Tr(G_\sigma\rho_{\mu\nu}),
\end{aligned}
    \label{eq:mixture-coeff-main}
\end{equation}
with $\rho_{\mu\nu}$ defined in Eq.~\eqref{eq:coordinate-conventions-main}, and
\begin{align}
    \Gamma^{(e)}_{\mu\nu\sigma}
    &=g(\nablaexp_{\partial_\mu}\partial_\nu,\partial_\sigma) \nonumber \\
    &=\Gamma^{(m)}_{\mu\nu\sigma}
    -T_{\mu\nu\sigma},
    \label{eq:exp-coords-main}
\end{align}
with
\begin{align}
    T_{\mu\nu\sigma} &= T(\partial_\mu,\partial_\nu,\partial_\sigma) \nonumber \\
    &=
    \frac{1}{2}\Tr\!\left(
        \rho\acomm{G_\mu}{\acomm{G_\nu}{G_\sigma}}
    \right).
    \label{eq:qac-coords}
\end{align}
We note that the mixture connection coefficients in Eq.~\eqref{eq:mixture-coeff-main} coincide, after symmetrization, with the ``Fisher part'' of the Bures connection introduced in Ref.~\cite{guan2026exploring}. That is, the mixture connection gives the lowest-order perturbative correction to the quantum Fisher information, and can be expressed in terms of second-order nonlinear response functions.

The symmetric part of the QAC tensor, on the other hand, is proportional to the ``intrinsic part'' of the Bures connection. As we show in Appendix~\ref{sec:the_connection_algebra}, the metric-compatible Bures-Levi-Civita connection can be reconstructed from linear combinations of  $\Gamma^{(m)}_{\mu\nu\sigma}$ and $T_{\mu\nu\sigma}$ consistent with the results of Ref.~\cite{guan2026exploring}.

Finally, we can apply what we have learned about the mixture and exponential connections to study the derivatives of the operator-valued QGT $Q_{\mu\nu}$ from Eq.~\eqref{eq:operator-qgt-main}. Taking derivatives, we can write
\begin{equation}
\begin{aligned}
    \partial_\sigma Q_{\mu\nu}
    &=
    (\partial_\sigma G_\mu)\rho G_\nu
    +C_{\sigma;\mu\nu}
    +G_\mu\rho(\partial_\sigma G_\nu),
    \\
    C_{\sigma;\mu\nu}
    &\equiv
    G_\mu\,\rho_\sigma\,G_\nu .
\end{aligned}
    \label{eq:qgt-core-derivative-main}
\end{equation}
Focusing on the middle term $C_{\sigma;\mu\nu}$, we can take a trace to define
\begin{equation}
\begin{aligned}
    Z_{\sigma;\mu\nu}
    &=
    \Tr(C_{\sigma;\mu\nu})
    =
    \Tr\!\left(G_\mu\,\rho_\sigma\,G_\nu\right).
\end{aligned}
    \label{eq:cubic-z-main}
\end{equation}
Since $\rho_\sigma$ has only off-diagonal matrix elements per Eqs.~\eqref{eq:coordinate-conventions-main} and \eqref{eq:orbit-generator-main}, $Z$ carries information about virtual transitions. Using our shorthand Eq.~\eqref{eq:lift-component-shorthand-main}, we can define the map of tangent vectors
\begin{equation}
Z(u,v,w)\equiv \Tr\!\left(G_v\,u(\rho)\,G_w\right).
\end{equation}
 The real part of $Z$ is the QAC tensor introduced in Eq.~\eqref{eq:qac-main}, 
\begin{equation}
    Z(u,v,w)
    =
    T(u,v,w)-\ii\Phi(u,v,w),
    \label{eq:cubic-t-phi-main}
\end{equation}
while $\Phi(u,v,w)$ is given by
\begin{equation}
    \Phi(u,v,w)
    =
    -\frac{\ii}{2}\Tr\!\left(
    \rho\acomm{G_u}{\comm{G_v}{G_w}}
    \right).
    \label{eq:phi-core-main}
\end{equation}

As we show in Appendix~\ref{sec:the_connection_algebra}, $Z_{\sigma;\mu\nu}$ is actually a tensor. Its real part, the QAC tensor, can be written as the exponential covariant derivative of the Bures metric. Analogously, the imaginary part $\Phi_{\sigma\mu\nu}\equiv \Phi(\partial_\sigma,\partial_\mu,\partial_\nu)$ is the exponential covariant derivative of the Uhlmann curvature,
\begin{equation}
\begin{aligned}
    \left(\nablaexp_{\partial_\sigma}g\right)_{\mu\nu}
    &=
    T_{\sigma\mu\nu},
    \\
    \left(\nablaexp_{\partial_\sigma}\mathcal F\right)_{\mu\nu}
    &=
    \Phi_{\sigma\mu\nu}.
\end{aligned}
    \label{eq:cubic-covariant-derivatives-main}
\end{equation}
Putting this together, we have
\begin{equation}
\begin{aligned}
    Z_{\sigma;\mu\nu}
    &=
    \left[\nablaexp_{\partial_\sigma}\!\left(g-\ii\,\mathcal{F}\right)\right]_{\mu\nu} \\
    &=\left[\nablaexp_{\partial_\sigma}\Tr Q\right]_{\mu\nu}.
    \label{eq:z-covariant-derivative-main}
\end{aligned}
\end{equation}
Thus, just like the QGT combines the metric and Uhlmann curvature into one complex tensor, $Z_{\sigma;\mu\nu}$ combines the QAC tensor and the covariant Uhlmann curvature dipole into a single complex quantum Amari-Chentsov (cQAC) tensor. 

As an example, let us compute $Z_{\sigma;\mu\nu}$ for perturbations to a thermal density matrix, evaluated at the point $\lambda=0$ along the orbit (i.e. at equilibrium). We take
\begin{equation}
\rho_0 = \frac{1}{\mathcal{Z}}\sum_{n} e^{-\beta E_n}\ket{n}\bra{n}, \label{eq:thermalrho}
\end{equation}
where $\beta=1/T$ is the inverse temperature, $\mathcal{Z}$ is the partition function, and $E_n$ is the unperturbed energy of the eigenstate $\ket{n}$. For deformations $G_\mu$ given by Eq.~\eqref{eq:sld-orbit-lift}, we can write at $\lambda=0$, 
\begin{equation}
    \bigl(G_\mu\bigr)_{mn}=
    -\ii t_{mn}
    (\mathcal{A}_\mu)_{mn},
    \label{eq:G-temp-general}
\end{equation}
with
\begin{equation}
\begin{aligned}
    t_{mn}&\equiv\tanh\!\left(\frac{\beta\omega_{mn}}{2}\right), \\
    \omega_{mn}&=E_m-E_n .
    \label{eq:t-conventions-main}
    \end{aligned}
\end{equation}
Inserting Eq.~\eqref{eq:G-temp-general} into our definition Eq.~\eqref{eq:cubic-z-main}, we find that the cQAC tensor evaluated at the point $\lambda=0$ on the orbit, i.e. at $\rho_0$, can be written
\begin{equation}
Z_{\sigma;\mu\nu}
    =
    \ii\sum_{a,b,k}
    t_{ab}(p_k-p_b)t_{ka}\,
    (\mathcal A_\mu)_{ab}(\mathcal A_\sigma)_{bk}(\mathcal A_\nu)_{ka}. \label{eq:z-nonzero-T-general}
\end{equation}
Note that the thermal factors ensure that matrix elements of $\mathcal{A}$ between states of equal energy never enter into the cQAC tensor. Taking the limit of zero temperature, we can write
\begin{equation}
\begin{aligned}
    \lim_{T\rightarrow 0} Z_{\sigma;\mu\nu}
    ={}&
    -\ii\sum_{a,b>0}\sgn(\omega_{ab})
    \\
    &\times
    \big[
    (\mathcal A_\nu)_{0a}(\mathcal A_\mu)_{ab}(\mathcal A_\sigma)_{b0}
    \\
    &\quad
    +
    (\mathcal A_\sigma)_{0a}(\mathcal A_\nu)_{ab}(\mathcal A_\mu)_{b0}
    \big],
\end{aligned}
    \label{eq:W-main}
\end{equation}
where $\ket{0}$ denotes the ground state, and we define $\sgn(0)=0$. Crucially, even though the zero-temperature density matrix is a projection onto a single pure state, the SLD generator Eq.~\eqref{eq:G-temp-general} still has nonvanishing matrix elements between excited states due to the $T\rightarrow 0$ limiting procedure; this ensures that Eq.~\eqref{eq:W-main} is not identically zero.

\subsection{The almost complex structure and J-rotated tensors}\label{sub:the_almost_complex_structure_and_conjugated_tensors}

To conclude our general discussion of the multi-state geometry of density matrices, we note that the orbits Eq.~\eqref{eq:orbit-param-main} that we are considering actually carry a natural (almost) complex structure that allows us to write tensors in complex coordinates~\cite{nakahara2018geometry}. To see this, note that we can write any density matrix as
\begin{equation}
    \rho=\sum_r p_r P_r,
    \label{eq:block-spectral-main}
\end{equation}
where $P_r=\sum_{\alpha}\ket{r,\alpha}\bra{r,\alpha}$ is a projection operator onto the set of states with the same probability $p_r$. Now consider a block-off-diagonal Hermitian operator $\mathcal{O}$ such that $P_r\mathcal{O}P_r=0$ for all $r$. In the eigenbasis of $\rho$, $\mathcal{O}$ is off-diagonal; its upper triangular block is the conjugate of the lower triangular block. We can thus view the lower and upper triangular blocks as, respectively, holomorphic and antiholomorphic ``components'' of $\mathcal{O}$. To formalize this, we can introduce an almost complex structure $J$, an operator that squares to $-\Id$ and represents ``multiplication by $\ii$'' for the holomorphic coordinates. Explicitly, we can define
\begin{equation}
    J_\rho \mathcal{O}
    =
    -\ii\sum_{r\neq s}\sgn(p_r-p_s)P_r\mathcal{O}P_s .
    \label{eq:J-main}
\end{equation}
Since $\mathcal{O}$ is off-diagonal, $J_\rho^2\mathcal{O}=-\mathcal{O}$. Also note that on our unitary orbits, the probabilities $p_r$ are constant, and only the projectors $P_r$ evolve. That means that the $\rho$ dependence of $J_{\rho}$ never mixes blocks with opposite sign, so $J_\rho$ is well defined along the entire orbit~\cite{kirillov2004lectures}.

To apply this to our geometric structures, we note that the SLD generators $G_u$ of Eq.~\eqref{eq:sld-spectral} are purely off-diagonal operators. Thus, for any $G_u$ we can define the action of $J_\rho$ as 
\begin{equation}
    (G_{Ju})_{mn}
    \equiv (J_\rho G_{u})_{mn}=
    -\ii\,\sgn(p_m-p_n)(G_u)_{mn}.
    \label{eq:J-lift-main}
\end{equation}

The almost complex structure $J_\rho$ plays a special geometric role because it is compatible with the Bures metric. More precisely, writing out the Bures metric in terms of the SLD generators using Eq.~\eqref{eq:bures-main}, we have
\begin{equation}
    g(u,v)
    =
    \frac{1}{2}\sum_{m,n}(p_m+p_n)
    (G_u)_{mn}(G_v)_{nm}.
    \label{eq:bures-spectral}
\end{equation}
Acting with the almost complex structure $J_\rho$, it follows from Eq.~\eqref{eq:J-lift-main} that
\begin{align}
    g(Ju,Jv)&=\frac{1}{2}\sum_{m,n}(p_m+p_n)
    (G_{Ju})_{mn}(G_{Jv})_{nm} \nonumber \\
&=-\frac{(-\ii)^2}{2}\sum_{m,n}(p_m+p_n)\sgn(p_m-p_n)^2 \nonumber \\
&\qquad\times(G_u)_{mn}(G_v)_{nm} \nonumber \\
    &=g(u,v).
    \label{eq:j-metric-invariance-main}
\end{align}
Thus, the almost complex structure $J_{\rho}$ gives the space of density matrix orbits a local Hermitian structure~\cite{nakahara2018geometry}.

While the almost complex structure $J$ and the Hermitian geometry that it implies are interesting in their own right, we will defer a full exploration of these points to future work. Here, we will focus on how $J_\rho$ acts on the cQAC tensor. We will see in Sec.~\ref{sec:rectification} that this serves as the bridge between geometry and nonlinear rectification. Acting with $J_\rho$ on the arguments of $Z$, we can define the tensor
\begin{equation}
    Z(J\partial_\sigma,J\partial_\mu,J\partial_\nu)
    =
    \widetilde S_{\sigma;\mu\nu}.
    \label{eq:triple-j-main}
\end{equation}
Note from Eq.~\eqref{eq:J-lift-main} that both $G_{Ju}$ and $G_u$ are Hermitian operators. As such, we can decompose Eq.~\eqref{eq:triple-j-main} into real and imaginary parts in analogy with Eq.~\eqref{eq:cubic-t-phi-main}. The real and imaginary parts do not mix under the action of $J$, yielding
\begin{equation}
    \widetilde S_{\sigma;\mu\nu}
    =
    T(J\partial_\sigma,J\partial_\mu,J\partial_\nu)
    -\ii\,\Phi(J\partial_\sigma,J\partial_\mu,J\partial_\nu).
    \label{eq:triple-j-tphi-main}
\end{equation}
 The tensor $\widetilde S_{\sigma;\mu\nu}$ is thus the cQAC tensor rotated by $J$. Since the action of $J$ is invertible, we can equally well regard $\widetilde S_{\sigma;\mu\nu}$ as the fundamental geometric object. This will be our perspective going forward. It will be useful to define the cyclic linear combination
  \begin{equation}
\begin{aligned}
    S_{\mu\sigma\nu}
    =
    \frac{1}{2}
    \left(
    \widetilde S_{\mu;\nu\sigma}
    +
    \widetilde S_{\nu;\sigma\mu}
    -
    \widetilde S_{\sigma;\mu\nu}
    \right),
\end{aligned}
    \label{eq:same-order-to-response-main}
\end{equation}
which allows us to express 
\begin{equation}
    \widetilde S_{\sigma;\mu\nu}
    =
    S_{\nu\mu\sigma}
    +
    S_{\sigma\nu\mu}.
    \label{eq:same-order-s-main}
\end{equation}

Through the cyclic linear combination $S_{\mu\sigma\nu}$, the J-rotated cQAC tensor can be related to the Christoffel coefficients we introduced in Sec.~\ref{sub:dual_connections_and_the_quantum_amari_chentsov_tensor} and Appendix~\ref{sec:the_connection_algebra}. In particular, defining
\begin{equation}
    T^J_{\sigma\mu\nu}
    \equiv
    \operatorname{Re}\widetilde S_{\sigma;\mu\nu}
    =
    T(J\partial_\sigma,J\partial_\mu,J\partial_\nu),
    \label{eq:tj-main}
\end{equation}
we have that the real part of $S_{\mu\sigma\nu}$ is given by 
\begin{equation}
    \operatorname{Re}S_{\mu\sigma\nu}
    =
    \frac{1}{2}
    \left(
    T^J_{\mu\nu\sigma}
    +
    T^J_{\nu\sigma\mu}
    -
    T^J_{\sigma\mu\nu}
    \right).
    \label{eq:s-response-contorsion-main}
\end{equation}
This is the J-rotation of the tensor that converts the mixture connection to the Levi-Civita connection that we derive in Appendix~\ref{sec:the_connection_algebra}. In the theory of metric-affine gravity, it is known as the distortion tensor~\cite{shimada2019metric,d2023inflation,iosifidis2023biconnection}. Similarly, defining the J-rotated Uhlmann curvature dipole
\begin{equation}
\Phi^J_{\sigma\mu\nu} = \Phi(J\partial_\sigma,J\partial_\mu,J\partial_\nu),
\end{equation}
we have
\begin{equation}
    \operatorname{Im}S_{\mu\sigma\nu}
    =
    -\frac{1}{2}
    \left(
    \Phi^J_{\mu\nu\sigma}
    +
    \Phi^J_{\nu\sigma\mu}
    -
    \Phi^J_{\sigma\mu\nu}
    \right).
    \label{eq:s-response-curvature-main}
\end{equation}
Equations~\eqref{eq:s-response-contorsion-main} and \eqref{eq:s-response-curvature-main} show that $S_{\mu\sigma\nu}$ combines the $J$-rotated distortion tensor and Uhlmann curvature dipole into a single complex tensor; we will accordingly refer to $S_{\mu\sigma\nu}$ as the \emph{complex distortion tensor}.

To conclude our general discussion of the multi-state geometric structures, let us examine $\widetilde S_{\sigma;\mu\nu}$ for a thermal density matrix Eq.~\eqref{eq:thermalrho}, evaluated at the equilibrium point $\lambda=0$ along the orbit. First, applying $J$ to the thermal SLD generators in Eq.~\eqref{eq:G-temp-general}, we find that for the thermal density matrix
\begin{align}
(G_{J\partial_\mu})_{mn} &= -t_{mn}\sgn(p_m-p_n)(\mathcal{A}_\mu)_{mn} \nonumber \\
&= |t_{mn}|(\mathcal{A}_\mu)_{mn}.
\end{align}
Inserting this into our definition Eq.~\eqref{eq:cubic-z-main} and using Eq.~\eqref{eq:triple-j-main} yields
\begin{equation}
\widetilde S_{\sigma;\mu\nu} = \sum_{a,b,k}
    |t_{ab}||p_k-p_b||t_{ka}|\,
    (\mathcal A_\mu)_{ab}(\mathcal A_\sigma)_{bk}(\mathcal A_\nu)_{ka}. \label{eq:stilde-nonzero-T-general}
\end{equation}
Taking the limit $T\rightarrow 0$, we find
\begin{equation}
\begin{aligned}
    \lim_{T\rightarrow 0}\widetilde S_{\sigma;\mu\nu}
    ={}&
    \sideset{}{'}\sum_{a,b>0}
    \big[
    (\mathcal A_\sigma)_{0a}(\mathcal A_\nu)_{ab}(\mathcal A_\mu)_{b0}
    \\
    &\quad
    +(\mathcal A_\nu)_{0a}(\mathcal A_\mu)_{ab}(\mathcal A_\sigma)_{b0}
    \big],
\end{aligned}
    \label{eq:triple-j-collapse-main}
\end{equation}
where we have introduced the notation $\sideset{}{'}\sum_{a_1,a_2,\dots}$ to indicate that any term in the sum where indexed states have coincident energy is omitted, i.e. $E_{a_i}\neq E_{a_j}$ for all $i\neq j$. Finally, comparing Eq.~\eqref{eq:triple-j-collapse-main} with Eq.~\eqref{eq:same-order-s-main} allows us to identify
\begin{equation}
    \lim_{T\rightarrow 0} S_{\alpha\beta\gamma}
    \equiv
    \sideset{}{'}\sum_{a,b>0}
    (\mathcal A_\alpha)_{0a}
    (\mathcal A_\beta)_{ab}
    (\mathcal A_\gamma)_{b0}.
    \label{eq:ordered-s-main}
\end{equation}

We see that even at zero temperature, the complex distortion tensor $S_{\alpha\beta\gamma}$ involves matrix elements between virtual excited states. It thus represents a many-body generalization of the ``multi-state'' geometry introduced recently for Bloch bands in Refs.~\cite{avdoshkin2024multi,jankowski2024quantized,mitscherling2024gauge,avdoshkin2023extrinsic}. The dependence on virtual excited states arises due to our zero-temperature limiting procedure in Eq.~\eqref{eq:W-main}. This means that the zero-temperature complex distortion tensor depends not only on the properties of the ground state. To connect this abstract many-body multi-state geometry to physical observables, we will now show that $\lim_{T\rightarrow 0} S_{\alpha\beta\gamma}$ plays a crucial role in determining the strength of second-order rectification response. In Table~\ref{tab:geometry-summary} we summarize the geometric objects that have been introduced in this section.

\begin{table*}
\caption{Summary of the multi-state geometric objects introduced in Sec.~\ref{sec:multi_state_geometry}. For each object we give its explicit expression along with an equation reference where its geometric meaning is discussed.}
\label{tab:geometry-summary}
\begin{ruledtabular}
\begin{tabular}{lll}
Object & Construction & Geometric meaning \\
\hline
$Q_{\mu\nu}$ & $G_\mu\rho G_\nu$ [Eq.~\eqref{eq:operator-qgt-main}] & QGT operator [Eq.~\eqref{eq:operator-qgt-trace-main}] \\
$g_{\mu\nu}$ & $\tfrac{1}{2}\Tr\left(\rho\acomm{G_\mu}{G_\nu}\right)$ [Eq.~\eqref{eq:bures-main}] & Bures metric [Eq.~\eqref{eq:operator-qgt-trace-main}] \\
$\mathcal{F}_{\mu\nu}$ & $-\tfrac{\ii}{2}\Tr\left(\rho\comm{G_\mu}{G_\nu}\right)$ [Eq.~\eqref{eq:curvature-form-main}] & Uhlmann curvature [Eq.~\eqref{eq:operator-qgt-trace-main}] \\
$T_{\sigma\mu\nu}$ & $\tfrac{1}{2}\Tr\left(\rho\acomm{G_\sigma}{\acomm{G_\mu}{G_\nu}}\right)$ [Eq.~\eqref{eq:qac-coords}] & QAC tensor [Eqs.~\eqref{eq:dual-difference-main} and \eqref{eq:cubic-covariant-derivatives-main}] \\
$\Phi_{\sigma\mu\nu}$ & $-\tfrac{\ii}{2}\Tr\left(\rho\acomm{G_\sigma}{\comm{G_\mu}{G_\nu}}\right)$ [Eq.~\eqref{eq:phi-core-main}] & Uhlmann curvature dipole [Eq.~\eqref{eq:cubic-covariant-derivatives-main}] \\
$Z_{\sigma;\mu\nu}$ & $\Tr\left(G_\mu\rho_\sigma G_\nu\right)=T_{\sigma\mu\nu}-\ii\Phi_{\sigma\mu\nu}$ [Eqs.~\eqref{eq:cubic-z-main} and \eqref{eq:cubic-t-phi-main}] & cQAC tensor [Eq.~\eqref{eq:z-covariant-derivative-main}] \\
$\widetilde S_{\sigma;\mu\nu}$ & $Z(J\partial_\sigma,J\partial_\mu,J\partial_\nu)$ [Eq.~\eqref{eq:triple-j-main}] & J-rotated cQAC tensor [Eq.~\eqref{eq:triple-j-tphi-main}] \\
$S_{\mu\sigma\nu}$ & Cyclic linear combination of $\widetilde S$ [Eq.~\eqref{eq:same-order-to-response-main}] & Complex distortion tensor [Eqs.~\eqref{eq:s-response-contorsion-main} and \eqref{eq:s-response-curvature-main}]
\end{tabular}
\end{ruledtabular}
\end{table*}

\section{Second-order rectification response in many-body systems}\label{sec:rectification}

Now we will apply our geometric framework to study nonlinear response and rectification. First, in Sec.~\ref{sub:orbits_and_geometric_amplitudes_from_time_dependent_perturbation_theory}, we specialize to density matrix orbits that arise from time-dependent perturbations to the equilibrium state of a material. This allows us to introduce the time-dependent and quasistatic cQAC tensors. Then, in Sec.~\ref{sub:general_zero_temperature_rectification_sum_rules}, we apply the formalism to second-order time-dependent perturbation theory, deriving our main result: a zero-temperature sum rule for the rectification response of a general insulating many-body system. Lastly, in Sec.~\ref{sub:physical_interpretation_of_linear_and_circular_rectification} we show how the sum rule manifests in bounded systems via the long-time change in observables after a short pulse.

\subsection{Orbits and geometric amplitudes from time-dependent perturbation theory}\label{sub:orbits_and_geometric_amplitudes_from_time_dependent_perturbation_theory}

In order to derive the link between the cQAC tensor and nonlinear rectification, we will now specialize to compute the geometric structures that arise from time-dependent perturbation theory building on Ref.~\cite{guan2026exploring}. We will focus on the particular case where a system with Hamiltonian $H_0$ and initial density matrix Eq.~\eqref{eq:thermalrho} is perturbed by 
\begin{equation}
    H_{\mathrm{src}}(t)=H_0+e^{\eta t} f^\alpha(t)C_\alpha, \label{eq:general-source-history-main}
\end{equation}
where the operator $C_\alpha$ is
\begin{equation}
    C_\alpha\equiv\dot B_\alpha=\ii\comm{H_0}{B_\alpha} + \mathcal{O}(f),
    \label{eq:general-source-history-main-1}
\end{equation}
$f^\alpha(t)$ are time-dependent external fields, and $\eta$ ensures that the perturbation vanishes adiabatically as $t\rightarrow -\infty$. We will take $\eta\rightarrow 0$ at the end of calculations. 

Using time-dependent perturbation theory, we can write the first-order change to the density matrix as
\begin{equation}
\begin{aligned}
    \rho_\mu(t;\tau)
    &\equiv \left.\frac{\delta \rho(t)}{\delta f^\mu(\tau)}\right|_{f=0} \\
    &=
    -\ii\,e^{\eta \tau}\Theta(t-\tau)\comm{C_\mu(\tau)}{\rho_0},
\end{aligned}
    \label{eq:source-tangent-main-operator}
\end{equation}
where $\Theta$ is a step function that enforces causality, and the time-dependence of the operator $C_\mu(\tau)$ is evaluated using the unperturbed Hamiltonian $H_0$. Taking matrix elements in the basis $\{\ket{n}\}$ of eigenstates of $H_0$ we can write
\begin{equation}
    \bigl(\rho_\mu(t;\tau)\bigr)_{mn}
    =
    -\ii\,e^{\eta \tau}\Theta(t-\tau)(p_n-p_m)
    e^{\ii\omega_{mn}\tau}(C_\mu)_{mn},\label{eq:source-tangent-main}
\end{equation}
where, as in Eq.~\eqref{eq:t-conventions-main}, $\omega_{mn}=E_m-E_n$ is the difference of eigenvalues of $H_0$. 

We will now connect perturbation theory to quantum geometry. We first note that time evolution of the density matrix is unitary, so $\rho(t)$ forms an orbit space just like in Eq.~\eqref{eq:orbit-param-main}. Since we work in a series expansion about $f=0$, we can relate the perturbation series for the density matrix to geometric structures evaluated at the base point $\lambda=\{f\}=0$ of the orbit, i.e. at the unperturbed equilibrium density matrix. For the remainder of this work, we will compute geometric structures at $\lambda=0$ unless otherwise specified.

We can thus use Eqs.~\eqref{eq:sld-lyapunov} and \eqref{eq:sld-spectral} to define time-dependent SLD generators $G_\mu(t;\tau)$, whose matrix elements in the unperturbed basis are
\begin{equation}
    \bigl(G_\mu(t;\tau)\bigr)_{mn}
    =
    -\ii\,e^{\eta \tau}\Theta(t-\tau)t_{mn}
    e^{\ii\omega_{mn}\tau}(C_\mu)_{mn}.
    \label{eq:source-lift-main}
\end{equation}
Note that matrix elements of $G_\mu(t;\tau)$ between states of equal energy vanish. For bounded systems, this follows from the fact that Eq.~\eqref{eq:general-source-history-main-1} ensures the equal-energy matrix elements of $C_\mu$ themselves vanish. For unbounded systems (or more precisely, unbounded operators $B_\mu$), equal-energy matrix elements of $G_\mu(t;\tau)$ are forced to vanish by the thermal factor $t_{mn}$ of Eq.~\eqref{eq:t-conventions-main}.

With the SLD generators and density matrix derivatives in hand, we can compute the geometric structures outlined in Sec.~\ref{sec:multi_state_geometry}. In particular, using only first-order perturbation theory and Eq.~\eqref{eq:z-nonzero-T-general}, we find that the time-dependent cQAC tensor $Z^C$ associated to the perturbation $C_\mu$ is
\begin{align}
    Z^C_{\sigma;\mu\nu}(t;\tau_\sigma,\tau_\mu,\tau_\nu)
    &=
    \Tr\!\left[
        G_\mu(t;\tau_\mu)\,
        \rho_\sigma(t;\tau_\sigma)\,
        G_\nu(t;\tau_\nu)
    \right] \nonumber\\
    &=
    \ii\,\Theta_\mu\Theta_\sigma\Theta_\nu
    \sum_{a,b,k}
    t_{ab}(p_k-p_b)t_{ka}
    \nonumber\\
    &\quad\times
    e^{\ii(\omega_{ab}\tau_\mu
        +\omega_{bk}\tau_\sigma
        +\omega_{ka}\tau_\nu)}
    \nonumber\\
    &\quad\times
    (C_\mu)_{ab}(C_\sigma)_{bk}(C_\nu)_{ka},
    \label{eq:source-cubic-main}
\end{align}
where we have introduced the notation $\Theta_\mu = e^{\eta \tau_\mu}\Theta(t-\tau_\mu)$. 

We will be most interested in this work in the quasistatic limit, where the perturbation $f^\mu(t)e^{\eta t} \rightarrow f^\mu e^{\eta t}$ becomes approximately independent of time aside from the adiabatic turning on. In the quasistatic limit, the chain rule for variational derivatives implies that
\begin{equation}
\begin{aligned}
\rho_\mu(t) &\equiv e^{-\eta t}\left.\frac{\partial \rho(t)}{\partial f^\mu}\right|_{f=0} \\
&= \int_{-\infty}^\infty \!\dd\tau\, e^{-\eta t}\rho_\mu(t;\tau),\label{eq:quasistatic-rho-mu}
\end{aligned}
\end{equation}
with the corresponding SLD generators
\begin{equation}
G_\mu(t) = \int_{-\infty}^\infty \!\dd\tau\, e^{-\eta t} G_\mu(t;\tau).\label{eq:quasistatic-sld}
\end{equation}
Combining Eqs.~\eqref{eq:quasistatic-sld} and \eqref{eq:quasistatic-rho-mu} with our expression Eq.~\eqref{eq:source-cubic-main} for the time-dependent cQAC tensor, we can evaluate the cQAC tensor in the quasistatic limit, $Z^{C,\mathrm{q.s.}}_{\sigma;\mu\nu}$. We find after a change of integration variables that
\begin{align}
    Z^{C,\mathrm{q.s.}}_{\sigma;\mu\nu}
    &\equiv
    \int_{-\infty}^{0}\!\dd\tau_\sigma
    \int_{-\infty}^{0}\!\dd\tau_\mu
    \nonumber\\
    &\quad
    \int_{-\infty}^{0}\!\dd\tau_\nu\,
    Z^C_{\sigma;\mu\nu}(0;\tau_\sigma,\tau_\mu,\tau_\nu)
    \nonumber\\
    &=
    \ii\sum_{a,b,k}
    t_{ab}(p_k-p_b)t_{ka}\,
    \frac{(C_\mu)_{ab}}{\eta+\ii\omega_{ab}}\,
    \nonumber\\
    &\quad\times
    \frac{(C_\sigma)_{bk}}{\eta+\ii\omega_{bk}}\,
    \frac{(C_\nu)_{ka}}{\eta+\ii\omega_{ka}}
    \nonumber\\
    &\xrightarrow{\eta\to0^+}
    \ii\sideset{}{'}\sum_{a,b,k}
    t_{ab}(p_k-p_b)t_{ka}\,
    (B_\mu)_{ab}(B_\sigma)_{bk}(B_\nu)_{ka}
    \nonumber\\
    &\equiv Z^B_{\sigma;\mu\nu}.
    \label{eq:source-cubic-to-b-main}
\end{align}
To take the $\eta\rightarrow 0$ limit in the last step, we made use of the fact from Eq.~\eqref{eq:general-source-history-main-1} that $(C_\mu)_{mn} = \ii\omega_{mn}(B_\mu)_{mn}$ for $\omega_{mn}\neq 0$, coupled with the fact that the $\eta\rightarrow 0$ limit ensures that equal-energy matrix elements do not appear in the sum. For bounded $B_\mu$ this is ensured automatically from the thermal factors. However, to emphasize the exclusion in the general case, we keep the $\sideset{}{'}{\sum}$ notation explicit. This will be important in Sec.~\ref{sec:application_to_free_fermions} when we consider operators $B$ with ill-defined diagonal matrix elements.

Equation~\eqref{eq:source-cubic-to-b-main} shows that the cQAC tensor associated to the perturbation $C_\mu$ in the quasistatic limit coincides with the cQAC tensor for a perturbation $B_\mu$ in the ``instantaneous'' limit. We can see this directly by noting that in the quasistatic limit, the orbit generator we obtain from Eqs.~\eqref{eq:orbit-generator-main} and \eqref{eq:quasistatic-sld} is
\begin{equation}
    (\mathcal A_\alpha)_{mn}
    =
    \frac{(C_\alpha)_{mn}}{\ii\omega_{mn}}
    =
    (B_\alpha)_{mn},
    \qquad E_m\neq E_n,
    \label{eq:amplitude-from-k-main}
\end{equation}
exactly as we would obtain from Eq.~\eqref{eq:orbit-param-main} with $U(\lambda)=e^{-\ii B_\alpha \lambda^\alpha}$.

Having thus seen how the cQAC tensor emerges naturally from time-dependent perturbations of density matrices, we will move on to explore how these geometric objects appear in response theory. In particular, we will see how the quasistatic cQAC tensor Eq.~\eqref{eq:source-cubic-to-b-main} determines the strength of second-order rectification response.

\subsection{General zero-temperature rectification sum rules}\label{sub:general_zero_temperature_rectification_sum_rules}

Now let us connect the many-body quantum geometry to the nonlinear rectification response. We consider a harmonic perturbation that couples to a bounded operator $B_\lambda$,
\begin{equation}
    H_\mathrm{src}(t)= H_0 + 
    \bigl(F^\lambda e^{-\ii\omega t}
    +F^{\lambda*}e^{\ii\omega t}\bigr)e^{\eta t}B_\lambda,
    \qquad \eta\to0^+ .
    \label{eq:primitive-drive-main}
\end{equation}
Note that this differs from the perturbation we considered in Sec.~\ref{sub:orbits_and_geometric_amplitudes_from_time_dependent_perturbation_theory} to compute the geometry associated to the operator $C_\lambda$. Here we consider the second-order DC ($\omega=0$) response of $\langle C_\sigma\rangle= \langle \dot{B}_\sigma\rangle$, which we can parametrize in terms of a response function
\begin{equation}
    \langle C_\sigma\rangle_{\mathrm{DC}}(\omega)
    =
    -\left[
    K^{\sigma\nu\lambda}(-\omega)
    +
    K^{\sigma\lambda\nu}(+\omega)
    \right]F^\nu F^{\lambda*}.
    \label{eq:primitive-rectified-main}
\end{equation}
The $\omega$ dependence of the left hand side denotes that the \emph{drive} is at frequency $\omega$, while Eq.~\eqref{eq:primitive-rectified-main} picks out the DC component of the response. As we show in Appendix~\ref{sec:general_rectification_response_derivation}, the response function contains both a ``paramagnetic'' piece determined by the second-order Kubo formula, as well as a ``diamagnetic'' piece originating from the explicit $F^\lambda$ dependence of $C_\sigma =\ii\comm{H_\mathrm{src}(t)}{B_\sigma}$. Combining the two contributions allows us to express the response function $K^{\sigma\nu\lambda}(-\omega)$ in terms of a spectral density~\cite{bradlyn2024spectral,matsyshyn2019nonlinear}, 
\begin{equation}
    -K^{\sigma\nu\lambda}(-\omega)
    =
    \frac{1}{\pi}
    \int_{-\infty}^{\infty}\!\dd\varpi\;
    \frac{\rho_K^{\sigma;\nu\lambda}(\varpi)}
    {\eta-\ii(\varpi-\omega)} .
    \label{eq:primitive-spectral-k-main}
\end{equation}
To define the spectral density, it is helpful to introduce the transition matrix element
\begin{equation}
    w_{ac}^{\sigma;\nu\lambda}
    =
    \pi(p_a-p_c)\,
    \comm{B_\nu}{B_{\sigma,\mathrm d}}_{ac}(B_\lambda)_{ca},
    \label{eq:primitive-line-weight-main}
\end{equation}
with
\begin{equation}
    B_{\sigma,\mathrm d}
    =
    \sum_{E_m=E_n}
    \ket{m}\!\bra{m}B_\sigma\ket{n}\!\bra{n},
    \label{eq:primitive-hat-main}
\end{equation}
the energy-diagonal part of the operator $B_\sigma$. In terms of Eq.~\eqref{eq:primitive-line-weight-main}, we can write the spectral density $\rho_K^{\sigma;\nu\lambda}(\varpi)$ as
\begin{equation}
\begin{aligned}
    \rho_K^{\sigma;\nu\lambda}(\varpi)
    ={}&
    \sum_{\omega_{ac}>0}
    \Bigl[
    w_{ac}^{\sigma;\nu\lambda}\,\delta(\varpi-\omega_{ac})
    \\
    &+
    \bigl(w_{ac}^{\sigma;\nu\lambda}\bigr)^{*}\,\delta(\varpi+\omega_{ac})
    \Bigr].
\end{aligned}
    \label{eq:primitive-rho-k-main}
\end{equation}
Note that in deriving Eqs.~\eqref{eq:primitive-line-weight-main} and \eqref{eq:primitive-rho-k-main}, we needed to make use of the boundedness of $B$ in order to ensure that the rectification response $K^{\sigma\nu\lambda}$ contained no bare $1/\eta$ divergences, and to ensure that $B_{\sigma,\mathrm d}$ has finite matrix elements. We will see in Sec.~\ref{sec:application_to_free_fermions} how to generalize the rectification response in the case where $B_\sigma$ is not bounded. Note that if there are no degeneracies between excited states of the system (or, more generally, if $B_\sigma$ is diagonal in every degenerate subspace; for a single fixed $\sigma$ we can always choose such a basis by diagonalizing $B_\sigma$ within each degenerate subspace), we can simplify Eq.~\eqref{eq:primitive-line-weight-main} for the line weight $w_{ac}^{\sigma;\nu\lambda}$. We can write in this case
\begin{equation}
    w_{ac}^{\sigma;\nu\lambda}
    \rightarrow
    \pi(p_a-p_c)
    \bigl((B_\sigma)_{cc}-(B_\sigma)_{aa}\bigr)
    (B_\nu)_{ac}(B_\lambda)_{ca}.
    \label{eq:primitive-line-weight-nondeg-main}
\end{equation}
Equation~\eqref{eq:primitive-line-weight-nondeg-main} shows that in this basis, the line weight takes the form of a generalized shift vector.

To find the total DC rectification response, we can consider a broadband source perturbation 
\begin{equation}
f^\lambda(t) = \int_0^\infty \!\dd\omega \left[\widetilde F^\lambda(\omega)e^{-\ii\omega t} +  \widetilde F^{\lambda *}(\omega)e^{\ii\omega t}\right] . \label{eq:general-f-t}
\end{equation}
The measured DC response $\langle C_\sigma\rangle_{\mathrm{DC}}^{\mathrm{tot}}$ is then given by integrating Eq.~\eqref{eq:primitive-rectified-main} over the drive frequency. Using the spectral representation Eq.~\eqref{eq:primitive-spectral-k-main}, we find
\begin{equation}
\begin{aligned}
    \langle C_\sigma\rangle_{\mathrm{DC}}^{\mathrm{tot}}
    ={}&\frac{1}{\pi}\int_0^\infty\!\dd\omega\int\!\dd\varpi\;
    \widetilde F^\nu(\omega)\widetilde F^{\lambda*}(\omega)\\
    &\times\left[
    \frac{\rho_K^{\sigma;\nu\lambda}(\varpi)}{\eta-\ii(\varpi-\omega)}
    +\frac{\rho_K^{\sigma;\lambda\nu}(-\varpi)}{\eta+\ii(\varpi-\omega)}
    \right].
\end{aligned}
    \label{eq:primitive-conjugate-kernel-main}
\end{equation}

We can simplify Eq.~\eqref{eq:primitive-conjugate-kernel-main} using the Plemelj formula
\begin{equation}
\frac{1}{\eta-\ii x} = \pi\delta(x) + \ii\,\mathrm{PV}\frac{1}{x},
\end{equation}
with $\mathrm{PV}$ denoting the principal value. This lets us separate the response $\langle C_\sigma\rangle_{\mathrm{DC}}^{\mathrm{tot}}$ into an ``absorptive'' and ``reactive'' part
\begin{equation}
\langle C_\sigma\rangle_{\mathrm{DC}}^{\mathrm{tot}} = \langle C_\sigma\rangle_{\mathrm{DC}}^{\mathrm{tot,abs}} + \langle C_\sigma\rangle_{\mathrm{DC}}^{\mathrm{tot,reac}},\label{eq:primitive-absorptive-reactive-main}
\end{equation}
with 
\begin{equation}
\begin{aligned}
   \langle C_\sigma\rangle_{\mathrm{DC}}^{\mathrm{tot,abs}}
    ={}&\int_0^\infty\!\dd\omega\;\widetilde F^\nu(\omega)\widetilde F^{\lambda*}(\omega)\\
    &\times\bigl[\rho_K^{\sigma;\nu\lambda}(\omega)+\rho_K^{\sigma;\lambda\nu}(-\omega)\bigr]\\
    ={}&\sum_{\omega_{ac}>0}
    \bigl[w_{ac}^{\sigma;\nu\lambda}+\bigl(w_{ac}^{\sigma;\lambda\nu}\bigr)^{*}\bigr]\\
    &\times\widetilde F^\nu(\omega_{ac})\,\widetilde F^{\lambda*}(\omega_{ac}),    
\end{aligned}
    \label{eq:primitive-absorptive}
\end{equation}
and 
\begin{equation}
\begin{aligned}
\langle C_\sigma\rangle_{\mathrm{DC}}^{\mathrm{tot,reac}}&= \frac{\ii}{\pi}\int_0^\infty\!\dd\omega\;\widetilde F^\nu(\omega)\widetilde F^{\lambda*}(\omega)\,
    \mathrm{PV}\!\int\!\dd\varpi\\
    &\qquad\times
    \frac{\rho_K^{\sigma;\nu\lambda}(\varpi)-\rho_K^{\sigma;\lambda\nu}(-\varpi)}{\varpi-\omega} \\
    &=\frac{\ii}{\pi}\int_0^\infty\!\dd\omega\;
    \widetilde F^\nu(\omega)\widetilde F^{\lambda*}(\omega)
    \\
    &\times\mathrm{PV}\!\sum_{\omega_{ac}>0}
    \Biggl[
    \frac{w_{ac}^{\sigma;\nu\lambda}-\bigl(w_{ac}^{\sigma;\lambda\nu}\bigr)^{*}}
    {\omega_{ac}-\omega}
    \\
    &\qquad\quad
    -\frac{\bigl(w_{ac}^{\sigma;\nu\lambda}\bigr)^{*}-w_{ac}^{\sigma;\lambda\nu}}
    {\omega_{ac}+\omega}
    \Biggr].
\end{aligned}
    \label{eq:primitive-reactive-lines-main}
\end{equation}
Crucially, from Eqs.~\eqref{eq:primitive-line-weight-main} and \eqref{eq:primitive-line-weight-nondeg-main}, we have that
\begin{equation}
    w_{ac}^{\sigma;\nu\lambda}
    -\bigl(w_{ac}^{\sigma;\lambda\nu}\bigr)^{*}
    =0,
\end{equation}
which implies that 
\begin{equation}
    \langle C_\sigma\rangle_{\mathrm{DC}}^{\mathrm{tot,reac}}
    =0.
    \label{eq:primitive-reactive-vanishes-main}
\end{equation}

Thus, from Eq.~\eqref{eq:primitive-absorptive}, the spectral density $\rho_K$ determines the total DC response.
As such, the absorptive response satisfies a sum rule at zero temperature. Let us assume we have an insulator with a nondegenerate ground state. Then, inspired by Ref.~\cite{avdoshkin2024multi}, we can introduce the ground state skewness
\begin{equation}
    c^{(3)}_{\lambda\sigma\nu}
    =
    \bra{0}
    \delta B_\lambda\,\delta B_\sigma\,\delta B_\nu
    \ket{0},
    \label{eq:primitive-c3-main}
\end{equation}
with $\delta B_\mu = B_\mu - (B_\mu)_{00}$. Integrating the spectral density and comparing with Eqs.~\eqref{eq:ordered-s-main} and \eqref{eq:amplitude-from-k-main} we find (see Appendix~\ref{sec:general_rectification_response_derivation} for details) that at zero temperature
\begin{equation}
    \lim_{T\rightarrow 0}\frac{1}{\pi}
    \int_0^\infty\!\dd\varpi\;
    \rho_K^{\sigma;\nu\lambda}(\varpi)
    =
    c^{(3)}_{\lambda\sigma\nu}
    -
    S_{\lambda\sigma\nu},
    \label{eq:primitive-c3-s-main}
\end{equation}
where $S_{\lambda\sigma\nu}$ is the complex distortion tensor Eq.~\eqref{eq:same-order-to-response-main} at zero temperature. Combining this with our definition Eq.~\eqref{eq:primitive-absorptive}, we find that the measured absorptive part of the rectification response satisfies the sum rule
\begin{equation}
    \boxed{\;
    \begin{aligned}
    &\lim_{T\rightarrow 0}\frac{1}{\pi}\sum_{\omega_{ac}>0}
    \bigl[w_{ac}^{\sigma;\nu\lambda}+\bigl(w_{ac}^{\sigma;\lambda\nu}\bigr)^{*}\bigr]
    \\
    &\qquad=2\bigl(c^{(3)}_{\lambda\sigma\nu}-S_{\lambda\sigma\nu}\bigr).
    \end{aligned}
    \;}
    \label{eq:primitive-absorptive-sumrule-main}
\end{equation}
This sum rule relates the total rectification response at zero temperature to a ground state skewness and a many-body, multi-state geometric object: the complex distortion tensor. This generalizes the multi-state geometric sum rule of Ref.~\cite{avdoshkin2024multi} to many-body systems and general perturbations. The sum rule establishes that many-body, multi-state geometry is experimentally accessible.

To conclude, it is also illuminating to separate the sum rule Eq.~\eqref{eq:primitive-absorptive-sumrule-main} into its symmetric and antisymmetric parts. From Eq.~\eqref{eq:primitive-rectified-main}, the symmetric part corresponds to the response to linear polarization, while the antisymmetric part gives the response to circular polarization. Introducing the notation
\begin{equation}
\begin{aligned}
    A^{(\nu\lambda)}
    &=\frac{1}{2}\left(A^{\nu\lambda}+A^{\lambda\nu}\right),
    \\
    A^{[\nu\lambda]}
    &=\frac{1}{2}\left(A^{\nu\lambda}-A^{\lambda\nu}\right),
\end{aligned}
    \label{eq:pairsym}
\end{equation}
and
\begin{equation}
\begin{aligned}
    S_{(\mu|\sigma|\nu)}
    &\equiv
    \frac{1}{2}\left(S_{\mu\sigma\nu}+S_{\nu\sigma\mu}\right),
    \\
    S_{[\mu|\sigma|\nu]}
    &\equiv
    \frac{1}{2}\left(S_{\mu\sigma\nu}-S_{\nu\sigma\mu}\right)
\end{aligned}
    \label{eq:s-middle-index-sym-main}
\end{equation}
for the symmetrization and antisymmetrization of pairs [Eq.~\eqref{eq:pairsym}] and triplets [Eq.~\eqref{eq:s-middle-index-sym-main}] of indices allows us to write
\begin{align}
    \lim_{T\rightarrow 0}\frac{1}{\pi}\int_0^\infty\!\dd\varpi\;\rho_K^{\sigma;(\nu\lambda)}(\varpi)
    &=
    c^{(3)}_{(\lambda|\sigma|\nu)}
    -S_{(\lambda|\sigma|\nu)},
    \label{eq:primitive-linear-channel-main}
    \\
    \lim_{T\rightarrow 0}\frac{1}{\pi}\int_0^\infty\!\dd\varpi\;\rho_K^{\sigma;[\nu\lambda]}(\varpi)
    &=
    c^{(3)}_{[\lambda|\sigma|\nu]}
    -S_{[\lambda|\sigma|\nu]}.
    \label{eq:primitive-circular-channel-main}
\end{align}
The total response sum rule Eq.~\eqref{eq:primitive-absorptive-sumrule-main} can similarly be decomposed into symmetric and antisymmetric parts. Note that in the special case of commuting sources $[B_\mu,B_\nu]=0$, the antisymmetric part of the cumulant vanishes and Eq.~\eqref{eq:primitive-circular-channel-main} depends only on the complex distortion tensor. This generalizes the nonlinear Hall effect sum rule of Ref.~\cite{matsyshyn2019nonlinear} to arbitrary responses in general many-body systems.

\subsection{Physical interpretation of the rectification response}\label{sub:physical_interpretation_of_linear_and_circular_rectification}

Because we derived the rectified response and sum rules for bounded observables, we must take care to properly interpret our results. In particular, because $C_\lambda=\dot{B}_\lambda$ at an operator level per Eq.~\eqref{eq:general-source-history-main-1}, the time-averaged DC current
\begin{equation}
\begin{aligned}
    \frac{1}{\mathcal T}\int_0^{\mathcal T}\!\dd t\,\langle C_\sigma\rangle(t)
    &=\frac{\langle B_\sigma\rangle(\mathcal T)-\langle B_\sigma\rangle(0)}{\mathcal T}
    \\
    &\xrightarrow{\;\mathcal T\to\infty\;}0
\end{aligned}
    \label{eq:primitive-no-dc-main}
\end{equation}
at a formal level by boundedness of $B_\sigma$. This means that, strictly speaking, there is no steady-state DC response for $\langle C_\sigma\rangle_{\mathrm{DC}}$. We should thus interpret $\langle C_\sigma\rangle_{\mathrm{DC}}^{\mathrm{tot}}$ computed in Eq.~\eqref{eq:primitive-conjugate-kernel-main} as the constant $\langle C_\sigma\rangle$ induced at short times. Integrating over time, we see then that this corresponds to a short-time linear growth in $\langle B_\sigma\rangle$ (i.e. an injection of $\langle B_\sigma\rangle$) which remains finite at long time outside of the perturbative regime. In that sense, we can interpret the spectral density $\rho_K$ in Eq.~\eqref{eq:primitive-rho-k-main} as the change in $B$ due to a constant Fermi's golden rule transition rate. 

To make this precise, let us consider the total induced change in $\langle B_\sigma \rangle$ due to a pulse $F^\lambda(t)$ of finite duration. Let $t_f$ denote the time at which the pulse turns off. Then we can define the time-averaged change $\Delta B_\sigma$ after long times as
\begin{equation}
    \Delta B_\sigma
    \equiv\lim_{\mathcal T\to\infty}\frac{1}{\mathcal T}
    \int_{t_f}^{t_f+\mathcal T}\!\!\dd t\,\langle B_\sigma\rangle(t)
    -\langle B_\sigma\rangle(-\infty).
    \label{eq:primitive-displacement-average-main}
\end{equation}
As we show in detail in Appendix~\ref{sec:general_rectification_response_derivation}, the time average can be written in terms of the time-dependent density matrix as
\begin{equation}
    \Delta B_\sigma=\sum_{E_m=E_n}
    \bigl[\rho_{mn}(\infty)-p_m\delta_{mn}\bigr](B_\sigma)_{nm},
    \label{eq:primitive-displacement-pop-main}
\end{equation}
where due to the time average only equal-energy matrix elements of the density matrix contribute. We can compute $\rho_{mn}(\infty)$ using second-order perturbation theory for the density matrix. Working for simplicity in a basis where $B_\sigma$ is diagonalized within degenerate subspaces, we find
\begin{equation}
\begin{aligned}
    \Delta B_\sigma=
    (2\pi)^2\sum_{\omega_{ac}>0}&(p_a-p_c)
    \bigl((B_\sigma)_{cc}-(B_\sigma)_{aa}\bigr)
    \\
    &\times\Bigl|\sum_\lambda\widetilde F^\lambda(\omega_{ac})(B_\lambda)_{ac}\Bigr|^{2}.
\end{aligned}
    \label{eq:primitive-displacement-main}
\end{equation}
Note that the right hand side of Eq.~\eqref{eq:primitive-displacement-main} is exactly the shift vector that appears in the line weight $w_{ac}^{\sigma;\nu\lambda}$ from Eq.~\eqref{eq:primitive-line-weight-nondeg-main}. Thus we may write
\begin{equation}
    \Delta B_\sigma
    =4\pi\sum_{\omega_{ac}>0}
    \widetilde F^\nu(\omega_{ac})\,\widetilde F^{\lambda*}(\omega_{ac})\,
    w_{ac}^{\sigma;\nu\lambda}.
    \label{eq:primitive-displacement-weight-main}
\end{equation}
Combining this observation with the definition of the spectral density in Eq.~\eqref{eq:primitive-rho-k-main} we have
\begin{equation}
    \Delta B_\sigma
    =4\pi\int_0^\infty\!\dd\omega\;
    \widetilde F^\nu(\omega)\,\widetilde F^{\lambda*}(\omega)\,
    \rho_K^{\sigma;\nu\lambda}(\omega).
    \label{eq:primitive-displacement-fold-main}
\end{equation}
Thus, the spectral density itself is directly measurable from the source frequency dependence of $\Delta B_\sigma$.

We can apply Eq.~\eqref{eq:primitive-displacement-fold-main} to extract part of the geometric sum rule Eq.~\eqref{eq:primitive-absorptive-sumrule-main} directly at zero temperature. In particular, we consider the change $\Delta B_\sigma$ due to an instantaneous pulse $F^\lambda(t) = F^\lambda \delta(t)$, which has a flat frequency spectrum $\widetilde F^\lambda(\omega)=F^\lambda/(2\pi)$. Inserting this into Eq.~\eqref{eq:primitive-displacement-main} and using the definition of the cumulant Eq.~\eqref{eq:primitive-c3-main} and complex distortion tensor Eq.~\eqref{eq:ordered-s-main}, we have
\begin{equation}
\begin{aligned}
    \lim_{T\rightarrow 0}\Delta B_\sigma
    &=\sum_{\nu\lambda}F^\nu F^\lambda
    \left(c^{(3)}_{\lambda\sigma\nu}-S_{\lambda\sigma\nu}\right)
    \\
    &=\sum_{\nu\lambda}F^\nu F^\lambda
    \left(c^{(3)}_{(\lambda|\sigma|\nu)}-S_{(\lambda|\sigma|\nu)}\right).
\end{aligned}
    \label{eq:primitive-instantaneous-main}
\end{equation}
Since $F^\nu$ is real for an instantaneous pulse, only the real, symmetric part $S_{(\lambda|\sigma|\nu)}$ of the complex distortion tensor enters into the instantaneous response. From Eq.~\eqref{eq:s-response-contorsion-main}, this is the J-rotated distortion tensor of the mixture connection. Thus when combined with a measurement of the cumulant $c^{(3)}_{\lambda\sigma\nu}$, which can be independently measured in equilibrium, the nonmetricity of the mixture connection is directly measurable from $\Delta B_\sigma$. 

Now that we have developed the general connection between rectification and many-body quantum geometry, we will turn to the practically interesting case of DC photocurrents. By taking $C_\sigma$ to be the current density and $B_\nu$ to be the position operator, we will be able to derive many-body geometric formulas for shift current sum rules in the length gauge.

\section{Geometry of shift current in bounded systems}\label{sec:application_shift_current_in_bounded_systems}

As a simple, physical example of our formalism, we can examine the many-body quantum geometry associated to shift current in systems with open boundary conditions. In this case, our perturbation couples to the many-body position operator,
\begin{equation}
    B_\lambda=X_\lambda,
\end{equation}
and we will measure the response of the charge current
\begin{equation}
    C_\sigma
    =\ii\comm{H(t)}{X_\sigma}
    =\ii\comm{H_0}{X_\sigma}
    \equiv J_\sigma.
    \label{eq:obc-dictionary-main}
\end{equation}
Note, importantly, that in this length-gauge picture there is no diamagnetic contribution to $C_\sigma$, due to the fact that all components of the position operator commute with each other.

On the geometry side, from Eq.~\eqref{eq:source-cubic-to-b-main} we have that the relevant quasistatic cQAC tensor is
\begin{equation}
    Z_{\sigma;\mu\nu}
    =
    \ii\sideset{}{'}\sum_{a,b,k}
    t_{ab}\,(p_k-p_b)\,t_{ka}\,
    (X_\mu)_{ab}(X_\sigma)_{bk}(X_\nu)_{ka}.
    \label{eq:position-cubic-main}
\end{equation}
The many-body multi-state geometric contribution to response at $T\rightarrow 0$ is determined by the complex distortion tensor
\begin{equation}
    \lim_{T\rightarrow 0}S_{\lambda\sigma\nu}
    =
    \sideset{}{'}\sum_{a,b>0}
    (X_\lambda)_{0a}(X_\sigma)_{ab}(X_\nu)_{b0}.
    \label{eq:obc-position-s-main}
\end{equation}

Following Sec.~\ref{sec:rectification}, we consider the DC response of the current to a perturbation coupled to $X_\mu$. In this case, the external fields $\widetilde F^\mu(\omega)$ are the Fourier components of an external electric field. The intensive DC current response can be written as
\begin{equation}
\begin{aligned}
\frac{1}{V}\langle J_\sigma \rangle_{\mathrm{DC}}&=\frac{1}{V}\sum_{\omega_{ac}>0}
    \bigl[w_{ac}^{\sigma;\nu\lambda}+\bigl(w_{ac}^{\sigma;\lambda\nu}\bigr)^{*}\bigr]
    \\
    &\qquad\times
    \widetilde F^\nu(\omega_{ac})\,\widetilde F^{\lambda*}(\omega_{ac})
    \\
    &\equiv \int_0^\infty \dd\omega\; \sigma_{\mathrm{shift}}^{\sigma;\nu\lambda}(\omega)\,\widetilde F^\nu(\omega)\,\widetilde F^{\lambda*}(\omega), \label{eq:dc-current-response}
\end{aligned}
\end{equation}
where $V$ is the volume of the system and the line weight $w_{ac}^{\sigma;\nu\lambda}$ is given by
\begin{equation}
\begin{aligned}
    w_{ac}^{\sigma;\nu\lambda}
    &=\pi(p_a-p_c)\comm{X_\nu}{X_{\sigma,\mathrm d}}_{ac}(X_\lambda)_{ca}
    \\
    &=\pi(p_a-p_c)
    \bigl((X_\sigma)_{cc}-(X_\sigma)_{aa}\bigr)
    (X_\nu)_{ac}(X_\lambda)_{ca}.
\end{aligned}
    \label{eq:obc-shift-weight-main}
\end{equation}
We see that the line weight takes the form of the shift vector contribution to the DC current~\cite{alexandradinata2024quantization,morimoto2016topological,aversa1995nonlinear,Resta_2024,vonbaltz1981theory,nastos2006optical,young2012first}. The positive-frequency shift conductivity $\sigma_{\mathrm{shift}}^{\sigma;\nu\lambda}(\omega)$ is, from Eq.~\eqref{eq:primitive-absorptive},
\begin{equation}
\sigma_{\mathrm{shift}}^{\sigma;\nu\lambda}(\omega) = \frac{1}{V}\sum_{\omega_{ac}>0}[w_{ac}^{\sigma;\nu\lambda}+(w_{ac}^{\sigma;\lambda\nu})^*]\delta(\omega - \omega_{ac}). \label{eq:shiftcon}
\end{equation}

From Eq.~\eqref{eq:primitive-absorptive-sumrule-main}, the shift conductivity satisfies a geometric sum rule at zero temperature. In particular, integrating Eq.~\eqref{eq:shiftcon} over positive frequency, we have
\begin{equation}
\begin{aligned}
    &\lim_{T\rightarrow 0}\frac{1}{\pi}\int_0^\infty \dd\omega\;\sigma_{\mathrm{shift}}^{\sigma;\nu\lambda}(\omega)
    \\
    &\quad=  \lim_{T\rightarrow0}\frac{1}{\pi V}\sum_{\omega_{ac}>0}
    \bigl[w_{ac}^{\sigma;\nu\lambda}+\bigl(w_{ac}^{\sigma;\lambda\nu}\bigr)^{*}\bigr]
    \\
    &\quad=
    \frac{2}{V}\bigl(c^{(3)}_{\lambda\sigma\nu}-S_{\lambda\sigma\nu}\bigr).
    \label{eq:obc-sum-rule-main}
    \end{aligned}
\end{equation}
Here the complex distortion tensor is given by Eq.~\eqref{eq:obc-position-s-main}, and $c^{(3)}_{\lambda\sigma\nu}$ is the third cumulant of position operators in the ground state, also known as the ground state skewness. Equation~\eqref{eq:obc-sum-rule-main} generalizes the multi-state rectification sum rule of Ref.~\cite{avdoshkin2024multi}. Separating the sum rule Eq.~\eqref{eq:obc-sum-rule-main} into its real and imaginary parts (which coincides with taking the symmetric and antisymmetric parts, respectively), we have
\begin{align}
    &\operatorname{Re}\,\lim_{T\rightarrow0}\frac{1}{\pi V}\sum_{\omega_{ac}>0}
    \bigl[w_{ac}^{\sigma;\nu\lambda}+\bigl(w_{ac}^{\sigma;\lambda\nu}\bigr)^{*}\bigr]
    \nonumber\\
    &\qquad=
    \frac{2}{V}\bigl(c^{(3)}_{\lambda\sigma\nu}-\operatorname{Re}S_{\lambda\sigma\nu}\bigr) \nonumber \\
    &\qquad= \frac{2}{V}\bigl(c^{(3)}_{(\lambda|\sigma|\nu)}-S_{(\lambda|\sigma|\nu)}\bigr),
    \label{eq:obc-skewness-sum-rule-main}
    \\
    &\operatorname{Im}\,\lim_{T\rightarrow0}\frac{1}{\pi V}\sum_{\omega_{ac}>0}
    \bigl[w_{ac}^{\sigma;\nu\lambda}+\bigl(w_{ac}^{\sigma;\lambda\nu}\bigr)^{*}\bigr]
    \nonumber\\
    &\qquad=
    -\frac{2}{V}\operatorname{Im}S_{\lambda\sigma\nu} \nonumber \\
    &\qquad = \frac{2\ii}{V}S_{[\lambda|\sigma|\nu]}.
    \label{eq:obc-circular-sum-rule-main}
\end{align}
In particular, because the position operators are mutually commuting, the skewness is totally symmetric. As such, the imaginary (circular) component of the sum rule directly measures $\operatorname{Im} S_{\lambda\sigma\nu}$.

The real part of the sum rule directly controls the accumulated electronic polarization $\delta P_\sigma = \Delta X_\sigma/V$ left behind by a finite-duration pulse, which is the true long-time observable in an open system (since a steady DC current cannot flow with open boundary conditions). In particular, using our results of Sec.~\ref{sub:physical_interpretation_of_linear_and_circular_rectification}, we find that after a pulse the accumulated polarization is
\begin{equation}
    \delta P_\sigma
    \equiv\frac{1}{V}\Delta X_\sigma
    =\frac{4\pi}{V}\sum_{\omega_{ac}>0}
    \widetilde F^\nu(\omega_{ac})\,\widetilde F^{\lambda*}(\omega_{ac})\,
    w_{ac}^{\sigma;\nu\lambda}.
    \label{eq:obc-polarization-main}
\end{equation}
If we now specialize to an instantaneous pulse, we have from Eq.~\eqref{eq:primitive-instantaneous-main} at zero temperature that
\begin{equation}
\delta P_\sigma \rightarrow\frac{1}{V}\sum_{\nu\lambda}F^\nu F^\lambda
    \left(c^{(3)}_{\lambda\sigma\nu}-\operatorname{Re}S_{\lambda\sigma\nu}\right),\label{eq:population-polarization-inst}
\end{equation}
where we used the fact that the skewness is totally symmetric, and that the symmetric part of $S_{\lambda\sigma\nu}$ is the real part. Equation~\eqref{eq:population-polarization-inst} shows that the polarization injected by an instantaneous pulse is a direct measure of the J-rotated distortion tensor $\operatorname{Re}S_{\lambda\sigma\nu}$ when combined with an independent static measurement of the cumulant.

We have thus derived a many-body shift current sum rule for finite systems (i.e. in the length gauge). Next, we apply our results to free fermion systems, where we can take the thermodynamic limit to make contact with existing literature. In particular, we will show how the cQAC tensor sheds new light on the sum rules of Ref.~\cite{avdoshkin2024multi}.

\section{Application to free fermions}\label{sec:application_to_free_fermions}

Let us now apply our analysis of the shift current to free fermion systems, to make contact with the majority of the literature. This will also allow us to take the thermodynamic limit and formally consider the shift current in infinite systems. First, in Sec.~\ref{sub:from_many_body_operators_to_single_particle_matrix_elements} we will evaluate the single-particle matrix elements of the many-body position operator that we will need to compute the geometry and response. Then, in Sec.~\ref{sub:thermodynamic_limit_blount} we will take the thermodynamic limit to express the shift conductivity in terms of momentum derivatives of the matrix elements. Putting this together with our results in Sec.~\ref{sec:application_shift_current_in_bounded_systems}, we will in Sec.~\ref{sub:the_single_particle_multi_state_geometric_sum_rule} derive the single-particle expression for the multi-state geometric sum rule. For systems with multiple occupied bands, we show that our cQAC approach naturally organizes the multiband corrections mentioned in Ref.~\cite{avdoshkin2024multi}. Finally, in Sec.~\ref{sub:example_the_four_band_kane_mele_model}, we demonstrate our results numerically by computing the shift current and multi-state geometry in a generalized Kane-Mele model.

Note, importantly, that because we work in the length gauge by taking the thermodynamic limit of a system with open boundary conditions, we recover only the shift response from our DC Kubo formula. A more complete treatment that either works in periodic boundary conditions (or else carefully works in the thermodynamic limit with unbounded operators from the outset) would be necessary to derive the injection current response. We defer this treatment to future work.

\subsection{From many-body operators to single-particle matrix elements}\label{sub:from_many_body_operators_to_single_particle_matrix_elements}

To begin, recall that the position operator $X_\mu$ is a one-body operator. Introducing a basis $\{\ket{p}\}$ of single-particle eigenstates of energy $\varepsilon_p$ created (annihilated) by operators $c^\dag_p$ ($c_p$), we can write
\begin{equation}
X_\mu
    =
    \sum_{p,q}
    \langle p|X_\mu|q\rangle\,
    c_p^\dagger c_q. \label{eq:pos-1body}
    \end{equation}
Note that since we are working in a finite-sized system, Eq.~\eqref{eq:pos-1body} includes both diagonal and off-diagonal matrix elements. For the purpose of taking the thermodynamic limit, it is helpful to separate out the part of the position operator that is off-diagonal in energy. We define the transition matrix elements $\mathfrak r^\mu_{pq}$ and the off-diagonal position operator $X_{\mu,\mathrm{od}}$ as
\begin{equation}
\begin{aligned}
\mathfrak r^\mu_{pq}
    &\equiv
    \langle p|X_\mu|q\rangle,
    \qquad p\neq q,
    \\
X_{\mu,\mathrm{od}}
    &\equiv
    \sideset{}{'}\sum_{p,q}\mathfrak r^\mu_{pq}\,c^\dag_p c_q
    \equiv
    \sum_{p,q}(x_{\mathrm{od}})^\mu_{pq}\,c^\dag_p c_q. \label{eq:r-def}
\end{aligned}
\end{equation}
The transition matrix elements $\mathfrak r^\mu_{pq}$ are related to the one-body matrix elements $j^\mu_{pq}=\bra{p}J_\mu\ket{q}$ of the current operator by the specialization of Eq.~\eqref{eq:obc-dictionary-main}
\begin{equation}
j^\mu_{pq}
    =
    \ii\,(\varepsilon_p-\varepsilon_q)\,
    \mathfrak r^\mu_{pq}. \label{eq:j-one-body-def}
\end{equation} 

Note that when acting on a filled Fermi sea, a one-body operator like $X_\mu$ or $J_\mu$ creates single particle-hole pairs. In order to evaluate nonlinear response functions and the cQAC tensor, we need to be able to evaluate the action of products of one-body operators on the ground state. Let $\ket{\mathrm{FS}}$ denote the filled Fermi sea of an insulator, and let $\ket{a,i^{-1}}\equiv c^\dag_{a} c_{i}\ket{\mathrm{FS}}$ be a state with a single particle-hole excitation. Then the matrix elements we will need are
\begin{equation}
\begin{aligned}
    \langle a\,i^{-1}|X_\sigma|b\,j^{-1}\rangle
    &=
    \delta_{ij}\mathfrak r^\sigma_{ab}
    -
    \delta_{ab}\mathfrak r^\sigma_{ji},
    \quad
    (a,i)\neq(b,j),
    \\
    \langle a\,i^{-1}|X_\sigma|a\,i^{-1}\rangle
    &=
    \bra{\mathrm{FS}} X_\sigma\ket{\mathrm{FS}}
    +(X_\sigma)_{aa}-(X_\sigma)_{ii},
\end{aligned}
    \label{eq:ff-sc-main}
\end{equation}
which follow from the canonical anticommutation relations.

For any finite-sized system, the diagonal matrix elements of $X_\sigma$ appearing in Eqs.~\eqref{eq:pos-1body} and \eqref{eq:ff-sc-main} are finite and well defined. To take the thermodynamic limit, we will follow Refs.~\cite{blount1962formalism,aversa1995nonlinear,parker2019diagrammatic}. We will see that, in the quantities we are interested in, the diagonal matrix elements of the position operator will be replaced by (Berry) covariant derivatives with respect to crystal momentum.

\subsection{The thermodynamic limit: position matrix elements and the covariant derivative}\label{sub:thermodynamic_limit_blount}

We now specialize to an infinite, periodic, free fermion crystal. In this case, the single-particle eigenstates can be labeled by their band index $n$ and crystal momentum $\kk$, $\ket{q}\rightarrow \ket{n\kk}$. Using Bloch's theorem, we have
\begin{equation}
\ket{n\kk} = e^{\ii\kk\cdot X}\ket{u_{n\kk}}. \label{eq:unk}
\end{equation}
Recall also that sums over single-particle eigenstates $q$ become sums over band indices and integrals over the Brillouin zone (BZ). In terms of the states of Eq.~\eqref{eq:unk}, the resolution of the identity and the conversion of eigenstate sums read
\begin{equation}
\begin{aligned}
    \Id
    &=
    \sum_{n}\,
    \mathfrak{v}\!\int_{\mathrm{BZ}}\!\frac{\dd^d k}{(2\pi)^d}\,
    \ket{n\kk}\bra{n\kk},
    \\
    \frac{1}{V}\sum_{q}
    &\;\longrightarrow\;
    \sum_{n}\,
    \int_{\mathrm{BZ}}\!\frac{\dd^d k}{(2\pi)^d}\,,
\end{aligned}
    \label{eq:ff-k-structure-main}
\end{equation}
where $d$ is the dimensionality of the crystal, $\mathfrak{v}$ is the primitive unit cell volume, and $V$ is the volume of the crystal that is taken to infinity in the thermodynamic limit. Note that, when using Eq.~\eqref{eq:ff-k-structure-main} to evaluate expectation values and matrix elements of single-particle operators, we will use periodic boundary conditions at finite $V$. This is justified as long as we only consider intensive quantities constructed from operators that are well defined on the torus, such as off-diagonal position matrix elements or current densities. Equation~\eqref{eq:ff-k-structure-main} corresponds to the continuum normalization convention that leaves $\ket{u_{n\kk}}$ normalized to $1$ within a single unit cell. We can
evaluate the matrix elements of the position operator in the basis Eq.~\eqref{eq:unk} to find
\begin{equation}
\begin{aligned}
    \langle n\kk|X_\mu|m\kk'\rangle
    &=
    \frac{(2\pi)^d}{\mathfrak{v}}\big[\ii\,\delta_{nm}\,\partial_{k_\mu}\delta(\kk-\kk')
    \\
    &\quad
    +(A_\mu)_{nm}(\kk)\,\delta(\kk-\kk')\big],
\end{aligned}
    \label{eq:blount-position-main}
\end{equation}
where the Berry connection $A_\mu$ is given by
\begin{equation}
    (A_\mu)_{nm}
    =
    \ii\langle u_{n\kk}|\partial_{k_\mu} u_{m\kk}\rangle .\label{eq:berrydef}
\end{equation}
Comparing Eq.~\eqref{eq:blount-position-main} with Eq.~\eqref{eq:r-def}, we can introduce the $\kk$-dependent dipole matrix element
\begin{equation}
\begin{aligned}
    \mathfrak r^\mu_{(n\kk),(m\kk')}
    &=
    \frac{(2\pi)^d}{\mathfrak{v}}\delta(\kk-\kk')\,\mathfrak r^\mu_{nm}(\kk) \\
    &=\frac{(2\pi)^d}{\mathfrak{v}}\delta(\kk-\kk')(A_\mu)_{nm}(\kk),
    \qquad n\neq m. \label{eq:r-from-A}
    \end{aligned}
\end{equation}

From Eq.~\eqref{eq:blount-position-main}, we see that the position operator acts as a covariant derivative when expressed in the basis of Bloch states. In particular, let $\mathcal{O}$ be a $\kk$-diagonal one-body operator, and let
\begin{equation}
\bra{n\kk}\mathcal{O}\ket{m\kk'} \equiv \frac{(2\pi)^d}{\mathfrak{v}}\,\mathcal{O}_{nm}(\kk)\,\delta(\kk-\kk'). 
\end{equation}
Then we have
\begin{equation}
\begin{aligned}
\bra{n\kk}\comm{X_\mu}{\mathcal{O}}\ket{m\kk'} &\equiv \frac{(2\pi)^d}{\mathfrak{v}}\,\ii D_\mu \mathcal{O}_{nm}(\kk)\,\delta(\kk-\kk'), \\
D_\mu \mathcal{O}_{nm}(\kk)&=\left(\partial_\mu \mathcal{O}_{nm}(\kk) - \ii\comm{A_\mu(\kk)}{\mathcal{O}(\kk)}_{nm}\right), \label{eq:cov-deriv-def}
\end{aligned}
\end{equation}
where $D_\mu$ is the Berry-covariant derivative. 

We can use Eq.~\eqref{eq:cov-deriv-def} to evaluate the line weights Eq.~\eqref{eq:obc-shift-weight-main} and shift conductivity for free fermions in the thermodynamic limit. Introducing the shorthand
\begin{equation}
\begin{aligned}
(\mathfrak r^\nu_{\;;\sigma})_{ac}(\kk)
    &\equiv (D_\sigma x^\nu_{\mathrm{od}})_{ac}(\kk)-\ii\comm{x^\nu_{\mathrm{od}}}{x^\sigma_{\mathrm{od}}}_{ac}(\kk),
\end{aligned}
    \label{eq:ff-covariant-r-main}
\end{equation}
we show in Appendix~\ref{sec:free_fermion_reduction_formulas} that
\begin{equation}
\begin{aligned}
    w_{ac}^{\sigma;\nu\lambda}(\kk)
    &=
    -\ii\pi[f_a(\kk)-f_c(\kk)]\,
    (\mathfrak r^\nu_{\;;\sigma})_{ac}(\kk)\,
    \mathfrak r^\lambda_{ca}(\kk),
\end{aligned}
    \label{eq:ff-general-weight-main}
\end{equation}
where $f_a(\kk)$ is the Fermi function for the single-particle energy $\varepsilon_a(\kk)$. 
Substituting into Eq.~\eqref{eq:shiftcon} yields the free fermion shift conductivity
\begin{equation}
\begin{aligned}
    \sigma_{\mathrm{shift}}^{\sigma;\nu\lambda}(\omega)
    ={}&
    \int_{\mathrm{BZ}}\!\frac{\dd^d k}{(2\pi)^d}
    \sum_{\omega_{ac}(\kk)>0}
    \bigl[w_{ac}^{\sigma;\nu\lambda}(\kk)
    +\bigl(w_{ac}^{\sigma;\lambda\nu}(\kk)\bigr)^{*}\bigr]
    \\
    &\times
    \delta\bigl(\omega-\omega_{ac}(\kk)\bigr).
\end{aligned}
    \label{eq:ff-shift-conductivity-main}
\end{equation}
Equations~\eqref{eq:ff-general-weight-main} and \eqref{eq:ff-shift-conductivity-main} agree with the results of Refs.~\cite{aversa1995nonlinear,ahn2022riemannian,ahn2020lowfrequency,avdoshkin2024multi,Resta_2024}. In particular, 
note that if we work in a basis where the Berry connection is diagonal when restricted to each degenerate subspace of the Hamiltonian, then we can rewrite Eq.~\eqref{eq:ff-covariant-r-main} as
\begin{equation}
(\mathfrak r^\nu_{\;;\sigma})_{ac}(\kk)
    \rightarrow \partial_\sigma\mathfrak{r}^\nu_{ac}-\ii\left[(A_\sigma)_{aa}-(A_\sigma)_{cc}\right]\mathfrak{r}^\nu_{ac}, \label{eq:r-semicolon-simplified}
\end{equation}
for $\varepsilon_a(\kk)\neq\varepsilon_c(\kk)$. This makes direct contact with the Fermi's golden rule result for the free fermion shift current found in the literature~\cite{ahn2020lowfrequency,Resta_2024}.

Next, armed with these results, we can now derive the general multi-state geometric sum rule for the shift conductivity and connect it to the free fermion cQAC tensor. 

\subsection{The single-particle multi-state geometric sum rule}\label{sub:the_single_particle_multi_state_geometric_sum_rule}

To tie the shift conductivity Eq.~\eqref{eq:ff-shift-conductivity-main} to quantum geometry, we consider an insulator at zero temperature. For such a system, we can compute the ground state skewness cumulant Eq.~\eqref{eq:primitive-c3-main}. As we show in Appendix~\ref{sec:free_fermion_reduction_formulas}, we can use Eqs.~\eqref{eq:ff-sc-main} and \eqref{eq:blount-position-main} to write the skewness per unit volume as
\begin{equation}
\begin{aligned}
    \frac{1}{V}\,c^{(3),\mathrm{ff}}_{\lambda\sigma\nu}
    ={}&
    \int_{\mathrm{BZ}}\!\frac{\dd^d k}{(2\pi)^d}
    \Biggl[\;
    \sum_{i\in\mathrm{occ}}
    \sideset{}{'}\sum_{a,b\in\mathrm{unocc}}
    \mathfrak r^\lambda_{ia}\mathfrak r^\sigma_{ab}
    \mathfrak r^\nu_{bi}
    \\
    &-
    \sum_{a\in\mathrm{unocc}}
    \sideset{}{'}\sum_{i,j\in\mathrm{occ}}
    \mathfrak r^\lambda_{ia}\mathfrak r^\sigma_{ji}
    \mathfrak r^\nu_{aj}
    \\
    &+
    \ii
    \sum_{i\in\mathrm{occ}}
    \sum_{a\in\mathrm{unocc}}
    \mathfrak r^\lambda_{ia}\,
    (\mathfrak r^\nu_{\;;\sigma})_{ai}
    \Biggr].
\end{aligned}
    \label{eq:ff-c3-main}
\end{equation}
We can recognize the last sum in Eq.~\eqref{eq:ff-c3-main} as proportional to the frequency integral of the shift conductivity in a zero-temperature insulator. The first two sums are exactly the free fermion expression for the complex distortion tensor Eq.~\eqref{eq:obc-position-s-main}. The free fermion cQAC tensor is, from Eq.~\eqref{eq:position-cubic-main},
\begin{align}
    \frac{1}{V}\,Z^{\mathrm{ff}}_{\sigma;\mu\nu}
    &=
    \int_{\mathrm{BZ}}\!\frac{\dd^d k}{(2\pi)^d}
    \Biggl[\,
    \ii
    \sum_{\alpha\beta\gamma}
    t_{\alpha\beta}t_{\gamma\alpha}
    (1-f_\alpha)\Delta f_{\gamma\beta}
    \nonumber\\
    &\qquad\qquad\times
    \mathfrak r^\mu_{\alpha\beta}
    \mathfrak r^\sigma_{\beta\gamma}
    \mathfrak r^\nu_{\gamma\alpha}
    \nonumber\\
    &\quad
    -\ii
    \sum_{\alpha\beta\gamma}
    t_{\alpha\beta}t_{\beta\gamma}
    f_\beta\Delta f_{\alpha\gamma}\,
    \mathfrak r^\mu_{\alpha\beta}
    \mathfrak r^\nu_{\beta\gamma}
    \mathfrak r^\sigma_{\gamma\alpha}
    \Biggr],
    \label{eq:ff-z-main}
\end{align}
where $\Delta f_{\gamma\beta}=f_\gamma-f_\beta$. Taking the zero-temperature limit, acting with the almost complex structure map $J$, and taking the cyclic sum yields
\begin{equation}
\begin{aligned}
    S^{\mathrm{ff}}_{\lambda\sigma\nu}
    ={}&
    S^{\mathrm{part}}_{\lambda\sigma\nu}
    +
    S^{\mathrm{hole}}_{\lambda\sigma\nu},
    \\
    \frac{1}{V}\,S^{\mathrm{part}}_{\lambda\sigma\nu}
    ={}&
    \int_{\mathrm{BZ}}\!\frac{\dd^d k}{(2\pi)^d}
    \sum_{i\in\mathrm{occ}}
    \sideset{}{'}\sum_{a,b\in\mathrm{unocc}}
    \mathfrak r^\lambda_{ia}\mathfrak r^\sigma_{ab}
    \mathfrak r^\nu_{bi},
    \\
    \frac{1}{V}\,S^{\mathrm{hole}}_{\lambda\sigma\nu}
    ={}&
    -\int_{\mathrm{BZ}}\!\frac{\dd^d k}{(2\pi)^d}
    \sum_{a\in\mathrm{unocc}}
    \sideset{}{'}\sum_{i,j\in\mathrm{occ}}
    \mathfrak r^\lambda_{ia}\mathfrak r^\sigma_{ji}
    \mathfrak r^\nu_{aj}.
\end{aligned}
    \label{eq:ff-s-main}
\end{equation}
$S^{\mathrm{ff}}_{\lambda\sigma\nu}$ is the multi-state geometric correction to the shift current sum rule noted in Ref.~\cite{avdoshkin2024multi}, which computed it explicitly for a system with one occupied band. Note that $S^\mathrm{part}_{\lambda\sigma\nu}$ vanishes when there is only one unoccupied band. Similarly, $S^\mathrm{hole}_{\lambda\sigma\nu}$ vanishes when there is only one occupied band. As such, $S^\mathrm{hole}_{\lambda\sigma\nu}$ represents the multiband corrections mentioned in Ref.~\cite{avdoshkin2024multi}. 

Combining Eqs.~\eqref{eq:ff-shift-conductivity-main} and \eqref{eq:ff-c3-main}, we thus have the geometric sum rule
\begin{equation}
    \lim_{T\rightarrow0}
    \frac{1}{\pi}\int_0^\infty\!\dd\omega\;
    \sigma_{\mathrm{shift}}^{\sigma;\nu\lambda}(\omega)
    =
    \frac{2}{V}\bigl(
        c^{(3),\mathrm{ff}}_{\lambda\sigma\nu}
        -
        S^{\mathrm{ff}}_{\lambda\sigma\nu}
    \bigr).
    \label{eq:ff-sum-rule-main}
\end{equation}

As an illustrative example, we will look at Eq.~\eqref{eq:ff-sum-rule-main} applied to an explicit tight-binding model. In order to ensure that both terms in Eq.~\eqref{eq:ff-s-main} for the multi-state geometric correction are nonzero, our model must have at least four bands, two occupied and two unoccupied. We will thus turn to generalizations of the Kane-Mele model.

\subsection{Example: the four-band generalized Kane-Mele model}\label{sub:example_the_four_band_kane_mele_model}

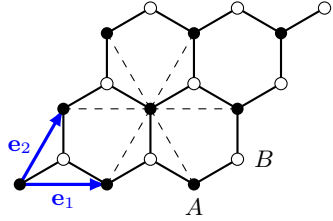
\begin{figure}[!ht]
\centering
\begin{tikzpicture}[scale=1.15]
  \foreach \m in {0,1,2}{
    \foreach \n in {0,1,2}{
      \coordinate (A\m\n) at ({\m + 0.5*\n + 0.5}, {0.86603*\n + 0.28868});
      \coordinate (B\m\n) at ({\m + 0.5*\n + 1.0}, {0.86603*\n + 0.57735});
  }}
  \foreach \m in {0,1,2}{ \foreach \n in {0,1,2}{
      \draw[black, thick] (A\m\n) -- (B\m\n);
  }}
  \foreach \m in {0,1,2}{ \foreach \n in {1,2}{
      \pgfmathtruncatemacro{\nm}{\n-1}
      \draw[black, thick] (A\m\n) -- (B\m\nm);
  }}
  \foreach \m in {1,2}{ \foreach \n in {0,1,2}{
      \pgfmathtruncatemacro{\mm}{\m-1}
      \draw[black, thick] (A\m\n) -- (B\mm\n);
  }}
  \draw[dashed] (A11) -- (A21);
  \draw[dashed] (A11) -- (A01);
  \draw[dashed] (A11) -- (A12);
  \draw[dashed] (A11) -- (A10);
  \draw[dashed] (A11) -- (A20);
  \draw[dashed] (A11) -- (A02);
  \draw[-latex, very thick, blue] (A00) -- (A10)
    node[midway, below] {$\mathbf e_1$};
  \draw[-latex, very thick, blue] (A00) -- (A01)
    node[midway, left] {$\mathbf e_2$};
  \foreach \m in {0,1,2}{ \foreach \n in {0,1,2}{
      \fill (A\m\n) circle (0.07);
      \draw[fill=white] (B\m\n) circle (0.07);
  }}
  \node[below=2pt] at (A20) {$A$};
  \node[right=3pt] at (B20) {$B$};
\end{tikzpicture}
\caption{%
Lattice structure for the generalized Kane-Mele model in 
Eq.~\eqref{eq:km-model-main}. The $A$ and $B$ sublattice
sites are depicted by filled and open circles, respectively. The primitive lattice vectors $\mathbf e_1=a(1,0)$,
$\mathbf e_2=a(1/2,\sqrt3/2)$ are shown as blue arrows, with $a$ the lattice constant, which we set to $1$ for convenience.
Solid lines denote the nearest-neighbor bonds. We show the six next-nearest-neighbor bonds
of a representative $A$ site with dashed lines.
}
\label{fig:km-lattice}
\end{figure}

Let us now apply our results to compute the shift conductivity, cumulant, and complex distortion tensor for an example free fermion model. We consider a generalized Kane-Mele-type model~\cite{kane2005quantum} on the honeycomb lattice. The lattice structure is depicted in Fig.~\ref{fig:km-lattice}. We assume we have one spinful $p_z$-type orbital on each lattice site. We use a multi-index $i=(\mathbf{R}_i,\Pi_i)$ to denote the unit cell index $\mathbf{R}_i$ and sublattice $\Pi_i=A$ or $B$, and denote by $c_{i\alpha}$ the annihilation operator for an electron at site $i$ with spin $\alpha$. In this notation, we take for our Hamiltonian
\begin{equation}
\begin{aligned}
    H ={}&
    t\sum_{\langle ij\rangle} c_{i\alpha}^\dagger c_{j\alpha}^{\phantom\dagger}
    +\ii\lambda_{\mathrm{SO}}\!\!\sum_{\langle\!\langle ij\rangle\!\rangle}\!\!
    \nu_{ij}\, c_{i\alpha}^\dagger (s_z)_{\alpha\beta}\, c_{j\beta}^{\phantom\dagger}
    \\
    &+\ii\lambda_{\mathrm{R}}\sum_{\langle ij\rangle}
    c_{i\alpha}^\dagger\,
    \bigl[\bigl(\mathbf{s}\times\hat{\mathbf d}_{ij}\bigr)_z\bigr]_{\alpha\beta}\,
    c_{j\beta}^{\phantom\dagger}
    \\
    &+ m\sum_{i} c_{i\alpha}^\dagger (s_z)_{\alpha\beta}\, c_{i\beta}^{\phantom\dagger}
    \;+\; \lambda_M H_M,
\end{aligned}
    \label{eq:km-model-main}
\end{equation}
where $\langle ij\rangle$ denotes a sum over nearest-neighbor bonds and $\langle\!\langle ij\rangle\!\rangle$ denotes a sum over next-nearest-neighbor bonds. 
Additionally, $t$ is the nearest-neighbor hopping, $\lambda_{\mathrm{SO}}$ is the Kane-Mele-type spin-orbit coupling strength, $\lambda_{\mathrm{R}}$ is the Rashba spin-orbit coupling strength, 
$\hat{\mathbf{d}}_{ij}$ is the unit vector pointing from site $i$ to site $j$, 
$\nu_{ij}=+1$ $(-1)$ when the shortest path from site $j$ to site $i$ turns counterclockwise (clockwise), and $m$ is a Zeeman field. The mirror-symmetry-breaking perturbation $H_M$ is defined as~\cite{monaco2020spin}
\begin{equation}
\begin{aligned}
    H_M ={}&
    -\frac{1}{2\ii}\sum_{i\in A}
    \bigl[
    c_{i+\mathbf e_1,\alpha}^\dagger (s_x)_{\alpha\beta}\,
    c_{i\beta}^{\phantom\dagger}
    +c_{i+\mathbf e_2,\alpha}^\dagger (s_y)_{\alpha\beta}\,
    c_{i\beta}^{\phantom\dagger}
    \bigr]
    \\
    &-\frac{1}{2\ii}\sum_{i\in B}
    \bigl[
    -c_{i+\mathbf e_2,\alpha}^\dagger (s_y)_{\alpha\beta}\,
    c_{i\beta}^{\phantom\dagger}
    \\
    &\qquad\quad
    +c_{i+\mathbf e_1-\mathbf e_2,\alpha}^\dagger (s_z)_{\alpha\beta}\,
    c_{i\beta}^{\phantom\dagger}
    \bigr]
    \;+\;\mathrm{h.c.}.
\end{aligned}
    \label{eq:km-mirror-term-main}
\end{equation}

\begin{figure*}[!ht]
\centering
\includegraphics[width=\textwidth]{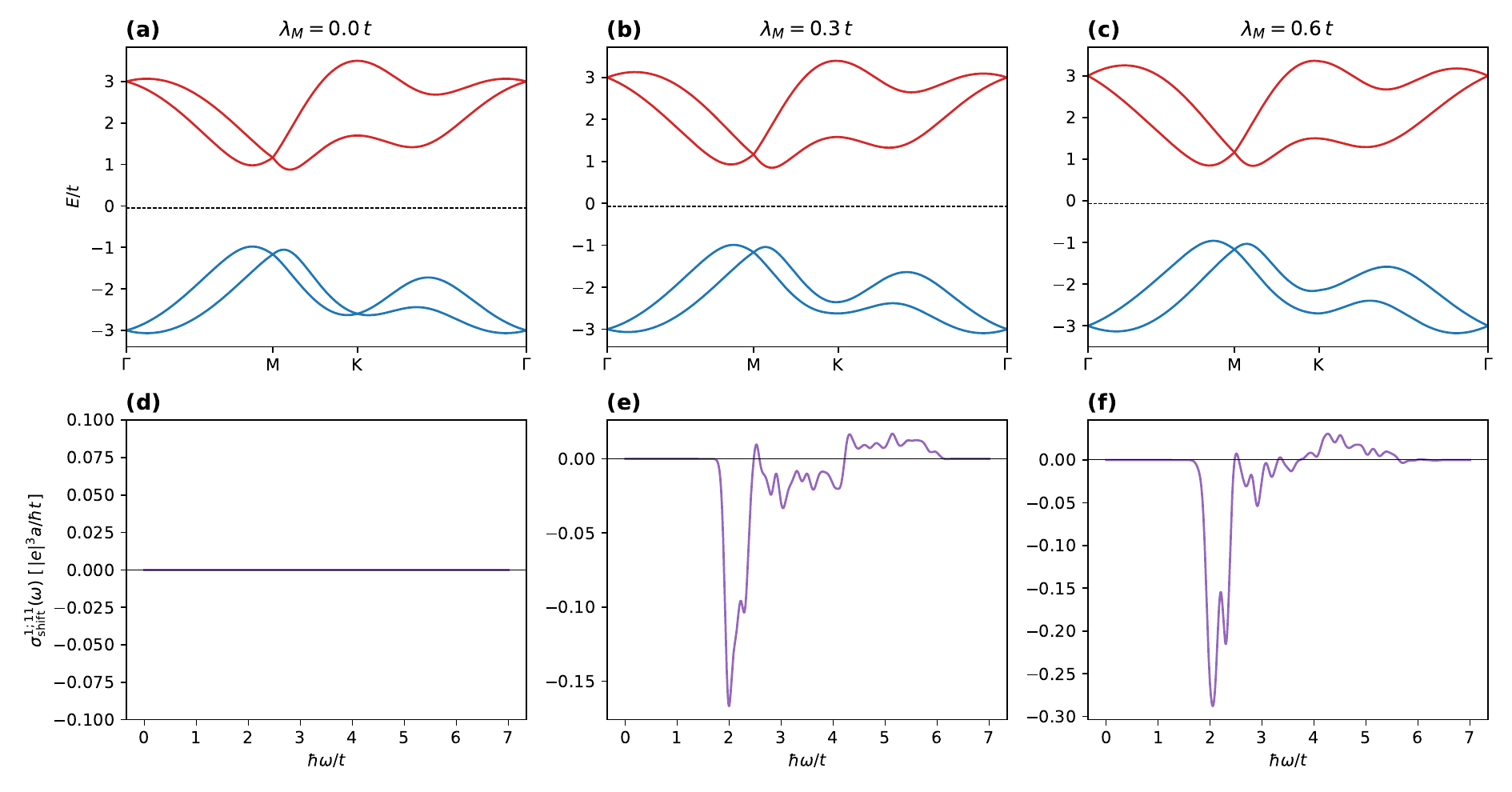}
\caption{%
Band structure [(a)--(c)] and shift conductivity
$\sigma_{\mathrm{shift}}^{1;11}(\omega)$ [(d)--(f), in units of
$|e|^3a/\hbar t$ with $a$ the lattice constant] of the generalized
Kane-Mele model Eq.~\eqref{eq:km-model-main} at $m=0$. Energy spectra are shown along the path
$\Gamma$--M--K--$\Gamma$ in the BZ, for mirror-breaking strengths
$\lambda_M=0$, $0.3t$, $0.6t$ ($\lambda_{\mathrm{SO}}=0.5t$,
$\lambda_{\mathrm{R}}=0.3t$, half filling). The two occupied (blue) and
two unoccupied (red) bands remain gapped at half filling for all $\lambda_M$ shown.
 The shift conductivity is symmetry-forbidden at $\lambda_M=0$ as seen in (d), and increases in magnitude as $\lambda_M$ increases [(e), (f)]. Here and in
Fig.~\ref{fig:km-sumrule}, tensor components are taken along the
normalized reciprocal lattice directions per Eq.~\eqref{eq:recip-dirs}.
}
\label{fig:km-bands-conductivity}
\end{figure*}

We compute the shift conductivity in axes adapted to the reciprocal lattice. Introducing primitive reciprocal lattice vectors $\mathbf{b}^a$, $a=1,2$, we compute 
\begin{equation}\label{eq:recip-dirs}
\sigma^{a;bc}_\mathrm{shift}(\omega) = \hat{b}^a_\sigma\hat{b}^b_\nu\hat{b}^c_\lambda\sigma^{\sigma;\nu\lambda}_\mathrm{shift}(\omega),
\end{equation}
with $\hat{\mathbf{b}}^a =\mathbf{b}^a/|\mathbf{b}^a|$. In Fig.~\ref{fig:km-bands-conductivity} we show the energy spectrum and the linear component $\sigma^{1;11}_\mathrm{shift}(\omega)$ of the shift conductivity computed using Eq.~\eqref{eq:ff-shift-conductivity-main} for Eq.~\eqref{eq:km-model-main} at half filling and zero temperature. We fix $\lambda_{\mathrm{SO}}=0.5t$, $\lambda_{\mathrm{R}}=0.3t$, and $m=0$ and vary the strength $\lambda_M$ of the mirror-symmetry-breaking perturbation. For $\lambda_M=0$ the Hamiltonian is mirror symmetric and $\sigma^{1;11}_\mathrm{shift}(\omega)=0$ by symmetry. In fact, with mirror symmetry, $\sigma^{\sigma;\nu\lambda}(\omega)=0$ identically, whether or not time-reversal symmetry is broken. For $\lambda_M\neq 0$ we see that the shift conductivity has a sharp peak at the band edge $\hbar\omega\approx 2t$ where interband transitions become possible, before eventually returning to zero for $\hbar\omega \approx 6.5t$ as the frequency approaches the maximum interband transition energy. 

\begin{figure*}[!ht]
\centering
\includegraphics[width=\textwidth]{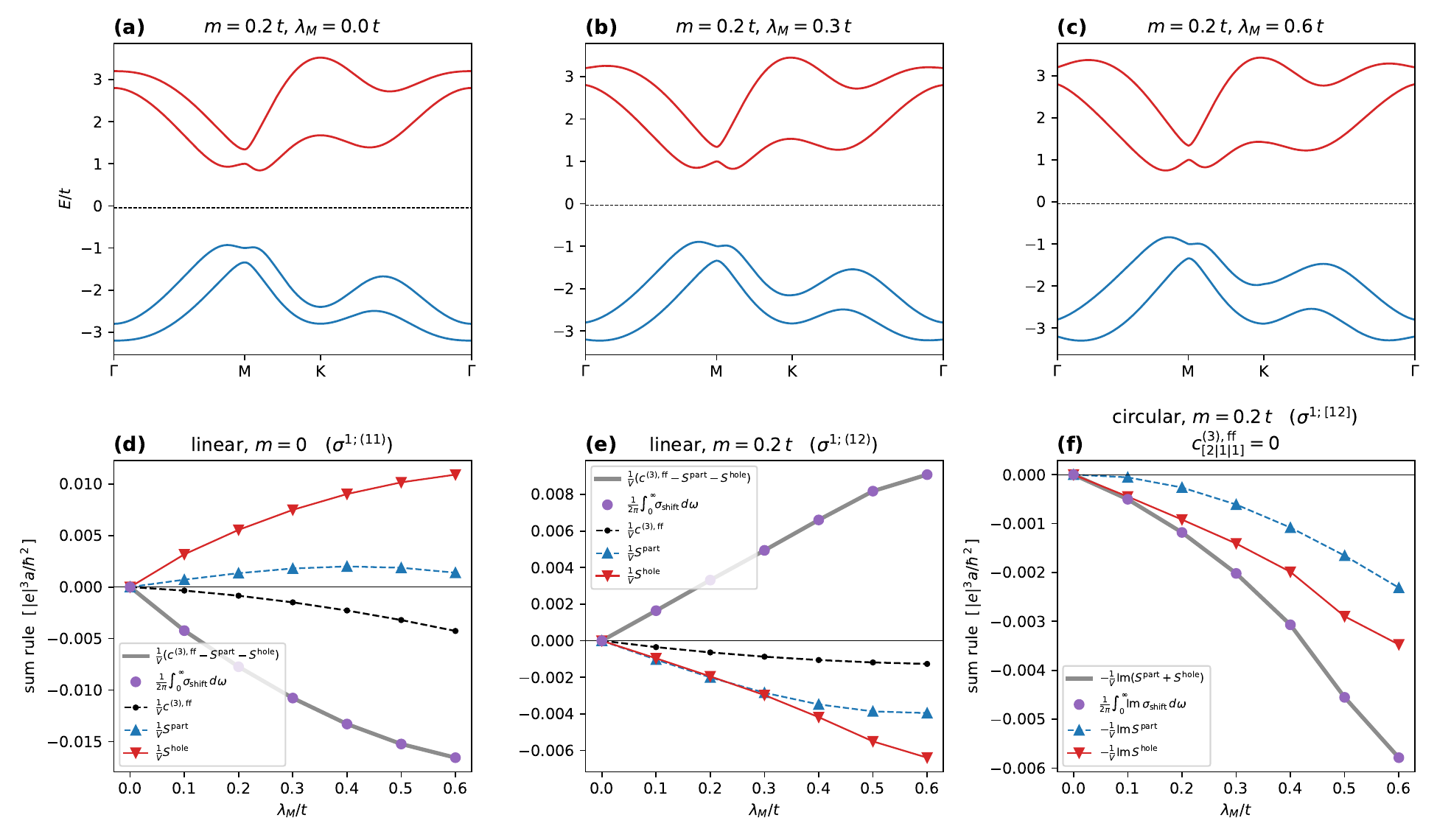}
\caption{%
The geometric sum rule Eq.~\eqref{eq:ff-sum-rule-main} evaluated for the generalized
Kane-Mele model as a function of mirror-breaking strength $\lambda_M$ (all
sum-rule curves in units of $|e|^3a/\hbar^2$, with $a$ the lattice
constant). (a)--(c) show the spectrum of the model in the presence of a time-reversal symmetry breaking Zeeman field  ($m=0.2t$).
(d)--(f) show the corresponding frequency-integrated shift conductivity
$\tfrac{1}{2\pi}\int_0^\infty\dd\omega\,\sigma_{\mathrm{shift}}$ [points, evaluated by summing the line weights in Eq.~\eqref{eq:shiftcon}]
together with the cumulant density $\tfrac{1}{V}c^{(3),\mathrm{ff}}$
and the multi-state geometric contributions $\tfrac{1}{V}S^{\mathrm{part}}$ and
$\tfrac{1}{V}S^{\mathrm{hole}}$. The cumulant density was computed using Eq.~\eqref{eq:ff-c3-projector-app}, while the components of the complex distortion tensor were computed using Eq.~\eqref{eq:ff-s-main}. We show the sum rules for the linear $\sigma^{1;(11)}$ shift conductivity at
$m=0$ [(d)], the linear off-diagonal $\sigma^{1;(12)}$ shift conductivity at $m=0.2t$
[(e)], and the circular antisymmetric $\sigma^{1;[12]}$ at $m=0.2t$
[(f)]. From Eq.~\eqref{eq:obc-circular-sum-rule-main}, the cumulant does not contribute to the antisymmetric part of the sum rule and so the
integrated $\sigma^{1;[12]}$ measures
$-\tfrac{1}{V}\operatorname{Im}S^{\mathrm{ff}}$ directly. The sum rule holds numerically up to discretization error
for every $\lambda_M$. 
}
\label{fig:km-sumrule}
\end{figure*}

Finally, in Fig.~\ref{fig:km-sumrule} we examine the different components of the sum rule Eq.~\eqref{eq:ff-sum-rule-main}. In Figs.~\ref{fig:km-sumrule}(a)--(c), we show the band structure for our model with $m=0.2t$ as $\lambda_M$ is varied from $0$ to $0.6t$. In Figs.~\ref{fig:km-sumrule}(d)--(f), we show three components of the sum rule as a function of $\lambda_M$. Figure~\ref{fig:km-sumrule}(d) shows the integrated shift conductivity $\tfrac{1}{2\pi}\int_0^\infty\dd\omega\,\sigma^{1;(11)}_\mathrm{shift}(\omega)$ in purple for $m=0$. Since the system is time-reversal symmetric for $m=0$, only the symmetrized (linear polarization) components of the shift conductivity are nonzero. The black circles show the value of the cumulant $\tfrac{1}{V}c^{(3),\mathrm{ff}}_{111}$, the blue triangles the value of $\tfrac{1}{V}S^\mathrm{part}_{111}$, and the red triangles the value of $\tfrac{1}{V}S^\mathrm{hole}_{111}$. The right-hand side of the sum rule is shown in gray; we see that the sum rule holds. Additionally, we see that the dominant contribution to the shift conductivity integral is $S^\mathrm{hole}_{111}$, the multiple-occupied-band contribution to the complex distortion tensor.

In Fig.~\ref{fig:km-sumrule}(e) we show, at $m=0.2t$, the symmetric part $\sigma^{1;(12)}_\mathrm{shift}$ of the integrated shift conductivity (corresponding to the response to perpendicular linearly polarized beams), along with the symmetrized $(1|1|2)$ components [Eq.~\eqref{eq:s-middle-index-sym-main}] of the cumulant and the complex distortion tensor. Again, we see that $S^\mathrm{hole}_{(1|1|2)}$ is a sizable contribution to the total integrated response. It is also noteworthy that the geometric contribution to the response dominates over the cumulant contribution in this case, to the point that the total sum rule integral has the opposite sign of the cumulant. Lastly, because both time-reversal and mirror symmetry are broken with $m,\lambda_M\neq 0$, the antisymmetric component of the shift conductivity can also be nonzero. In Fig.~\ref{fig:km-sumrule}(f) we show the antisymmetric part $\sigma^{1;[12]}_\mathrm{shift}$ of the integrated shift conductivity (corresponding to the response to circularly polarized beams), along with the antisymmetrized $[2|1|1]$ components of the complex distortion tensor. Because of the antisymmetrization, the cumulant drops out of the sum rule and the integrated conductivity is determined solely from the antisymmetric part of the complex distortion tensor, i.e. the J-rotated Uhlmann curvature dipole of Eq.~\eqref{eq:s-response-curvature-main}.

These calculations validate our free fermion sum rule Eq.~\eqref{eq:ff-sum-rule-main}. Furthermore, we see that the new $S^\mathrm{hole}_{\lambda\sigma\nu}$ contributions to the multi-state geometric term in the sum rule play a significant role in determining the sign and magnitude of the total shift current response. 

\section{Nonzero temperature}\label{sec:nonzero_T}

Let us now examine the fate of our geometric sum rule for insulators at nonzero temperature. Returning to the general case of Sec.~\ref{sec:rectification}, we can examine the sum rule for the total rectification response Eq.~\eqref{eq:primitive-absorptive}. Integrating the total response function over source frequencies gives the general sum rule

\begin{equation}
\begin{aligned}
\frac{1}{\pi}&\int_0^\infty\!\dd\varpi\;\bigl[\rho_K^{\sigma;\nu\lambda}(\varpi)+\rho_K^{\sigma;\lambda\nu}(-\varpi)\bigr] \\
    =\frac{1}{\pi}&\sum_{\omega_{ac}>0}
    \bigl[w_{ac}^{\sigma;\nu\lambda}+\bigl(w_{ac}^{\sigma;\lambda\nu}\bigr)^{*}\bigr]
    \\
    ={}&\sum_{\omega_{ac}>0}
    (p_a-p_c)
    \Bigl[
    \comm{B_\nu}{B_{\sigma,\mathrm d}}_{ac}(B_\lambda)_{ca}
    \\
    &\qquad\qquad
    +\comm{B_{\sigma,\mathrm d}}{B_\lambda}_{ca}(B_\nu)_{ac}
    \Bigr],
\end{aligned}
    \label{eq:thermal-lehmann-main}
\end{equation}
where we made use of the general forms Eq.~\eqref{eq:primitive-rho-k-main} for the spectral density and Eq.~\eqref{eq:primitive-line-weight-main} for the line weights. Equation~\eqref{eq:thermal-lehmann-main} is a DC rectification sum rule valid for any (bounded) quantum system at nonzero temperature. 

However, unlike at zero temperature, the thermal sum rule Eq.~\eqref{eq:thermal-lehmann-main} does not split directly into a sum of cumulants and multi-state geometric tensors, even for gapped systems. The obstruction comes from the Boltzmann factors in the sum rule. We can still expand the commutator $[B_\nu,B_{\sigma,\mathrm{d}}]_{ac}=[(B_\sigma)_{cc}-(B_\sigma)_{aa}](B_{\nu})_{ac}$ in a basis where $B_{\sigma}$ is diagonal in every degenerate subspace. But inserting this into Eq.~\eqref{eq:thermal-lehmann-main}, we cannot reorganize the diagonal matrix elements into an expectation value when $T\neq 0$. This means we cannot write Eq.~\eqref{eq:thermal-lehmann-main} in terms of $\delta B_\sigma$ when $T\neq 0$. 

Similarly, we can evaluate the complex distortion tensor $S_{\lambda\sigma\nu}(T)$ of Eq.~\eqref{eq:same-order-to-response-main} directly at nonzero $T$. We can start from the $T\neq 0$ J-rotated cQAC tensor in Eq.~\eqref{eq:stilde-nonzero-T-general} with $\mathcal{A}_\mu = B_\mu$, 
\begin{equation}
\begin{aligned}
\widetilde S_{\sigma;\lambda\nu}(T) &= \sum_{a,b,k}
    |t_{ab}||p_k-p_b||t_{ka}|\,
    (B_\lambda)_{ab}(B_\sigma)_{bk}(B_\nu)_{ka}, \\
    &=\sum_{a,b,k}(p_b+p_k)|t_{ab}||t_{bk}||t_{ka}| \\
    &\qquad\times
    (B_\lambda)_{ab}(B_\sigma)_{bk}(B_\nu)_{ka},\label{eq:thermal-j-cqac}
\end{aligned}
\end{equation}
where $t_{ab}$ was defined in Eq.~\eqref{eq:t-conventions-main} and we have used that for thermal probabilities
\begin{equation}
    |p_a-p_c|
    =
    (p_a+p_c)\,\bigl|t_{ac}\bigr|.
    \label{eq:thermal-detailed-balance-main}
\end{equation}
Finally, taking cyclic sums of Eq.~\eqref{eq:thermal-j-cqac} yields
\begin{equation}
\begin{aligned}
    S_{\lambda\sigma\nu}(T)
    &=\frac{1}{2}\left(\widetilde S_{\lambda;\nu\sigma}
    +
    \widetilde S_{\nu;\sigma\lambda}
    -
    \widetilde S_{\sigma;\lambda\nu}\right) \\
    &=\sideset{}{'}\sum_{a,b,c}
    p_a\,
    \bigl|t_{ab}\,t_{bc}\,t_{ca}\bigr|\,
    (B_\lambda)_{ab}
    (B_\sigma)_{bc}
    (B_\nu)_{ca}.
    \label{eq:s-thermal-main}
\end{aligned}
\end{equation}
Equation~\eqref{eq:s-thermal-main} is the general form of the complex distortion tensor at nonzero temperature. We see that, in contrast to the matrix elements of $B$ that appear in the sum rule Eq.~\eqref{eq:thermal-lehmann-main}, $S_{\lambda\sigma\nu}(T)$ contains a product of three hyperbolic tangents. As such, it will not directly appear in any bare response function. We note, however, that at low temperatures the sum rule Eq.~\eqref{eq:thermal-lehmann-main} differs from its $T=0$ form $2(c^{(3)}_{\lambda\sigma\nu} - S_{\lambda\sigma\nu})$ by exponentially small corrections for gapped systems with nondegenerate ground states, since the thermal probabilities are exponentially suppressed for $a,c\neq 0$. We expect this to be true also in the thermodynamic limit provided the gap to the ground state remains nonzero and the operators in Eq.~\eqref{eq:s-thermal-main} have suitably bounded off-diagonal matrix elements. By contrast, the rate at which $S_{\lambda\sigma\nu}(T)$ approaches its zero-temperature limit is controlled not by the ground state energy gap, but instead by the size of the smallest many-body energy difference due to the intermediate state sum in Eq.~\eqref{eq:s-thermal-main}.  Therefore we expect that for temperatures much less than the energy gap the zero-temperature sum rules of Sec.~\ref{sub:general_zero_temperature_rectification_sum_rules} will still be practically useful, although they are not accurate probes of \emph{nonzero} temperature quantum geometry.

Note also that the dependence of $S_{\lambda\sigma\nu}(T)$ on excited state energy differences has implications for the dependence of $S_{\lambda\sigma\nu}(T)$ on Hamiltonian parameters. In particular, while Eq.~\eqref{eq:s-thermal-main} is a continuous function of energies for any bounded $B$ and $T\neq 0$, as $T\rightarrow 0$ the terms in the sum develop discontinuities as a function of energy. If a parameter in the Hamiltonian drives two excited states towards a degeneracy, then exactly at the degeneracy point the value of $S_{\lambda\sigma\nu}(T\rightarrow 0)$ can change discontinuously. 
 
Lastly, we note that like the Fisher information in linear response, $S_{\lambda\sigma\nu}(T)$ can only be extracted from sum rules weighted by these hyperbolic tangent factors. Following Ref.~\cite{guan2026exploring}, we expect that $S_{\lambda\sigma\nu}(T)$ can be extracted from a weighted sum rule for a correlation function that determines the full, two-frequency-dependent response $K^{\sigma\nu\lambda}(\omega_1,\omega_2)$. However, unlike the case for linear response, the matrix elements appearing in Eq.~\eqref{eq:s-thermal-main} do not have a definite sign. As such, we do not expect there to be general geometric bounds on rectification at nonzero temperature, in contrast to the linear case~\cite{onishi2025quantum,souza2025optical,balut2026fundamental}. We leave a full exploration of these issues to future work. 

\section{Conclusion}\label{sec:conclusion}
In this work, we have shown how the many-body quantum geometry of density matrices controls sum rules for DC rectification. Building on the work of Refs.~\cite{ji2025density,guan2026exploring,hetenyi2023fluctuations}, we used the symmetric logarithmic derivative to formulate multi-state quantum geometric quantities to characterize general many-body quantum systems at arbitrary temperature. We introduced the complex quantum Amari-Chentsov tensor (cQAC), which generalizes the notion of skewness from classical information geometry to the space of density matrices. From this we derived the complex distortion tensor. The real part of the complex distortion tensor captures the incompatibility between the Bures metric and the mixture connection associated to second-order response theory. Similarly, the imaginary part of the complex distortion tensor is derived from the J-rotated Uhlmann curvature dipole. 

We applied this formalism to study sum rules for second-order DC rectification in general systems. Using the second-order Kubo formula, we derived the spectral density and total response for an operator $C_\sigma=\dot{B}_\sigma$ to a perturbation coupled to $B_\lambda$. While the motivating example is DC shift current response in the length gauge, our results are generally applicable. We showed that for insulators, the integral of the total response over probe frequency is, at zero temperature, given by the difference between the third cumulant of $B$ in the ground state and the complex distortion tensor associated to the perturbation. This generalizes previous results on shift current sum rules by extending them to general interacting and/or disordered systems, as well as to more general nonlinear responses. We then directly applied our formalism to shift current, and considered an example free fermion model to demonstrate our results. In particular, we showed how our formalism naturally organized the contributions to the multi-state geometric correction to the shift current sum rule for multiband systems that were first identified in Ref.~\cite{avdoshkin2024multi}.

Lastly, we examined the fate of the rectification sum rule at nonzero temperature. We argued that for insulators, temperature corrections to the sum rule integrand are exponentially suppressed at low temperatures, implying that $T=0$ multi-state geometry as encoded through the complex distortion tensor is experimentally accessible. On one hand, our result shows that by measuring the probe frequency dependence of the shift current, one can directly measure the combinations of the complex distortion tensor and skewness that enter the sum rules for interacting or disordered insulators. Viewed in a different light, our result also shows that materials with enhanced photovoltaic response could be designed by increasing the magnitude of the complex distortion tensor and/or the ground state skewness, even in the presence of disorder and interactions. Since Eqs.~\eqref{eq:obc-skewness-sum-rule-main} and \eqref{eq:obc-circular-sum-rule-main} are both computable from low-energy effective models, they can be used to inform photovoltaic material design pipelines. Additionally, the multi-state geometric terms quantify how remote states that have been integrated out of the low-energy model can influence photovoltaic response. Intriguingly, however, we also showed that the splitting of the sum rule into cumulant and geometric contributions does not extend to nonzero temperature; temperature corrections to the complex distortion tensor are in general not small, since they are controlled by the level spacing of excited states.

Our work opens up several avenues for future investigation. First, it would be interesting to apply our formalism to study nonlinear viscoelastic response in many-body systems. Recently, Refs.~\cite{jain2026topological,jain2026nonlinear} showed that nonmetricity plays a role in second harmonic generation for free fermion viscoelastic response. Exploring the connection between that work and our general many-body results could be a fruitful avenue. Additionally, while we worked primarily in the length gauge, it is straightforward to extend our work to a velocity-gauge calculation to include injection as well as shift current effects for systems with periodic boundary conditions. For the shift current in particular, our analysis in Appendix~\ref{app:ff-blount} [before assuming a free fermion ground state, e.g. Eq.~\eqref{eq:blount-od-split-app}] lets us rewrite Eq.~\eqref{eq:obc-shift-weight-main} in terms of the off-diagonal matrix elements of the position operator and its commutators with the full position operator. Written in this way, one can use the formalism of Ref.~\cite{souza2000polarization} to rewrite our expression for the shift current in a form suitable to systems with periodic boundary conditions: off-diagonal matrix elements of the position operator are the Berry connection in the space of twisted boundary conditions, and commutators with the position operator give covariant derivatives with respect to the boundary condition twist. We defer a full exploration of the velocity gauge to a future work. Lastly, as we discussed in Sec.~\ref{sec:nonzero_T}, studying the full frequency dependence of the second-order response function could yield more general frequency-weighted sum rules for the complex distortion tensor.

\begin{acknowledgments}
We thank V.~V.~Albert and J.~E.~Moore for fruitful discussions.
The quantum information geometry theory development was supported by the National Science Foundation under Grant No.~DMR-2510219.
The application to response theory was supported by the U.S.
Department of Energy, Office of Basic Energy Sciences, Grant No.~DE-SC0026342.
\end{acknowledgments}

\section*{AI Disclosure}
Claude Fable 5 and OpenAI GPT 5.5 were used to prototype results and code during the preparation of this work.
All mathematical output was independently derived and checked by the author. All prose was written by the author.

\appendix

\section{The connection algebra}\label{sec:the_connection_algebra}

In this appendix, we examine the failure of metric compatibility for the exponential and mixture connections. In particular, we will first show that the QAC tensor captures the torsion of the exponential connection. Next, we show how to construct the Bures-Levi-Civita connection from the mixture and exponential connections combined with a contorsion tensor built from the QAC tensor. Finally we derive Eq.~\eqref{eq:z-covariant-derivative-main} to establish that the complex QAC tensor is indeed tensorial. 

To begin, let us consider covariant derivatives of the Bures metric. Since $g(v,w)$ is a scalar for any vector fields $v$, and $w$, we have that for any covariant derivative $\nabla$,
\begin{equation}
\begin{aligned}
\nabla_u[g(v,w)] &= u(g(v,w)) \\
&= (\nabla_u g)(v,w) + g(\nabla_u v,w) + g(v,\nabla_u w),
\end{aligned}
\end{equation}
where the second line follows from the product rule. For basis vector fields $g_{\nu\lambda} = g(\partial_\nu,\partial_\lambda)$, this gives the component-wise covariant derivative of the metric. Introducing the general Christoffel symbol (of the first kind)
\begin{equation}
\Gamma_{\mu\nu\lambda} = g(\nabla_{\partial_\mu} \partial_\nu,\partial_\lambda),
\end{equation}
\begin{equation}\label{eq:cov-deriv-of-metric-general}
\nabla_{\partial_\mu} g_{\nu\lambda} = \partial_\mu g_{\nu\lambda} - \Gamma_{\mu\nu\lambda} - \Gamma_{\mu\lambda\nu}.
\end{equation} 
Specifying to the mixture connection, and making use of the duality relations Eqs.~\eqref{eq:dual-product-main} and \eqref{eq:dual-difference-main}, we find that the mixture covariant derivative of the metric is
\begin{equation}
\begin{aligned}
    \left(\nablamixture_{\partial_\sigma}g\right)_{\mu\nu}
    &=
    \partial_\sigma g_{\mu\nu}
    -\Gamma^{(m)}_{\sigma\mu\nu}
    -\Gamma^{(m)}_{\sigma\nu\mu}
    \\
    &=
    \Gamma^{(e)}_{\sigma\nu\mu}
    -\Gamma^{(m)}_{\sigma\nu\mu}
    =
    -T_{\sigma\mu\nu},
\end{aligned}
    \label{eq:mixture-nonmetricity-app}
\end{equation}
where $T_{\sigma\mu\nu}$ is the QAC tensor of Eq.~\eqref{eq:qac-coords}. Thus, the QAC tensor measures by how much the mixture connection fails to be compatible with the Bures metric. Performing the analogous calculation for the exponential connection, we find
\begin{equation}
\begin{aligned}
    \left(\nablaexp_{\partial_\sigma}g\right)_{\mu\nu}
    &=
    \partial_\sigma g_{\mu\nu}
    -\Gamma^{(e)}_{\sigma\mu\nu}
    -\Gamma^{(e)}_{\sigma\nu\mu}
    \\
    &=
    \Gamma^{(m)}_{\sigma\mu\nu}
    -\Gamma^{(e)}_{\sigma\mu\nu}
    =
    T_{\sigma\mu\nu}.
\end{aligned}
    \label{eq:exp-nonmetricity-app}
\end{equation}
Thus, the exponential connection also fails to be metric compatible, by exactly the opposite amount. Combining Eqs.~\eqref{eq:mixture-nonmetricity-app} and \eqref{eq:exp-nonmetricity-app}, we can introduce a metric compatible connection
\begin{equation}
\begin{aligned}
\nabla^{(0)}_{\partial_\mu}
    &=
    \frac{1}{2}\left(\nablamixture_{\partial_\mu}+\nablaexp_{\partial_\mu}\right)
    \label{eq:midpoint-main} \\
\nabla^{(0)}_{\partial_\mu} g_{\nu\lambda} &= 0.
\end{aligned}
\end{equation}
Although $\nabla^{(0)}$ constructed in this way is metric compatible, it is not the Levi-Civita connection for the Bures metric, precisely because it has nonvanishing torsion. The components of the torsion tensor of a connection are given by
\begin{equation}
\mathrm{Tor}_{\mu\nu\lambda} = \Gamma_{\mu\nu\lambda} - \Gamma_{\nu\mu\lambda}.\label{eq:tordef}
\end{equation}
From Eq.~\eqref{eq:mixture-coeff-main}, we see that the mixture connection is explicitly torsion-free. However, the torsion of the exponential connection is
\begin{equation}
\begin{aligned}
\mathrm{Tor}^{(e)}_{\mu\nu\lambda} &= T_{\nu\mu\lambda} - T_{\mu\nu\lambda} \\
&=\frac{1}{2}\Tr\!\left(
        \rho\comm{\comm{G_\nu}{G_\mu}}{G_\lambda}
    \right).\label{eq:exp-torsion-app}
\end{aligned}
\end{equation}
The torsion of the exponential connection is thus determined by the antisymmetric part of the QAC tensor. From Eq.~\eqref{eq:midpoint-main}, we see that the torsion of $\nabla^{(0)}$ is half the torsion of the exponential connection. Using the definition of the Levi-Civita connection $\Gamma^{\mathrm{LC}}_{\mu\nu\sigma}$ as
\begin{equation}
    2\Gamma^{\mathrm{LC}}_{\mu\nu\sigma}
    =
    \partial_\mu g_{\nu\sigma}
    +\partial_\nu g_{\sigma\mu}
    -\partial_\sigma g_{\mu\nu},
    \label{eq:koszul-app}
\end{equation}
we can use Eq.~\eqref{eq:midpoint-main} to write
\begin{equation}
\begin{aligned}
    2\Gamma^{\mathrm{LC}}_{\mu\nu\sigma}
    ={}&
    \Gamma^{(0)}_{\mu\nu\sigma}+\Gamma^{(0)}_{\mu\sigma\nu}
    +\Gamma^{(0)}_{\nu\sigma\mu}+\Gamma^{(0)}_{\nu\mu\sigma}
    \\
    &-\Gamma^{(0)}_{\sigma\mu\nu}-\Gamma^{(0)}_{\sigma\nu\mu}
    \\
    ={}&
    2\Gamma^{(0)}_{\mu\nu\sigma}
    -\mathrm{Tor}^{(0)}_{\mu\nu\sigma}
    +\mathrm{Tor}^{(0)}_{\mu\sigma\nu}
    +\mathrm{Tor}^{(0)}_{\nu\sigma\mu}. \label{eq:LC-starting}
    \end{aligned}
    \end{equation}
Since the torsion of $\nabla^{(0)}$ is half the torsion of $\nabla^{(e)}$, we have
\begin{equation}
\begin{aligned}
    \mathrm{Tor}^{(0)}_{\mu\sigma\nu}
    +\mathrm{Tor}^{(0)}_{\nu\sigma\mu}
    -\mathrm{Tor}^{(0)}_{\mu\nu\sigma} & = \frac{1}{2}\left(T_{\mu\nu\sigma} - T_{\nu\mu\sigma}\right) \\
    &+\frac{1}{2}\left(T_{\sigma\mu\nu} - T_{\mu\sigma\nu}\right) \\
    & +\frac{1}{2}\left(T_{\sigma\nu\mu} - T_{\nu\sigma\mu}\right) \\
    &= T_{\sigma\mu\nu} - T_{\nu\sigma\mu},
\end{aligned}
\end{equation}
where we used the symmetry $T_{\mu\nu\sigma}=T_{\mu\sigma\nu}$ from Eq.~\eqref{eq:qac-coords}. Introducing the contorsion tensor
\begin{equation}
\begin{aligned}
 K_{\mu\nu\sigma}
    &=
    \frac{1}{2}\left(
        T_{\nu\sigma\mu}-T_{\sigma\nu\mu}
    \right)
    \\
    &=
    \frac{1}{4}\Tr\!\left(
        \rho\comm{\comm{G_\nu}{G_\sigma}}{G_\mu}
    \right).
\end{aligned}
    \label{eq:contorsion-app}
\end{equation}
We can thus write Eq.~\eqref{eq:LC-starting} as
\begin{equation}
    \Gamma^{\mathrm{LC}}_{\mu\nu\sigma}
    =
    \Gamma^{(0)}_{\mu\nu\sigma}
    -
    K_{\mu\nu\sigma}.
\end{equation}
We can further write the Levi-Civita connection entirely in terms of the mixture connection and the QAC tensor. Using the duality relation Eq.~\eqref{eq:exp-coords-main} combined with our definition Eq.~\eqref{eq:midpoint-main} of $\nabla^{(0)}_{\partial_\mu}$ gives
\begin{equation}
    \Gamma^{\mathrm{LC}}_{\mu\nu\sigma}
    =
    \Gamma^{(m)}_{\mu\nu\sigma}
    -\frac{1}{2}\left(
        T_{\mu\nu\sigma}+T_{\nu\sigma\mu}-T_{\sigma\mu\nu}
    \right).
    \label{eq:lc-full-app}
\end{equation}
Equation~\eqref{eq:lc-full-app} allows us to make a direct analogy between our results and those of Ref.~\cite{guan2026exploring}. As shown by those authors, the fully symmetrized Bures-Levi-Civita connection $\Gamma^{\mathrm{LC}}[f] = f^\mu f^\nu f^\sigma \Gamma^{\mathrm{LC}}_{\mu\nu\sigma}$ can be written as the sum of two terms, a ``Fisher'' term and an ``intrinsic'' term. Comparing with Eq.~\eqref{eq:lc-full-app} and our definitions Eqs.~\eqref{eq:mixture-coeff-main} and \eqref{eq:qac-coords}, the Fisher term is exactly $\Gamma^{(m)}[f]$, while the intrinsic term is $-T[f]/2$.

Equation~\eqref{eq:exp-nonmetricity-app} establishes that the QAC tensor is the exponential covariant derivative of the metric. We now show how to extend this to the full QGT $\Tr Q_{\mu\nu}$ from Eq.~\eqref{eq:operator-qgt-trace-main}. To do so, let us examine the Uhlmann curvature defined in Eq.~\eqref{eq:curvature-form-main}. Evaluating $\mathcal F_{\mu\nu} = \mathcal{F}(\partial_\mu, \partial_\nu)$ and taking derivatives we find
\begin{align}
    \partial_\sigma\mathcal F_{\mu\nu}
    ={}&
    -\frac{\ii}{2}\Tr\!\left(
        \rho_\sigma\comm{G_\mu}{G_\nu}
    \right)
    -\frac{\ii}{2}\Tr\!\left(
        \rho\comm{\partial_\sigma G_\mu}{G_\nu}
    \right) \nonumber\\
    &-\frac{\ii}{2}\Tr\!\left(
        \rho\comm{G_\mu}{\partial_\sigma G_\nu}
    \right).
    \label{eq:curvature-ordinary-derivative-app}
\end{align}
By analogy with Eq.~\eqref{eq:cov-deriv-of-metric-general}, we can define the exponential covariant derivative of $\mathcal{F}$ via the product rule,
\begin{equation}
\begin{aligned}
    \left(\nablaexp_{\partial_\sigma}\mathcal F\right)_{\mu\nu}
    &=
    \partial_\sigma\mathcal F_{\mu\nu}
    -
    \mathcal F(\nablaexp_{\partial_\sigma}\partial_\mu,\partial_\nu)
    \\
    &\quad
    -
    \mathcal F(\partial_\mu,\nablaexp_{\partial_\sigma}\partial_\nu).
\end{aligned}
    \label{eq:exp-curvature-covdef-app}
\end{equation}
To evaluate the last two terms, we can use Eq.~\eqref{eq:exponential-main} along with Eq.~\eqref{eq:g-from-exp} to find
\begin{equation}
\begin{aligned}
    \mathcal F(\nablaexp_{\partial_\sigma}\partial_\mu,\partial_\nu)
    &=
    -\frac{\ii}{2}\Tr\!\left(
        \rho\comm{\partial_\sigma G_\mu}{G_\nu}
    \right),
    \\
    \mathcal F(\partial_\mu,\nablaexp_{\partial_\sigma}\partial_\nu)
    &=
    -\frac{\ii}{2}\Tr\!\left(
        \rho\comm{G_\mu}{\partial_\sigma G_\nu}
    \right).
\end{aligned}
    \label{eq:exp-curvature-generator-terms-app}
\end{equation}
Finally, combining Eqs.~\eqref{eq:curvature-ordinary-derivative-app}--\eqref{eq:exp-curvature-generator-terms-app} yields
\begin{equation}
    \left(\nablaexp_{\partial_\sigma}\mathcal F\right)_{\mu\nu}
    =
    -\frac{\ii}{2}\Tr\!\left(
        \rho_\sigma\comm{G_\mu}{G_\nu}
    \right)
    =
    \Phi_{\sigma\mu\nu},
    \label{eq:exp-curvature-phi-app}
\end{equation}
where $\Phi_{\sigma\mu\nu}$ was defined in Eq.~\eqref{eq:phi-core-main}. Thus the exponential covariant derivative of the Uhlmann curvature yields the imaginary part of the cQAC tensor. That is, combining Eqs.~\eqref{eq:exp-curvature-phi-app} with Eq.~\eqref{eq:exp-nonmetricity-app} and Eq.~\eqref{eq:cubic-z-main} yields
\begin{equation}
\begin{aligned}
    Z_{\sigma;\mu\nu}
    &=
    \left(\nablaexp_{\partial_\sigma}g\right)_{\mu\nu}
    -\ii
    \left(\nablaexp_{\partial_\sigma}\mathcal F\right)_{\mu\nu}
    \\
    &=
    \left[\nablaexp_{\partial_\sigma}\!\left(g-\ii\,\mathcal F\right)\right]_{\mu\nu}
    =
    \left[\nablaexp_{\partial_\sigma}\Tr Q\right]_{\mu\nu},
\end{aligned}
    \label{eq:z-exponential-app}
\end{equation}
which is exactly Eq.~\eqref{eq:z-covariant-derivative-main}.

\section{Derivation of the general rectification response}\label{sec:general_rectification_response_derivation}

In this appendix, we will derive the general rectification sum rules presented in Sec.~\ref{sec:rectification}. We consider a system with initial thermal density matrix $\rho\propto e^{-\beta H_0}$ in terms of the unperturbed Hamiltonian $H_0$. We consider a time-dependent harmonic perturbation
\begin{equation}
    H'(t)=
    \bigl(F^\lambda e^{-\ii\omega t}
    +F^{\lambda*}e^{\ii\omega t}\bigr)e^{\eta t}B_\lambda,
    \qquad \eta>0.
    \label{eq:primitive-drive-app}
\end{equation}
Our goal is to compute the DC component of the response of the operator
\begin{equation}
C_\sigma
    =\dot B_\sigma
    =\ii\comm{H(t)}{B_\sigma}
\end{equation}
to this perturbation. A key observation is that $C_\sigma$ defined in this way depends explicitly on the perturbation $H'(t)$. Writing $H(t)=H_0+H'(t)$, we can separate $C_\sigma$ into an unperturbed part $C_\sigma^{(0)}$ and a ``diamagnetic'' contribution $C^{+}_\sigma(\omega,t)
    +C^{-}_\sigma(\omega,t)$ as
\begin{equation}
\begin{aligned}
 C_\sigma&=C^{(0)}_\sigma
    +C^{+}_\sigma(\omega,t)
    +C^{-}_\sigma(\omega,t),
    \\
    C^{(0)}_\sigma
    &\equiv\ii\comm{H_0}{B_\sigma},
    \\
    C^{+}_\sigma(\omega,t)
    &\equiv\ii\,F^{\mu}e^{(\eta-\ii\omega)t}\comm{B_\mu}{B_\sigma},
    \\
    C^{-}_\sigma(\omega,t)
    &\equiv\ii\,F^{\mu*}e^{(\eta+\ii\omega)t}\comm{B_\mu}{B_\sigma}.
\end{aligned}
    \label{eq:primitive-d-split-app}
\end{equation}
We will compute the rectified (zero total frequency) component of the response $\langle C_\sigma\rangle_\mathrm{rect}$ to second order in the perturbation $F^\lambda$, which can be parameterized as
\begin{equation}
\langle C_\sigma\rangle_\mathrm{rect} = -[K^{\sigma\nu\lambda}(-\omega) + K^{\sigma\lambda\nu}(+\omega)]F^\nu F^{\lambda*}, 
\end{equation}
where we have made the symmetry under the exchange $\nu\leftrightarrow\lambda$ of perturbing field indices explicit. The response function can be split into a ``paramagnetic'' and ``diamagnetic'' contribution,
\begin{equation}
    K^{\sigma\nu\lambda}
    =K^{\sigma\nu\lambda}_{\mathrm{para}}
    +K^{\sigma\nu\lambda}_{\mathrm{dia}}.
    \label{eq:primitive-k-split-app}
\end{equation}
The paramagnetic contribution $K^{\sigma\nu\lambda}_{\mathrm{para}}$ arises from the second-order response of $C^{(0)}_\sigma$ to the perturbation. The diamagnetic $K^{\sigma\nu\lambda}_{\mathrm{dia}}$ arises from the \emph{linear} response of $C^{+}_\sigma(\omega,t)
    +C^{-}_\sigma(\omega,t)$ to the perturbation. We will now compute both contributions.

\subsection{Paramagnetic response} 
\label{app:primitive-retarded-kernel}

To compute $K^{\sigma\nu\lambda}_{\mathrm{para}}$, we start from the second-order Kubo formula for $\delta\langle C^{(0)}_\sigma\rangle^{(2)}(t)$ in the time domain. In terms of $H'(t)$, we have~\cite{bradlyn2024spectral}
\begin{equation}
\begin{aligned}
    \delta\langle C^{(0)}_\sigma\rangle^{(2)}(t)
    ={}&
    (-\ii)^2
    e^{-2\eta t}\int_{-\infty}^{t}\!\dd t_1
    \int_{-\infty}^{t_1}\!\dd t_2
    \\
    &\times
    \Tr\!\left[
        \rho_0
        \comm{\comm{(C^{(0)}_\sigma)(t)}{H'(t_1)}}{H'(t_2)}
    \right],
\end{aligned}
    \label{eq:primitive-retarded-app}
\end{equation}
where the time evolution of operators in the trace is evaluated using the unperturbed evolution operator. Because $\delta\langle C^{(0)}_\sigma\rangle^{(2)}(t)$ is quadratic in $H'(t)$, it contains three Fourier harmonics: a $-2\omega$ oscillation proportional to $(F^\nu F^\lambda)^*$, a $+2\omega$ oscillation proportional to $(F^\nu F^\lambda)$, and a zero-frequency rectified component proportional to $(F^\nu F^{\lambda*})$. Isolating the rectified component from Eq.~\eqref{eq:primitive-retarded-app}, we can write
\begin{equation}
\begin{aligned}
    \langle C^{(0)}_\sigma\rangle_{\mathrm{rect}}
    &=
    -\left[
    K_{\mathrm{para}}^{\sigma\nu\lambda}(-\omega)
    +
    K_{\mathrm{para}}^{\sigma\lambda\nu}(+\omega)
    \right]F^\nu F^{\lambda*},
\end{aligned}
    \label{eq:primitive-dc-kernel-app}
\end{equation}
with
\begin{equation}
\begin{aligned}
    &K_{\mathrm{para}}^{\sigma\nu\lambda}(-\omega)
    \\
    &=e^{-2\eta t}\int_{-\infty}^{t}\!\dd t_1
    \int_{-\infty}^{t_1}\!\dd t_2\;
    e^{(\eta-\ii\omega)t_1}\,e^{(\eta+\ii\omega)t_2}
    \\
    &\quad\times
    \Tr\!\left[
        \rho_0
        \comm{\comm{(C^{(0)}_\sigma)(t)}{B_\nu(t_1)}}{B_\lambda(t_2)}
    \right].
\end{aligned}
    \label{eq:Kpara-time-domain}
\end{equation}
To simplify Eq.~\eqref{eq:Kpara-time-domain}, we can expand the nested commutator to get four terms, into which we insert complete sets of eigenstates of $H_0$ into the trace to isolate the time dependence. With the help of the double integral
\begin{equation}
\begin{aligned}
    &\int_{-\infty}^{t}\!\dd t_1\,e^{(\eta+\ii A)t_1}
    \int_{-\infty}^{t_1}\!\dd t_2\,e^{(\eta+\ii B)t_2}
    \\
    &=
    \frac{e^{(2\eta+\ii(A+B))t}}
    {(\eta+\ii B)\bigl(2\eta+\ii(A+B)\bigr)},
\end{aligned}
    \label{eq:primitive-master-integral-app}
\end{equation}
we find
\begin{equation}
K_{\mathrm{para}}^{\sigma\nu\lambda}(-\omega)=T_1-T_2-T_3+T_4,
\label{eq:primitive-t-sum-app}
\end{equation}
with
\begin{equation}
\begin{aligned}
    T_1
    &=
    \sum_{abc}
    \frac{
        p_a (C^{(0)}_\sigma)_{ab}(B_\nu)_{bc}(B_\lambda)_{ca}
    }
    {
        (2\eta-\ii\omega_{ab})
        \bigl(\eta-\ii(\omega_{ac}-\omega)\bigr)
    },
    \\
    T_2
    &=
    \sum_{abc}
    \frac{
        p_a (B_\nu)_{ab}(C^{(0)}_\sigma)_{bc}(B_\lambda)_{ca}
    }
    {
        (2\eta-\ii\omega_{bc})
        \bigl(\eta-\ii(\omega_{ac}-\omega)\bigr)
    },
    \\
    T_3
    &=
    \sum_{abc}
    \frac{
        p_a (B_\lambda)_{ab}(C^{(0)}_\sigma)_{bc}(B_\nu)_{ca}
    }
    {
        (2\eta-\ii\omega_{bc})
        \bigl(\eta-\ii(\omega_{ba}-\omega)\bigr)
    },
    \\
    T_4
    &=
    \sum_{abc}
    \frac{
        p_a (B_\lambda)_{ab}(B_\nu)_{bc}(C^{(0)}_\sigma)_{ca}
    }
    {
        (2\eta-\ii\omega_{ca})
        \bigl(\eta-\ii(\omega_{ba}-\omega)\bigr)
    } .
\end{aligned}
    \label{eq:primitive-t-terms-app}
\end{equation}

To further simplify $K_{\mathrm{para}}^{\sigma\nu\lambda}(-\omega)$, we can take matrix elements of the relation $C_{\sigma}^{(0)} = \ii \comm{H_0}{B_\sigma}$ from Eq.~\eqref{eq:primitive-d-split-app} to find
\begin{equation}
    \frac{(C^{(0)}_\sigma)_{ab}}{2\eta-\ii\omega_{ab}}
    =
    \frac{\ii\omega_{ab}}{2\eta-\ii\omega_{ab}}(B_\sigma)_{ab}
    \xrightarrow{\eta\to0^+}
    -(B_\sigma-B_{\sigma,\mathrm{d}})_{ab},
    \label{eq:primitive-conversion-app}
\end{equation}
where $B_{\sigma,\mathrm{d}}$ is the energy-diagonal projection of $B_\sigma$ defined in Eq.~\eqref{eq:primitive-hat-main}.

Using Eq.~\eqref{eq:primitive-conversion-app} and defining
\begin{equation}
    D^{\sigma\nu}\equiv\comm{B_\sigma-B_{\sigma,\mathrm{d}}}{B_\nu},
    \label{eq:primitive-d-def-app}
\end{equation}
we can rewrite Eqs.~\eqref{eq:primitive-t-sum-app} and~\eqref{eq:primitive-t-terms-app} as
\begin{equation}
\begin{aligned}
    K_{\mathrm{para}}^{\sigma\nu\lambda}(-\omega)
    =
    -\sum_{a,c}p_a
    \bigg[
    &
    \frac{
        D^{\sigma\nu}_{ac}(B_\lambda)_{ca}}
        {\eta-\ii(\omega_{ac}-\omega)}
    \\
    &-
    \frac{
        (B_\lambda)_{ac}D^{\sigma\nu}_{ca}}
        {\eta-\ii(\omega_{ca}-\omega)}
    \bigg].
\end{aligned}
    \label{eq:primitive-reduced-kernel-app}
\end{equation}

\subsection{The diamagnetic contribution}
\label{app:primitive-diamagnetic}

We now turn our attention to the diamagnetic response. Inspired by our separation of $C_\sigma$ into unperturbed, positive, and negative frequency components in Eq.~\eqref{eq:primitive-d-split-app}, we can introduce the positive and negative frequency components of the perturbing Hamiltonian,
\begin{equation}
\begin{aligned}
    H'(t)&=H'_{+}(t)+H'_{-}(t),
    \\
    H'_{+}(t)&=F^{\lambda}e^{(\eta-\ii\omega)t}B_\lambda,
    \\
    H'_{-}(t)&=F^{\lambda*}e^{(\eta+\ii\omega)t}B_\lambda.
\end{aligned}
    \label{eq:primitive-drive-split-app}
\end{equation}
The diamagnetic part of the rectified response has two contributions: the linear response of $\langle C^+_\sigma\rangle(t)$ to $H'_{-}$, and the complementary linear response of $\langle C^-_\sigma\rangle(t)$ to $H'_{+}$. Using the Kubo formula, we can thus write 
\begin{equation}
\begin{aligned}
    \langle C_\sigma\rangle^{\mathrm{dia}}_{\mathrm{rect}}\,e^{2\eta t}
    ={}&-\ii\int_{-\infty}^{t}\!\dd t'\,
    \bigl\langle\comm{(C^{+}_\sigma)(t)}{(H'_{-})(t')}\bigr\rangle_0
    \\
    &-\ii\int_{-\infty}^{t}\!\dd t'\,
    \bigl\langle\comm{(C^{-}_\sigma)(t)}{(H'_{+})(t')}\bigr\rangle_0 .
\end{aligned}
    \label{eq:primitive-dia-linear-response-app}
\end{equation}
Inserting a complete set of states to find the Lehmann representation of the Kubo formula, we have
\begin{equation}
\begin{aligned}
    \langle C_\sigma\rangle^{\mathrm{dia}}_{\mathrm{rect}}
    ={}&-\sum_{ac}(p_a-p_c)
    \Biggl[
    \frac{\comm{B_\sigma}{B_\lambda}_{ac}(B_\nu)_{ca}}
    {\eta-\ii(\omega_{ac}+\omega)}
    \\
    &\qquad
    +\frac{\comm{B_\sigma}{B_\nu}_{ac}(B_\lambda)_{ca}}
    {\eta-\ii(\omega_{ac}-\omega)}
    \Biggr]F^\nu F^{\lambda*}.
\end{aligned}
    \label{eq:primitive-dia-residue-app}
\end{equation}
Next, we can rewrite the four terms in Eq.~\eqref{eq:primitive-dia-residue-app} in terms of a single density matrix eigenvalue $p_a$ by relabeling dummy indices $a\leftrightarrow c$. Doing so, we can write $\langle C_\sigma\rangle^{\mathrm{dia}}_{\mathrm{rect}}$ in a form directly paralleling Eq.~\eqref{eq:primitive-dc-kernel-app},
\begin{equation}
\begin{aligned}
    \langle C_\sigma\rangle^{\mathrm{dia}}_{\mathrm{rect}}
    &=-\bigl[K_{\mathrm{dia}}^{\sigma\nu\lambda}(-\omega)
    +K_{\mathrm{dia}}^{\sigma\lambda\nu}(+\omega)\bigr]F^\nu F^{\lambda*},
\end{aligned}
\end{equation}
with
\begin{equation}
\begin{aligned}
    K_{\mathrm{dia}}^{\sigma\nu\lambda}(-\omega)
    &=\sum_{a,c}p_a
    \bigg[
    \frac{\comm{B_\sigma}{B_\nu}_{ac}(B_\lambda)_{ca}}
    {\eta-\ii(\omega_{ac}-\omega)}
    \\
    &\qquad\quad
    -\frac{(B_\lambda)_{ac}\comm{B_\sigma}{B_\nu}_{ca}}
    {\eta-\ii(\omega_{ca}-\omega)}
    \bigg].
    \end{aligned}
    \label{eq:primitive-dia-reduced-app}
\end{equation}
Comparing with Eq.~\eqref{eq:primitive-reduced-kernel-app}, we see that the diamagnetic and paramagnetic responses have similar form. Adding Eqs.~\eqref{eq:primitive-reduced-kernel-app} and \eqref{eq:primitive-dia-reduced-app}, we see that the commutators combine to yield for the full response
\begin{equation}
\begin{aligned}
    K^{\sigma\nu\lambda}(-\omega)
    =
    -\sum_{a,c}p_a
    \bigg[
    &
    \frac{
        \tilde{D}^{\sigma\nu}_{ac}(B_\lambda)_{ca}}
        {\eta-\ii(\omega_{ac}-\omega)}
    \\
    &-
    \frac{
        (B_\lambda)_{ac}\tilde{D}^{\sigma\nu}_{ca}}
        {\eta-\ii(\omega_{ca}-\omega)}
    \bigg],
\end{aligned}
    \label{eq:total-kernel-app}
\end{equation}
where $\tilde{D}^{\sigma\nu}$ is given by
\begin{equation}
\tilde{D}^{\sigma\nu} = - \comm{B_{\sigma,\mathrm{d}}}{B_\nu}.
\label{eq:primitive-dtilde-def-app}
\end{equation}

\subsection{The spectral density}
\label{app:primitive-spectral-density}

The simple pole structure of the response in Eq.~\eqref{eq:total-kernel-app} hints at the existence of a simpler spectral density representation. To make this manifest, let us first note from Eq.~\eqref{eq:total-kernel-app} that within a degenerate subspace $\omega_{ac}=0$ we have that
\begin{equation}
\begin{aligned}
    &\sum_{E_a=E_c} p_a
    \bigg[
    \frac{\tilde{D}^{\sigma\nu}_{ac}(B_\lambda)_{ca}}
    {\eta-\ii(\omega_{ac}-\omega)}
    -
    \frac{(B_\lambda)_{ac}\tilde{D}^{\sigma\nu}_{ca}}
    {\eta-\ii(\omega_{ca}-\omega)}
    \bigg]
    \\
    &=
    \sum_{E_a=E_c}
    \frac{p_a}{\eta+\ii\omega}
    \left[
    \tilde{D}^{\sigma\nu}_{ac}(B_\lambda)_{ca}
    -
    (B_\lambda)_{ac}\tilde{D}^{\sigma\nu}_{ca}
    \right]
    \\
    &=0,
\end{aligned}
\end{equation}
where we used the fact that $p_a=p_c$ if $E_a=E_c$ for a thermal state. This means that by collecting terms we can rewrite Eq.~\eqref{eq:total-kernel-app} as a restricted sum over states $a,c$ with $\omega_{ac}>0$. We find
\begin{equation}
\begin{aligned}
    &K^{\sigma\nu\lambda}(-\omega)
    \\
    &=
    -\sum_{\omega_{ac}>0}
    \Biggl\{
    p_a
    \bigg[
    \frac{\tilde{D}^{\sigma\nu}_{ac}(B_\lambda)_{ca}}
    {\eta-\ii(\omega_{ac}-\omega)}
    -
    \frac{(B_\lambda)_{ac}\tilde{D}^{\sigma\nu}_{ca}}
    {\eta-\ii(-\omega_{ac}-\omega)}
    \bigg]
    \\
    &\qquad
    +p_c
    \bigg[
    \frac{\tilde{D}^{\sigma\nu}_{ca}(B_\lambda)_{ac}}
    {\eta-\ii(-\omega_{ac}-\omega)}
    -
    \frac{(B_\lambda)_{ca}\tilde{D}^{\sigma\nu}_{ac}}
    {\eta-\ii(\omega_{ac}-\omega)}
    \bigg]
    \Biggr\}.
    \label{eq:primitive-pair-kernel-app}
\end{aligned}
\end{equation}
Next, we can collect the coefficients with the same denominators. Introducing the line weight (transition matrix element)
\begin{equation}
    w_{ac}^{\sigma;\nu\lambda}
    =
    \pi(p_a-p_c)\comm{B_\nu}{B_{\sigma,\mathrm d}}_{ac}(B_\lambda)_{ca},
    \label{eq:primitive-residue-r-app}
\end{equation}
we can re-express Eq.~\eqref{eq:primitive-pair-kernel-app} as 
\begin{equation}
\begin{aligned}
    &K^{\sigma\nu\lambda}(-\omega)
    \\
    &=
    -\frac{1}{\pi}
    \sum_{\omega_{ac}>0}
    \left[
    \frac{w_{ac}^{\sigma;\nu\lambda}}
    {\eta-\ii(\omega_{ac}-\omega)}
    +
    \frac{\bigl(w_{ac}^{\sigma;\nu\lambda}\bigr)^{*}}
    {\eta-\ii(-\omega_{ac}-\omega)}
    \right].
\end{aligned}
\end{equation}

Introducing the spectral density
\begin{equation}
\begin{aligned}
    \rho_K^{\sigma;\nu\lambda}(\varpi)
    &=
    \sum_{\omega_{ac}>0}
    \Bigl[
    w_{ac}^{\sigma;\nu\lambda}\,\delta(\varpi-\omega_{ac})
    \\
    &\qquad
    +
    \bigl(w_{ac}^{\sigma;\nu\lambda}\bigr)^{*}
    \delta(\varpi+\omega_{ac})
    \Bigr],
\end{aligned}
    \label{eq:primitive-rho-k-app}
\end{equation}
we can see by performing the explicit integration that
\begin{equation}
\begin{aligned}
    &-\frac{1}{\pi}
    \int_{-\infty}^{\infty}\!\dd\varpi\,
    \frac{\rho_K^{\sigma;\nu\lambda}(\varpi)}
    {\eta-\ii(\varpi-\omega)}
    \\
    &=
    -\frac{1}{\pi}
    \sum_{\omega_{ac}>0}
    \left[
    \frac{w_{ac}^{\sigma;\nu\lambda}}
    {\eta-\ii(\omega_{ac}-\omega)}
    +
    \frac{\bigl(w_{ac}^{\sigma;\nu\lambda}\bigr)^{*}}
    {\eta-\ii(-\omega_{ac}-\omega)}
    \right]
    \\
    &=
    K^{\sigma\nu\lambda}(-\omega).
\end{aligned}
    \label{eq:primitive-cauchy-app}
\end{equation}

Finally, we note that Eq.~\eqref{eq:primitive-residue-r-app} takes a particularly simple form if we work in a basis such that $B_\sigma$ is diagonal within every degenerate subspace of $H_0$. In such a basis, we can write
\begin{equation}
    \comm{B_\nu}{B_{\sigma,\mathrm d}}_{ac}
    =
    \bigl((B_\sigma)_{cc}-(B_\sigma)_{aa}\bigr)(B_\nu)_{ac}.
    \label{eq:primitive-d-split-ac-app}
\end{equation}
Inserting this into Eq.~\eqref{eq:primitive-residue-r-app} yields
\begin{equation}
\begin{aligned}
    w_{ac}^{\sigma;\nu\lambda}
    =
    {}&
    \pi(p_a-p_c)
    \bigl((B_\sigma)_{cc}-(B_\sigma)_{aa}\bigr)
    (B_\nu)_{ac}(B_\lambda)_{ca}.
\end{aligned}
    \label{eq:primitive-line-weight-app}
\end{equation}
This shows that in a basis adapted to $B_\sigma$, the line weight takes the form of a generalized shift vector. 

\subsection{The zero-temperature moment}
\label{app:primitive-zero-t}

Because the spectral density Eq.~\eqref{eq:primitive-rho-k-app} is given by a sum of delta functions, it satisfies a general sum rule
\begin{equation}
\begin{aligned}
\frac{1}{\pi}
    \int_0^\infty\!\dd\varpi\;
    \rho_K^{\sigma;\nu\lambda}(\varpi)
    &= \frac{1}{\pi} \sum_{\omega_{ac}>0} w_{ac}^{\sigma;\nu\lambda} \\
    &=\sum_{\omega_{ac}>0}(p_a-p_c)\comm{B_\nu}{B_{\sigma,\mathrm d}}_{ac}(B_\lambda)_{ca}.\label{eq:spectral-sum-general-T-app}
\end{aligned}
\end{equation}
In the limit of zero temperature, we can re-express this sum rule in terms of equilibrium and geometric properties of the system. Let us assume we have a gapped, nondegenerate ground state that we denote by $\ket{0}$.  Taking $T\rightarrow 0$ in Eq.~\eqref{eq:spectral-sum-general-T-app} gives $p_a=0, p_c=\delta_{c0}$ in the restricted sum, leaving
\begin{equation}
    \lim_{T\rightarrow 0}\frac{1}{\pi}
    \int_0^\infty\!\dd\varpi\;
    \rho_K^{\sigma;\nu\lambda}(\varpi)
    =
    -\sum_{a>0}
    \comm{B_\nu}{B_{\sigma,\mathrm d}}_{a0}(B_\lambda)_{0a}.
    \label{eq:primitive-zero-t-raw-app}
\end{equation}
To proceed further, we will work in a basis where $B_\sigma$ is diagonal in any degenerate subspace of $H_0$ (this will simplify the notation in the manipulations that follow, though our final result is basis-independent). In such a basis, we have 
\begin{equation}
\begin{aligned}
    &\lim_{T\rightarrow 0}\frac{1}{\pi}\int_0^\infty\!\dd\varpi\;\rho_K^{\sigma;\nu\lambda}(\varpi)
    \\
    &=
    \sum_{a>0}
    \bigl((B_\sigma)_{aa}-(B_\sigma)_{00}\bigr)
    (B_\lambda)_{0a}(B_\nu)_{a0}.
    \label{eq:primitive-zero-t-split-app}
\end{aligned}
\end{equation}

Now let us introduce the ground state cumulant
\begin{equation}
    c^{(3)}_{\lambda\sigma\nu}
    =
    \bra{0}
    \delta B_\lambda\,\delta B_\sigma\,\delta B_\nu
    \ket{0},
    \label{eq:primitive-c3-app}
\end{equation}
with $\delta B_\mu = B_\mu - (B_\mu)_{00}$. Writing out the matrix elements at zero temperature, we have in this basis that
\begin{equation}
\begin{aligned}
    c^{(3)}_{\lambda\sigma\nu}
    ={}&\sum_{ab}[(B_{\lambda})_{0a}-\delta_{a0}(B_\lambda)_{00}][(B_{\sigma})_{ab}-\delta_{ab}(B_\sigma)_{00}]
    \\
    &\quad\times[(B_{\nu})_{b0}-\delta_{b0}(B_\nu)_{00}]
    \\
    ={}&\sum_{a,b>0}(B_{\lambda})_{0a}[(B_{\sigma})_{ab}-\delta_{ab}(B_\sigma)_{00}](B_{\nu})_{b0}
    \\
    ={}&\sum_{a>0}
    \bigl((B_\sigma)_{aa}-(B_\sigma)_{00}\bigr)
    (B_\lambda)_{0a}(B_\nu)_{a0}
    \\
    &+
    \sum_{\substack{a,b>0\\a\neq b}}(B_\lambda)_{0a}(B_\sigma)_{ab}(B_\nu)_{b0}.
\end{aligned}
\end{equation}

Recalling our definition of the complex distortion tensor at zero temperature,
\begin{equation}
    S_{\lambda\sigma\nu}
    =
    \sideset{}{'}\sum_{a,b>0}
    (B_\lambda)_{0a}(B_\sigma)_{ab}(B_\nu)_{b0},
    \label{eq:primitive-s-loop-energy-app}
\end{equation}
from Eq.~\eqref{eq:ordered-s-main}, we see that we can write
\begin{equation}
\sum_{a>0}
    \bigl((B_\sigma)_{aa}-(B_\sigma)_{00}\bigr)
    (B_\lambda)_{0a}(B_\nu)_{a0} = c^{(3)}_{\lambda\sigma\nu} - S_{\lambda\sigma\nu}.
\end{equation}
Comparing with the right hand side of the sum rule Eq.~\eqref{eq:primitive-zero-t-split-app}, we find the geometric sum rule
\begin{equation}
    \lim_{T\rightarrow 0}\frac{1}{\pi}
    \int_0^\infty\!\dd\varpi\;
    \rho_K^{\sigma;\nu\lambda}(\varpi)
    =
    c^{(3)}_{\lambda\sigma\nu}
    -
    S_{\lambda\sigma\nu}.
    \label{eq:primitive-c3-sum-rule-app}
\end{equation}

\subsection{Pulse-induced displacement}
\label{app:primitive-displacement}

In this section, we compute the accumulated change in $\langle B_\sigma\rangle$ due to the perturbation $V(t)=f^\lambda(t)B_\lambda$ with $f(t)$ defined in Eq.~\eqref{eq:general-f-t}. We consider a pulse $f^\lambda(t)$ that turns off after some time $t_f$. Then we can work in the interaction picture, where time evolution of operators is governed by the unperturbed Hamiltonian, and the interaction picture density matrix $\tilde\rho(t)$ satisfies
\begin{equation}
\frac{d\tilde{\rho}}{dt} = -\ii \comm{\tilde{V}(t)}{\tilde \rho}. \label{eq:int_schr}
\end{equation}
In particular, this means that $\tilde{\rho}(t>t_f)=\tilde{\rho}(\infty)$ is constant. We can thus write
\begin{equation}
    \langle B_\sigma\rangle(t)
    =\sum_{m,n}\tilde\rho_{mn}(\infty)\,(B_\sigma)_{nm}\,e^{-\ii\omega_{mn}t},
    \qquad t>t_f .
    \label{eq:primitive-post-pulse-app}
\end{equation}
However, even at long times $\langle B_\sigma\rangle(t)$ will still have coherent oscillations. In order to extract the average accumulated change in $\langle B_\sigma\rangle$, we can take a time average,
\begin{equation}
    \Delta B_\sigma
    \equiv\lim_{\mathcal T\to\infty}\frac{1}{\mathcal T}
    \int_{t_f}^{t_f+\mathcal T}\!\!\dd t\,\langle B_\sigma\rangle(t)
    -\langle B_\sigma\rangle(-\infty).
    \label{eq:primitive-displacement-average-app}
\end{equation}
Using the fact that
\begin{equation}
    \frac{1}{\mathcal T}\int_{t_f}^{t_f+\mathcal T}\!\!\dd t\;e^{-\ii\omega_{mn}t}
    =
    \begin{cases}
    1, & \omega_{mn}=0,\\[4pt]
    e^{-\ii\omega_{mn}t_f}\,
    \frac{1-e^{-\ii\omega_{mn}\mathcal T}}{\ii\omega_{mn}\mathcal T},
    & \omega_{mn}\neq0 ,
    \end{cases}
    \label{eq:primitive-window-average-app}
\end{equation}
we see that energy off-diagonal terms in Eq.~\eqref{eq:primitive-displacement-average-app} average to zero as $\mathcal{T}\rightarrow\infty$, and we are left with
\begin{equation}
    \Delta B_\sigma=\sum_{E_m=E_n}
    \bigl[\tilde\rho_{mn}(\infty)-p_m\delta_{mn}\bigr](B_\sigma)_{nm}.
    \label{eq:primitive-displacement-pop-app}
\end{equation}
Working in the basis where $B_\sigma$ is diagonal in every degenerate subspace simplifies Eq.~\eqref{eq:primitive-displacement-pop-app} to
\begin{equation}
\Delta B_\sigma=\sum_{a}
    \bigl[\tilde\rho_{aa}(\infty)-p_a\bigr](B_\sigma)_{aa}.
    \label{eq:primitive-displacement-pop-app-nondegen}
\end{equation}

We can evaluate Eq.~\eqref{eq:primitive-displacement-pop-app} perturbatively order-by-order in $V(t)$. First, note that
\begin{equation}
    \tilde V_{mn}(t)=\sum_\lambda f^\lambda(t)\,(B_\lambda)_{mn}\,e^{\ii\omega_{mn}t}.
    \label{eq:primitive-vbar-app}
\end{equation}
We can integrate the interaction-picture Schr\"{o}dinger equation \eqref{eq:int_schr} order-by-order, writing
\begin{equation}
\tilde{\rho} = \rho_0 + \rho^{(1)}+\rho^{(2)}+\dots.
\end{equation}
To first order, we find after carrying out the time integral that
\begin{equation}
    \rho^{(1)}_{mn}(\infty)
    =-2\pi\ii\,(p_n-p_m)\sum_\lambda\widetilde F^\lambda(\omega_{mn})(B_\lambda)_{mn},
    \label{eq:primitive-first-order-app}
\end{equation}
where $\widetilde F^\lambda(\omega)= \widetilde F^{\lambda*}(-\omega)$ are the Fourier coefficients of $f(t)$ from Eq.~\eqref{eq:general-f-t}. Importantly, since $p_n=p_m$ whenever $E_n=E_m$ for a thermal density matrix, Eq.~\eqref{eq:primitive-first-order-app} does not contribute to the shift $\Delta B_\sigma$ in Eq.~\eqref{eq:primitive-displacement-pop-app}. The first contribution then comes at second order. We can write the second-order correction $\rho^{(2)}_{mn}(\infty)$ as
\begin{equation}
\begin{aligned}
    \rho^{(2)}_{mn}(\infty)
    =-\int_{-\infty}^{\infty}\!\dd t_1\int_{-\infty}^{t_1}\!\dd t_2\,
    \bigl[\tilde V(t_1),[\tilde V(t_2),\rho_0]\bigr]_{mn},
\end{aligned}
    \label{eq:primitive-tdpt-app}
\end{equation}
where the matrix elements of the double commutator are
\begin{equation}
\begin{aligned}
    &\bigl[\tilde V(t_1),[\tilde V(t_2),\rho_0]\bigr]_{mn}
    \\
    &=\sum_k\bigl[
    (p_n-p_k)\,\tilde V_{mk}(t_1)\tilde V_{kn}(t_2)
    \\
    &\qquad+(p_m-p_k)\,\tilde V_{mk}(t_2)\tilde V_{kn}(t_1)\bigr].
\end{aligned}
    \label{eq:primitive-tdpt-diag-app}
\end{equation}

To proceed, we examine the diagonal matrix elements in Eq.~\eqref{eq:primitive-tdpt-diag-app}. The population factors are identical and we can combine terms. Inserting this into Eq.~\eqref{eq:primitive-tdpt-app} and exchanging $t_1\leftrightarrow t_2$ in the second term removes the domain restriction on the time integral. We thus have
\begin{equation}
\begin{aligned}
    \rho^{(2)}_{aa}(\infty)
    =-\sum_k(p_a-p_k)
    &\int_{-\infty}^{\infty}\!\dd t_1\,\tilde V_{ak}(t_1)
    \\
    &\times\int_{-\infty}^{\infty}\!\dd t_2\,\tilde V_{ka}(t_2).
\end{aligned}
    \label{eq:primitive-tdpt-factor-app}
\end{equation}

Introducing the shorthand
\begin{equation}
    \int\dd t\,\tilde V_{mk}(t)
    =2\pi\sum_\lambda\widetilde F^\lambda(\omega_{mk})(B_\lambda)_{mk}
    \equiv\bigl(\tilde F\!\cdot\!B\bigr)_{mk},
    \label{eq:primitive-fweighted-app}
\end{equation}
we can insert Eq.~\eqref{eq:primitive-tdpt-factor-app} into Eq.~\eqref{eq:primitive-displacement-pop-app-nondegen} to find
\begin{equation}
    \Delta B_\sigma=\sum_{a\neq c}(p_c-p_a)\,(B_\sigma)_{aa}
    \Bigl|\bigl(\tilde F\!\cdot\!B\bigr)_{ac}\Bigr|^{2},
    \label{eq:primitive-displacement-pairsum-app}
\end{equation}
where we used the fact that $B$ is Hermitian and $\widetilde F^\lambda(-\omega) = \widetilde F^{\lambda*}(\omega)$.

Finally, we can reorganize terms to restrict the sum to positive $\omega_{ac}>0$. Using
\begin{equation}
\begin{aligned}
    &(p_c-p_a)(B_\sigma)_{aa}+(p_a-p_c)(B_\sigma)_{cc}
    \\
    &\qquad=(p_c-p_a)\bigl((B_\sigma)_{aa}-(B_\sigma)_{cc}\bigr),
\end{aligned}
    \label{eq:primitive-pair-combine-app}
\end{equation}
combined with the fact that $\Bigl|\bigl(\tilde F\!\cdot\!B\bigr)_{ac}\Bigr|^{2}$ is invariant under $a\leftrightarrow c$, we can write
\begin{equation}
\begin{aligned}
    \Delta B_\sigma=
    4\pi\sum_{\omega_{ac}>0}
    &\widetilde F^\nu(\omega_{ac})\,\widetilde F^{\lambda*}(\omega_{ac})\,
    w^{\sigma;\nu\lambda}_{ac}.
\end{aligned}
    \label{eq:primitive-displacement-weight-app}
\end{equation}
Although we derived Eq.~\eqref{eq:primitive-displacement-weight-app} in the basis where $B_\sigma$ is diagonal in every degenerate subspace, the final result is basis independent. We see that the total change $\Delta B_\sigma$ depends only on the line weight Eq.~\eqref{eq:primitive-line-weight-app}.

\section{Free fermion reduction formulas}\label{sec:free_fermion_reduction_formulas}

In this appendix, we will derive the free fermion, thermodynamic limit expressions for position operator matrix elements, line weights, the shift conductivity, and the geometric sum rule that appear in Sec.~\ref{sec:application_to_free_fermions}. First, in Sec.~\ref{app:ff-reduction} we derive the finite-sized matrix elements Eq.~\eqref{eq:ff-sc-main} for the position operator for a free fermion ground state. Next, in Sec.~\ref{app:ff-blount} we move to the thermodynamic limit and derive the free fermion line weight Eq.~\eqref{eq:ff-general-weight-main}. Lastly, in Sec.~\ref{app:ff-sum-rule} we derive the expressions for the free fermion cumulant density and complex distortion tensor that appear in the sum rule in Eq.~\eqref{eq:ff-sum-rule-main}.

\subsection{From many-body matrix elements to single-particle sums}\label{app:ff-reduction}

Our goal is to evaluate matrix elements of the position operator $X_\sigma$ between eigenstates of a free fermion Hamiltonian. To do so, we make use of Eq.~\eqref{eq:pos-1body}, which expresses the position operator in the basis of single-particle energy eigenstates. Introducing the shorthand $(X_\sigma)_{qp}=\bra{q}X_\sigma\ket{p}$, we can use $\{c_p,c^\dag_q\}=\delta_{pq}$ to derive that
\begin{equation}
\begin{aligned}
    \comm{X_\sigma}{c^\dagger_p}
    &=
    \sum_{q}(X_\sigma)_{qp}\,c^\dagger_q.
\end{aligned}
\end{equation}
Applying this commutation relation iteratively for a general $N$-particle eigenstate yields
\begin{equation}
\begin{aligned}
    X_\sigma\, c^\dagger_{p_1}\!\cdots c^\dagger_{p_N}\ket{0}
    &=
    \sum_{k=1}^{N}\sum_{q}
    (X_\sigma)_{q p_k}
    \\
    &\qquad\times
    c^\dagger_{p_1}\!\cdots c^\dagger_{q}\cdots c^\dagger_{p_N}\ket{0},
\end{aligned}
    \label{eq:ff-leibniz-app}
\end{equation}
where $\ket{0}$ is the vacuum state which is annihilated by $X_\sigma$.

Our key observation is that every many-body eigenstate of a free fermion Hamiltonian can be written in the form used in Eq.~\eqref{eq:ff-leibniz-app}, i.e. every eigenstate has the form
\begin{equation}
\ket{A} = \prod_{p\in A}c^\dag_p\ket{0}, \label{eq:ketA}
\end{equation}
with energy
\begin{equation}
E_A = \sum_{p\in A}\varepsilon_p.
\end{equation}
Since $X_\sigma$ is a one-body operator, it cannot have matrix elements between eigenstates that differ by more than a single particle-hole pair. Thus, for two eigenstates $\ket{A}$ and $\ket{B}$ we have from Eq.~\eqref{eq:ff-leibniz-app}
\begin{equation}
\begin{aligned}
    \langle A|X_\sigma|A\rangle
    &=
    \sum_{p\in A}(X_\sigma)_{pp},
    \\
    \langle A|X_\sigma|B\rangle
    &=\begin{cases}
    \pm(X_\sigma)_{pq}, &
    \ket{A}=\pm c^\dagger_p c_q\ket{B},
    \\
    0, & \text{otherwise}.
    \end{cases}
\end{aligned}
    \label{eq:ff-sc-general-app}
\end{equation}

Because we will be analyzing matrix elements of one-body operators, it will be helpful to analyze the decomposition of a general two-particle-hole pair state. In particular, let $\ket{A}$ be as in Eq.~\eqref{eq:ketA}. Then for $p\neq q, j\in A, b\not\in A$ we can use the canonical anticommutation relations to write
\begin{equation}
    c^\dagger_p c_q\,\bigl(c^\dagger_b c_j\ket{A}\bigr)
    =
    \delta_{qb}\,c^\dagger_p c_j\ket{A}
    -\delta_{pj}\,c^\dagger_b c_q\ket{A}
    +\ket{A'},
    \label{eq:ff-pair-product-app}
\end{equation}
where $\ket{A'}$ is a state with two particle-hole pairs. Equation~\eqref{eq:ff-pair-product-app} will allow us to immediately evaluate matrix elements of one-body operators between states that differ by at most two particle-hole pair excitations.
Our main object of interest will be the action of products of position operators on free fermion eigenstates, including the $T=0$ ground state. Since $X_\mu$ is a one-body operator, acting with $X_\mu$ on $\ket{A}$ creates a superposition of particle-hole pairs. As in Sec.~\ref{sub:from_many_body_operators_to_single_particle_matrix_elements}, let us denote by
\begin{equation}
\ket{ai^{-1}} = c^\dag_a c_i\ket{A}
\end{equation}
the state with a single particle-hole pair ($a\not\in A, i\in A$). For the purposes of evaluating response functions and geometric objects, we will need to evaluate averages of products of at most three position operators. As such, we will need
\begin{equation}
\begin{aligned}
    &\langle a\,i^{-1}|X_\sigma|b\,j^{-1}\rangle
    =
    \sum_{p,q}
    (X_\sigma)_{pq}
    \langle a\,i^{-1}|
    c_p^\dagger c_q^{\phantom\dagger}
    c_b^\dagger c_j^{\phantom\dagger}
    |A\rangle. \label{eq:ff-ph-matrix-elem}
\end{aligned}
\end{equation}
To simplify Eq.~\eqref{eq:ff-ph-matrix-elem}, we can use Eq.~\eqref{eq:ff-pair-product-app} to write
\begin{equation}
\begin{aligned}
    &\langle a\,i^{-1}|X_\sigma|b\,j^{-1}\rangle \\
    &=
    \sum_{p\neq q}\bra{ai^{-1}}(X_\sigma)_{pq}\left(\delta_{qb}\ket{pj^{-1}} - \delta_{pj}\ket{bq^{-1}}  \right) \\
    &\quad+
    \delta_{ab}\delta_{ij}
    \bigl[\bra{A}X_\sigma\ket{A}
    +(X_\sigma)_{aa}-(X_\sigma)_{ii}\bigr] \\
    &=
    \delta_{ij}\,\mathfrak r^\sigma_{ab}
    -
    \delta_{ab}\,\mathfrak r^\sigma_{ji}
    \\
    &\qquad+
    \delta_{ab}\delta_{ij}
    \bigl[\bra{A}X_\sigma\ket{A}
    +(X_\sigma)_{aa}-(X_\sigma)_{ii}\bigr],
\end{aligned}
    \label{eq:ff-sc-app}    
\end{equation}
where we used Eq.~\eqref{eq:ff-sc-general-app} to evaluate the diagonal $p=q$ elements, and we extend $\mathfrak{r}^\sigma_{aa}\equiv 0$ for diagonal entries. Equation~\eqref{eq:ff-sc-app} is valid for any finite-sized free fermion state $\ket{A}$. In particular, taking $\ket{A}=\ket{\mathrm{FS}}$, the zero-temperature ground state, yields Eq.~\eqref{eq:ff-sc-main}. While the $\mathfrak{r}$-dependent terms also exist in the thermodynamic limit, care must be taken to properly handle the diagonal matrix elements in infinite systems. To do so, we will now examine the matrix elements of the position operator for Bloch states in the thermodynamic limit.  

\subsection{The thermodynamic limit and covariant derivatives}\label{app:ff-blount}

In this section, we will use Eq.~\eqref{eq:blount-position-main} for the single-particle matrix elements of the position operator in order to derive Eq.~\eqref{eq:ff-general-weight-main} for the free fermion line weight entering the shift conductivity. First, let us examine the general form Eq.~\eqref{eq:obc-shift-weight-main} for the line weight. We see that the fundamental objects we need are matrix elements of $X_\mu$ (which we analyzed in Sec.~\ref{sub:thermodynamic_limit_blount}) as well as matrix elements of $\comm{X_\nu}{X_{\sigma,\mathrm{d}}}$, where $X_{\sigma,\mathrm{d}}$ is the energy-diagonal part of the position operator. Defining $X_{\sigma,\mathrm{d}}$ directly in the thermodynamic limit requires some care. However, we can sidestep this issue in evaluating the commutator. First, let us recall from Eq.~\eqref{eq:r-def} the energy-off-diagonal component of $X_\sigma$,
\begin{equation}
\begin{aligned}
    X_{\sigma,\mathrm{od}}
    &=
    X_\sigma - X_{\sigma,\mathrm d}.
\end{aligned}
\end{equation}
Using Eq.~\eqref{eq:r-def}, the matrix elements of $X_{\sigma,\mathrm{od}}$ between single-particle states are well defined and given in our continuum normalization convention by
\begin{equation}
\begin{aligned}
    &\langle n\kk|X_{\sigma,\mathrm{od}}|m\kk'\rangle
    =\frac{(2\pi)^d}{\mathfrak{v}}\,(x_{\mathrm{od}})^\sigma_{nm}(\kk)\,\delta(\kk-\kk'),
    \\
    &(x_{\mathrm{od}})^\sigma_{nm}(\kk)
    =\begin{cases}
    \mathfrak r^\sigma_{nm}(\kk), & \varepsilon_n(\kk)\neq \varepsilon_m(\kk) \\
    0, & \text{otherwise}
    \end{cases}.\label{eq:xod-matrix-elems}
\end{aligned}
\end{equation}
The key observation is that, on the complete Hilbert space, $\comm{X_\nu}{X_\sigma}=0$. Practically speaking, this is the constraint that the total Berry curvature of all bands vanish for the low-energy effective Wannier model used to approximate the physics. Using $\comm{X_\nu}{X_\sigma}=0$ lets us write
\begin{equation}
\begin{aligned}
    \comm{X_\nu}{X_{\sigma,\mathrm d}}
    &=
    -\comm{X_\nu}{X_{\sigma,\mathrm{od}}}.
\end{aligned}
    \label{eq:blount-od-split-app}
\end{equation}
Since $X_{\sigma,\mathrm{od}}$ is a one-body, $\kk$-diagonal operator, we can use Eq.~\eqref{eq:cov-deriv-def} to directly evaluate $\comm{X_\nu}{X_{\sigma,\mathrm{od}}}$. We thus have
\begin{equation}
\langle a\kk|\comm{X_\nu}{X_{\sigma,\mathrm d}}|c\kk'\rangle
    = -\frac{(2\pi)^d}{\mathfrak{v}}\,\ii\,(D_\nu x^\sigma_{\mathrm{od}})_{ac}\,\delta(\kk-\kk'). \label{eq:xxd-comm-step-1}
\end{equation}
This shows that $\comm{X_\nu}{X_{\sigma,\mathrm d}}$ is a well-defined $\kk$-diagonal one-body operator regardless of boundary conditions.

We can make use of the vanishing commutator of position operators to further reduce Eq.~\eqref{eq:xxd-comm-step-1}. In particular, the matrix elements of the commutator of position operators are the non-Abelian Berry curvature, and so
\begin{equation}
\begin{aligned}
0&=\bra{n\kk}\comm{X_\nu}{X_\sigma}\ket{m\kk'} \\
&=\ii\frac{(2\pi)^d}{\mathfrak{v}}\delta(\kk-\kk')\left[\partial_\nu A_\sigma - \partial_\sigma A_\nu-\ii\comm{A_\nu}{A_\sigma}\right]_{nm}.
\end{aligned}
\end{equation}  
Taking off-diagonal matrix elements $n=a, m=c, \varepsilon_a(\kk)\neq \varepsilon_c(\kk)$, and using Eq.~\eqref{eq:r-from-A}, we have~\cite{aversa1995nonlinear}
\begin{equation}
\begin{aligned}
(D_\sigma x^\nu_{\mathrm{od}})_{ac} - (D_\nu x^\sigma_{\mathrm{od}})_{ac}
    &=
    \ii\,\comm{x^\nu_{\mathrm{od}}}{x^\sigma_{\mathrm{od}}}_{ac}.
\end{aligned}
    \label{eq:blount-curl-app}
\end{equation}
Combining the identity Eq.~\eqref{eq:blount-curl-app} with our expression Eq.~\eqref{eq:blount-od-split-app} yields
\begin{equation}
\begin{aligned}
    \langle a\kk|\comm{X_\nu}{X_{\sigma,\mathrm d}}|c\kk'\rangle
    &=
    -\frac{(2\pi)^d}{\mathfrak{v}}\,\ii\,(D_\nu x^\sigma_{\mathrm{od}})_{ac}\,\delta(\kk-\kk')
    \\
    &=
    -\frac{(2\pi)^d}{\mathfrak{v}}\Bigl[
    \ii\,(D_\sigma x^\nu_{\mathrm{od}})_{ac}
    \\
    &\qquad
    +\comm{x^\nu_{\mathrm{od}}}{x^\sigma_{\mathrm{od}}}_{ac}
    \Bigr]\,\delta(\kk-\kk')
    \\
    &=
    -\frac{(2\pi)^d}{\mathfrak{v}}\,\ii\,(\mathfrak r^\nu_{\;;\sigma})_{ac}\,\delta(\kk-\kk'),
\end{aligned}
    \label{eq:blount-commutator-app}
\end{equation}
for $\varepsilon_a(\kk)\neq\varepsilon_c(\kk)$, where we have used the shorthand of Eq.~\eqref{eq:ff-covariant-r-main}. 

Finally, we can insert Eqs.~\eqref{eq:xod-matrix-elems} and \eqref{eq:blount-commutator-app} directly into the first equality of Eq.~\eqref{eq:obc-shift-weight-main}. Using the fact that both $X_\lambda$ and $\comm{X_\nu}{X_{\sigma,\mathrm{d}}}$ are well-defined one-body operators, we find
\begin{equation}
\begin{aligned}
    w_{ac}^{\sigma;\nu\lambda}(\kk)
    &=
    \pi
    \mathfrak{v}\!\int_{\mathrm{BZ}}\!\frac{\dd^d k'}{(2\pi)^d}\bigl[f_a(\kk)-f_c(\kk')\bigr]
    \\
    &\qquad\times
    \langle a\kk|\comm{X_\nu}{X_{\sigma,\mathrm d}}|c\kk'\rangle\,
    \mathfrak r^\lambda_{ca}(\kk')
    \\
    &=
    -\ii\pi\bigl[f_a(\kk)-f_c(\kk)\bigr]\,
    (\mathfrak r^\nu_{\;;\sigma})_{ac}(\kk)\,\mathfrak r^\lambda_{ca}(\kk),
\end{aligned}
    \label{eq:blount-weight-app}
\end{equation}
which is Eq.~\eqref{eq:ff-general-weight-main}.

\subsection{The cumulant, complex distortion tensor, and the sum rule}\label{app:ff-sum-rule}

In this section, we derive the free fermion expression for the cumulant density Eq.~\eqref{eq:ff-c3-main} at zero temperature. We start by considering a large but finite system. To begin, let us first separate the variation $\delta X_\mu$ into its diagonal and off-diagonal parts,
\begin{equation}
\delta X_\mu = X_{\mu,\mathrm{od}} + (X_{\mu,\mathrm{d}}-\langle X_\mu\rangle).\label{eq:dx-split-app}
\end{equation}
Now, since $\ket{\mathrm{FS}}$ is the gapped ground state of our Hamiltonian, it must be an eigenstate of $X_{\mu,\mathrm{d}}$; that is
\begin{equation}
\begin{aligned}
    X_{\sigma,\mathrm d}|\mathrm{FS}\rangle
    &=
    \langle X_\sigma\rangle|\mathrm{FS}\rangle. \label{eq:xd-eigenstate-app}
\end{aligned}
\end{equation}
Combining Eqs.~\eqref{eq:dx-split-app} and \eqref{eq:xd-eigenstate-app}, we find that the action of $\delta X_\mu$ on the ground state $\ket{\mathrm{FS}}$ is fully determined by the matrix elements of $X_{\mu,\mathrm{od}}$,
\begin{equation}
\begin{aligned}
    \delta X_\sigma|\mathrm{FS}\rangle
    &=
    X_{\sigma,\mathrm{od}}|\mathrm{FS}\rangle
    \\
    &=
    \sum_{i\in\mathrm{occ}}\sum_{a\in\mathrm{unocc}}
    \mathfrak r^\sigma_{ai}\,
    c_a^\dagger c_i|\mathrm{FS}\rangle .
\end{aligned}
    \label{eq:ff-first-slot-app}
\end{equation}

Now let us turn to the full cumulant density $c^{(3)}_{\lambda\sigma\nu}/V =\bra{\mathrm{FS}}
    \delta X_\lambda\,\delta X_\sigma\,\delta X_\nu
    \ket{\mathrm{FS}}/V$. Using Eq.~\eqref{eq:ff-first-slot-app} for the two position operators acting on $\ket{\mathrm{FS}}$ gives
\begin{equation}
\begin{aligned}
    c^{(3)}_{\lambda\sigma\nu}/V
    ={}&
    \frac{1}{V}\langle\mathrm{FS}|X_{\lambda,\mathrm{od}}\delta X_{\sigma}X_{\nu,\mathrm{od}}|\mathrm{FS}\rangle
    \\
    &=
    \frac{1}{V}\Big(\langle\mathrm{FS}|X_{\lambda,\mathrm{od}}X_{\sigma,\mathrm{od}}X_{\nu,\mathrm{od}}|\mathrm{FS}\rangle \\
    &+\langle\mathrm{FS}|X_{\lambda,\mathrm{od}}(X_{\sigma,\mathrm{d}}-\langle X_\sigma\rangle){X_{\nu,\mathrm{od}}}|\mathrm{FS}\rangle\Big) \\
    &=\frac{1}{V}\Big(\langle\mathrm{FS}|X_{\lambda,\mathrm{od}}X_{\sigma,\mathrm{od}}X_{\nu,\mathrm{od}}|\mathrm{FS}\rangle \\
    &+\langle\mathrm{FS}|X_{\lambda,\mathrm{od}}\comm{X_{\sigma,\mathrm d}}{X_{\nu,\mathrm{od}}}|\mathrm{FS}\rangle\Big),
\end{aligned}
    \label{eq:ff-c3-od-split-app}
\end{equation}
where in the last equality we used Eq.~\eqref{eq:xd-eigenstate-app} to replace $\langle X_\sigma \rangle$ with the action of $X_{\sigma,\mathrm{d}}$ on $\ket{\mathrm{FS}}$.

In deriving Eq.~\eqref{eq:ff-c3-od-split-app}, we have managed to express the cumulant in a way where diagonal entries of the position operator appear only through a commutator, which can be evaluated in the thermodynamic limit. In particular, writing $X_{\sigma,\mathrm{d}}= X_\sigma - X_{\sigma,\mathrm{od}}$, we find from Eqs.~\eqref{eq:cov-deriv-def}, \eqref{eq:r-def}, and \eqref{eq:ff-covariant-r-main} that
\begin{equation}
\begin{aligned}
    \langle a\kk|\comm{X_{\sigma,\mathrm d}}{X_{\nu,\mathrm{od}}}|i\kk'\rangle
    &=
    \langle a\kk|\comm{X_\sigma}{X_{\nu,\mathrm{od}}}|i\kk'\rangle
    \\
    &\quad
    -\frac{(2\pi)^d}{\mathfrak{v}}\,
    \comm{x^\sigma_{\mathrm{od}}}{x^\nu_{\mathrm{od}}}_{ai}\,
    \delta(\kk-\kk')
    \\
    &=
    \frac{(2\pi)^d}{\mathfrak{v}}
    \Bigl[
    \ii\,(D_\sigma x^\nu_{\mathrm{od}})_{ai}
    \\
    &\qquad
    -\comm{x^\sigma_{\mathrm{od}}}{x^\nu_{\mathrm{od}}}_{ai}
    \Bigr]\,\delta(\kk-\kk')
    \\
    &=
    \frac{(2\pi)^d}{\mathfrak{v}}\,
    \ii\,(\mathfrak r^\nu_{\;;\sigma})_{ai}\,
    \delta(\kk-\kk').
\end{aligned}
    \label{eq:ff-commutator-covariant-app}
\end{equation}
This shows that $c^{(3)}_{\lambda\sigma\nu}$ can be expressed entirely in terms of matrix elements of well-defined operators. Inserting Eq.~\eqref{eq:ff-commutator-covariant-app} into Eq.~\eqref{eq:ff-c3-od-split-app} and using the matrix elements Eqs.~\eqref{eq:ff-ph-matrix-elem} and \eqref{eq:r-def}, we get Eq.~\eqref{eq:ff-c3-main}.

We note that the same matrix elements of $X_{\mu,\mathrm{od}}$ appear in the nonzero-temperature cQAC tensor in Eq.~\eqref{eq:position-cubic-main}. Since Eq.~\eqref{eq:ff-ph-matrix-elem} is valid for any eigenstate, we can use it to reduce the cQAC tensor to products of single-particle matrix elements. Using the factorization of single-particle thermal probabilities, we can combine Eqs.~\eqref{eq:position-cubic-main}, \eqref{eq:ff-ph-matrix-elem} and \eqref{eq:r-def} to obtain Eq.~\eqref{eq:ff-z-main}.

To conclude, we can also make contact with Refs.~\cite{avdoshkin2023extrinsic,avdoshkin2024multi,mitscherling2024gauge,patankar2018resonance} and rewrite Eq.~\eqref{eq:ff-c3-main} in terms of gauge-invariant traces of projectors. To do so, let us introduce the single-particle projector
\begin{equation}
P(\kk) = \sum_{i\in\mathrm{occ}}\ket{u_i(\kk)}\bra{u_i(\kk)}
\end{equation}
onto occupied states. Using the definitions Eqs.~\eqref{eq:berrydef} and \eqref{eq:r-from-A} of the Berry connection and dipole matrix elements, we can write the matrix elements of the first derivative of $P(\kk)$ between Bloch states as
\begin{equation}
    (\partial_\mu P)_{nm}
    =\begin{cases}
\ii\mathfrak r^\mu_{nm}, & n\in\mathrm{occ}, m\in\mathrm{unocc} \\
-\ii\mathfrak r^\mu_{nm}, & m\in\mathrm{occ}, n\in\mathrm{unocc} \\
0, & \text{otherwise}
    \end{cases}.
    \label{eq:ff-proj-dict-app}
\end{equation}
Taking another derivative of Eq.~\eqref{eq:ff-proj-dict-app} and using the chain rule, we can also find the matrix elements of the second derivatives $\partial_\sigma\partial_\nu P(\kk)$. In particular, focusing on the off-diagonal matrix elements with $a$ unoccupied and $i$ occupied, we have
\begin{equation}
\begin{aligned}
    (\partial_\sigma\partial_\nu P)_{ai}={}& \partial_\sigma\left[\left(\partial_\nu P\right)_{ai}\right]-\bra{\partial_\sigma u_a}\partial_\nu P\ket{u_i} \\
    &- \bra{u_a}\partial_\nu P\ket{\partial_\sigma u_i} \\
    ={}&
    -\ii\,\partial_\sigma\mathfrak r^\nu_{ai}
    -\sum_{b\in\mathrm{unocc}}(A_\sigma)_{ab}\,\mathfrak r^\nu_{bi}
    \\
    &+\sum_{j\in\mathrm{occ}}\mathfrak r^\nu_{aj}\,(A_\sigma)_{ji}.\label{eq:ff-proj-d2P-app-step-1}
\end{aligned}
\end{equation}
To further simplify Eq.~\eqref{eq:ff-proj-d2P-app-step-1}, it is simplest to work in a basis where $A_\sigma$ is diagonal within degenerate subspaces (as usual, our final results will be gauge-covariant and thus independent of this basis choice). We can then separate the sums over $A_\sigma$ into their diagonal and off-diagonal parts. Using Eqs.~\eqref{eq:r-semicolon-simplified} and \eqref{eq:xod-matrix-elems} we find
\begin{equation}
\begin{aligned}
    (\partial_\sigma\partial_\nu P)_{ai}={}&
    -\ii\,(\mathfrak r^\nu_{\;;\sigma})_{ai}
    -\sum_{b\in\mathrm{unocc}}(x^\sigma_{\mathrm{od}})_{ab}\,\mathfrak r^\nu_{bi}
    \\
    &+\sum_{j\in\mathrm{occ}}\mathfrak r^\nu_{aj}\,(x^\sigma_{\mathrm{od}})_{ji}.
\end{aligned}
    \label{eq:ff-proj-d2P-app}
\end{equation}

We can now combine Eqs.~\eqref{eq:ff-proj-dict-app} and \eqref{eq:ff-proj-d2P-app} to find an expression for the skewness. In particular, taking the trace over occupied states at a fixed $\kk$ (which we denote by $\operatorname{tr}$), we find
\begin{equation}
\begin{aligned}
    \ii\operatorname{tr}\!\left[P\,(\partial_\lambda P)(\partial_\sigma\partial_\nu P)\right]
    ={}&
    \sum_{i\in\mathrm{occ}}\sum_{a,b\in\mathrm{unocc}}
    \mathfrak r^\lambda_{ia}(x^\sigma_{\mathrm{od}})_{ab}\mathfrak r^\nu_{bi}
    \\
    &-
    \sum_{a\in\mathrm{unocc}}\sum_{i,j\in\mathrm{occ}}
    \mathfrak r^\lambda_{ia}(x^\sigma_{\mathrm{od}})_{ji}\mathfrak r^\nu_{aj}
    \\
    &+
    \ii\sum_{i\in\mathrm{occ}}\sum_{a\in\mathrm{unocc}}
    \mathfrak r^\lambda_{ia}\,
    (\mathfrak r^\nu_{\;;\sigma})_{ai},
\end{aligned}
    \label{eq:ff-proj-pointwise-app}
\end{equation}
where we have used the fact that $\partial_\mu P$ has vanishing matrix elements between occupied states. If we replace the sums over matrix elements of $x^\sigma_{\mathrm{od}}$ with primed (coincident energy terms excluded) sums over $\mathfrak{r}^\sigma$, we see that the right hand side of Eq.~\eqref{eq:ff-proj-pointwise-app} coincides with the integrand in our expression Eq.~\eqref{eq:ff-c3-main} for the skewness. We thus have
\begin{equation}
    \frac{1}{V}\,c^{(3),\mathrm{ff}}_{\lambda\sigma\nu}
    =
    -\int_{\mathrm{BZ}}\!\frac{\dd^d k}{(2\pi)^d}\,
    \operatorname{Im}\operatorname{tr}\!\left[P\,(\partial_\lambda P)(\partial_\sigma\partial_\nu P)\right],
    \label{eq:ff-c3-projector-app}
\end{equation}
which agrees with Ref.~\cite{avdoshkin2024multi}. Equation~\eqref{eq:ff-c3-projector-app} is directly amenable to numerical calculations, since it is written in terms of gauge-invariant projectors. We use Eq.~\eqref{eq:ff-c3-projector-app} to compute the skewness for the Kane-Mele model in Sec.~\ref{sub:example_the_four_band_kane_mele_model}. 

\bibliography{refs-rectification}

\end{document}